\documentclass[a4paper,fleqn,usenatbib]{mnras}

\usepackage[T1]{fontenc}
\usepackage{ae,aecompl}

\usepackage{graphicx}	
\usepackage{amsmath}	

\usepackage{amssymb}	
\usepackage{cuted}
\usepackage{lipsum}
\usepackage{bm}
\usepackage{lmodern}
\usepackage{pdflscape}
\usepackage{rotating}
\usepackage{adjustbox}
\usepackage{capt-of}
\immediate\write18{cp `kpsewhich apacite.bst` .}
\let\oldhref\href
\renewcommand{\href}[2]{\oldhref{#1}{\hbox{#2}}}

\usepackage{color}
\definecolor{orange}{rgb}{1,0.5,0}
\newcommand{\red}[1]{\textcolor{red}{{#1}}}

\title[Tianlai Dish Pathfinder Array]{The Tianlai Dish Pathfinder Array: design, operation and performance of a prototype transit radio interferometer}

\author[Fengquan Wu et al.]{
Fengquan Wu$^{1}$,
Jixia Li$^{1,2}$,
Shifan Zuo$^{1,2,3}$,
Xuelei Chen$^{1,2,4}$, 
Santanu Das$^{5,6}$,
\newauthor
John P. Marriner$^{5}$,
Trevor M. Oxholm$^{6}$, 
Anh Phan$^{6}$,
Albert Stebbins$^{5}$.
\newauthor
Peter T. Timbie$^{6}$\thanks{E-mail: pttimbie@wisc.edu},
Reza Ansari$^{7}$,
Jean-Eric Campagne$^{7}$,
Zhiping Chen$^{8}$,
\newauthor
Yanping Cong$^{1,2}$,
Qizhi Huang$^{1,2}$,
Juhun Kwak$^{6}$,
Yichao Li$^{9}$,
Tao Liu $^{8}$,
Yingfeng Liu$^{1,2}$,
\newauthor
Chenhui Niu$^1$,
Calvin Osinga$^{6}$,
Olivier Perdereau$^{7}$,
Jeffrey B. Peterson$^{10}$,
\newauthor
John Podczerwinski$^{6}$,
Huli Shi$^{1}$,
Gage Siebert$^{6}$,
Shijie Sun$^{1,2}$,
Haijun Tian$^{11}$,
\newauthor
Gregory S. Tucker$^{12}$,
Qunxiong Wang$^{11}$,
Rongli Wang$^{8}$,
Yougang Wang$^{1}$,
Yanlin Wu$^{6}$,
\newauthor
Yidong Xu$^{1}$,
Kaifeng Yu$^{1,2}$,
Zijie Yu$^{1,2}$,
Jiao Zhang$^{13}$,
Juyong Zhang$^{8}$,
\newauthor
Jialu Zhu$^{8}$
\\
$^{1}$ National Astronomical Observatory, Chinese Academy of Sciences, 20A Datun Road, Beijing 100101, P. R. China\\
$^{2}$ University of Chinese Academy of Sciences, Beijing 100049, P. R. China\\
$^{3}$ Department of Astronomy and Tsinghua Center for Astrophysics, Tsinghua University, Beijing 100084, P.R.China\\
$^{4}$ Center of High Energy Physics, Peking University, Beijing 100871, P. R. China\\
$^{5}$ Fermi National Accelerator Laboratory, P.O. Box 500, Batavia IL 60510-5011, USA\\
$^{6}$ Department of Physics, University of Wisconsin Madison, 1150 University Ave, Madison WI 53703, USA\\
$^{7}$ Universit\'e Paris-Saclay, CNRS/IN2P3, IJCLab, 91405 Orsay, France\\
$^{8}$ Hangzhou Dianzi University, 115 Wenyi Rd., Hangzhou 310018, P. R. China\\
$^{9}$ Department of Physics and Astronomy, University of the Western Cape, Robert Sobukwe Road, Belville 7535,\\
Republic of South Africa\\
$^{10}$ Department of Physics, Carnegie Mellon University, 5000 Forbes Avenue, Pittsburgh, PA 15213, USA\\
$^{11}$ China Three Gorges University, Yichang 443002, P. R. China\\
$^{12}$ Department of Physics, Brown University, 182 Hope St., Providence, RI 02912, USA\\
$^{13}$ College of Physics and Electronic Engineering, Shanxi University, Taiyuan, Shanxi 030006, P. R. China
}

\date{Accepted XXX. Received YYY; in original form ZZZ}

\usepackage{}
\pubyear{2020}

\begin{document}
\maketitle

\begin{abstract}
The Tianlai Dish Pathfinder Array is a radio interferometer designed to test techniques for 21~cm intensity mapping in the post-reionization universe as a means for measuring large-scale cosmic structure. It performs drift scans of the sky at constant declination. We describe the design, calibration, noise level, and stability of this instrument based on the analysis of about $\sim 5 \%$ of 6,200 hours of on-sky observations through October, 2019. Beam pattern determinations using drones and the transit of bright sources are in good agreement, and compatible with electromagnetic simulations. Combining all the baselines, we make maps around bright sources and show that the array behaves as expected. A few hundred hours of observations at different declinations have been used to study the array geometry and pointing imperfections, as well as the instrument noise behaviour. We show that the system  temperature is below 80~K for most feed antennas, and that noise fluctuations decrease as expected with integration time, at least up to a few hundred seconds. Analysis of long integrations, from 10 nights of observations of the North Celestial Pole, yielded visibilities with amplitudes of 20-30~mK, consistent with the expected signal from the NCP radio sky with $<10\,$mK precision for $1 ~\mathrm{MHz} \times 1~ \mathrm{min}$ binning.  Hi-pass filtering the spectra to remove smooth spectrum signal yields a residual consistent with zero signal at the $0.5\,$mK level.
\end{abstract}

\begin{keywords}
galaxies: evolution -- large-scale structure -- 21-cm
\end{keywords}



\section{Introduction} 
\label{sec:introduction}
This paper describes the first astronomical observations by the Tianlai Dish Pathfinder Array.  The instrument is an array of 16, 6-meter, on-axis dish antennas operated as a radio interferometer and is co-located with the Tianlai Cylinder Pathfinder Array, an interferometric array of 3 cylinder reflectors in Xinjiang, China \citep{Tianlai, Tianlai2021}. These complementary designs were chosen specifically for testing approaches to 21~cm intensity mapping. Both arrays saw first light in 2016. 

21~cm intensity mapping is a technique for measuring the large scale structure of the universe using the redshifted 21~cm line from neutral hydrogen gas (HI) \citep{Liu&Shaw2020,Morales&Wyithe2010}. It is an example of the general case of line intensity mapping \citep{Kovetz2019}, in which spectral lines from any species, such as CO and CII, are used to make three-dimensional, ``tomographic'' maps of large cosmic volumes.  21~cm intensity mapping is used to study the formation of the first objects during the Cosmic Dawn and the Epoch of Reionization ($6\lesssim z \lesssim 50$) and for addressing other cosmological questions with observations in the post-reionization epoch ($z \lesssim 6$), such as constraining inflation models \citep{Xu2016} and the equation of state of dark energy \citep{Xu2015}. In the latter epoch, the approach provides an attractive alternative to galaxy redshift surveys.  It measures the collective emission from many haloes simultaneously, both bright and faint, rather than cataloging just the brightest objects. As a result, the required angular resolution is relaxed as individual galaxies do not need to be resolved. By observing with wide-band receivers one simultaneously obtains signals over a range of redshifts and can construct a tomographic map.  The primary analysis tool for cosmological measurements is the three-dimensional power spectrum of the underlying dark matter, and intensity mapping provides a natural means to compute this spectrum over a range of wave numbers, $k$, in which the perturbations are in the linear regime. Of particular interest in the power spectrum are the baryon acoustic oscillation (BAO) features, which can be used as a cosmic ruler for studying the expansion rate of the universe as a function of redshift.

So far, the 21~cm signal has been detected with intensity mapping by two instruments:  the Green Bank Telescope (GBT) \citep{Masui2013,Switzer2013} and the Parkes Observatory \citep{Anderson2018} by cross-correlating intensity maps with galaxy redshift surveys.

While HI intensity mapping is being used out to a redshift of $\sim 50$ to study the EoR and Cosmic Dawn by a variety of instruments, including LOFAR \citep{LOFAR2013}, MWA \citep{MWA2013}, HERA \citep{HERA2017}, PAPER \citep{PAPER2010}, and LWA \citep{LWA2018}, this paper focuses on measurements of the post-reionization epoch.  Several dedicated instruments have been constructed, or are under development, to detect the 21~cm signal from this epoch using intensity mapping, ultimately without the need for cross-correlation with other surveys: Tianlai \citep{2012IJMPS..12..256C,Tianlai,Li2020a,Li2020b}, CHIME \citep{CHIME}, HIRAX \citep{HIRAX}, BINGO \citep{BINGO} and OWFA \citep{OWFA}. Other instruments being designed and built to test the technique include BMX \citep{BMX2020} and PAON-4 \citep{PAON4_Zhang_2016, Ansari2019}. These 21~cm instruments have several features in common: with the exception of BINGO, they are all interferometers, achieving modest angular resolution at modest cost;  they have large numbers of receivers in order to provide enough mapping speed to detect the faint 21~cm signal; and the arrays are laid out in a compact arrangement in order to provide sensitivity at the relatively large scales ($0.5\gtrsim k \gtrsim 0.05$) where the BAO features appear in the power spectrum.

Although the 21~cm intensity mapping approach has been used for over a decade, it faces significant challenges.  The most significant is the fact that the 21~cm signal is roughly 4 orders of magnitude dimmer than foreground emission (primarily synchrotron radiation) from Galactic and extragalactic radio sources. Analysis techniques for extracting the 21~cm signal generally rely on the fact that foreground emission is a slowly-varying function of frequency while the 21~cm spectrum has structure arising from the large-scale distribution of matter along the line of sight \citep{Liu&Shaw2020}.   However, instrumental effects, such as the delay term, can introduce structure into the spectrum of otherwise smooth foregrounds.  In particular, the spatial angular dependence of the antenna patterns is also frequency dependent and, in a process called `mode-mixing,' couples the angular dependence of the bright foregrounds into frequency dependence that masquerades as cosmic 21~cm structure. In addition, although the 21~cm signal is unpolarized, the bright foregrounds are partially polarized, and frequency-dependent instrumental `leakage' of Stokes Q, U, and V into I introduces another type of foreground with a complicated spectrum. Faraday rotation in the interstellar medium creates further spectral structure in the polarization signal. Removing mode-mixing effects requires detailed understanding and measurement of the frequency-dependent gain patterns of the antennas and of the gain and phase of the instrument's electronic response:  i.e. calibration. 
To determine the scale of the calibration challenge, \citep{Shaw2014} performed detailed simulations of the CHIME interferometer's ability to measure the HI signal in the presence of foregrounds.  
They showed it is necessary to know the beamwidth of the antennas to $0.1\%$ and the electronic gain to $1\%$ within each minute of observation to recover the unbiased power spectrum of the HI signal.

Unlike most radio interferometers, 
all currently operating or proposed post-EoR HI intensity mapping instruments observe by drift scanning the sky.  This observing strategy allows for large sky coverage using simple and inexpensive instrument designs but requires new types of calibration strategies. Tracking instruments can calibrate continuously on bright sources in, or near, the field they are mapping. Drift scanning instruments like Tianlai must wait for bright sources to pass through the field, or attempt to calibrate on dimmer sources. 
Therefore, much of the discussion in this paper focuses on measuring the instrument stability and performing the calibration of the Tianlai Dish Array.

Future intensity mapping interferometers may include thousands of antennas \citep{Ansari2018,Slosar2019} in order to increase sensitivity. The only currently feasible way to correlate this many signals is by using fast Fourier transform algorithms (i.e. an `FFT correlator') that require that the antenna patterns from the dishes be identical.  Anticipating the need for such advances, this paper also characterizes the uniformity of the antenna patterns in the Tianlai Dish Array.

This paper uses a small fraction (a few hundred hours) of the data collected to date in order to quantify basic performance characteristics of the Tianlai Dish Array.  Future analysis of this data set will focus on making and cleaning maps of the NCP region, for which we have 3,700 hours of observations.

This paper is organized as follows:  Section \ref{sec:instrument}
describes the Tianlai Dish Array;  Section \ref{sec:observations} 
describes the observations carried out by the Tianlai Dish Array from 2016--2019; Section \ref{sec:beams} compares the measured antenna patterns with simulations and evaluates their uniformity;  Section \ref{sec:analysis} provides an overview of the data analysis process; Section \ref{sec:calibration} describes the gain and phase calibration process and Section \ref{sec:sensitivity_originally_Tnoise} presents the results of this calibration in terms of noise level, and sensitivity vs. integration time; Section \ref{sec:maps} presents maps of bright calibration sources; Data from several constant declination scans, corresponding to a total observation time of more than 100 hours, have been used for the analysis presented in sections \ref{sec:calibration} and \ref{sec:maps}. Section \ref{sec:NCP} describes results of long integrations on the North Celestial Pole (NCP), using data from 10 nights of January, 2018, conclusions and plans for the future appear in Section \ref{sec:conclusion}.

\section{Instrument}
\label{sec:instrument}

The objective of the Tianlai program is to make a 21~cm intensity mapping survey of the northern sky. At present the Tianlai program is in its Pathfinder stage, which aims to test the technology for making 21~cm intensity mapping observations with an interferometer array. The Pathfinder comprises two arrays, one consisting of dish antennas and the other of cylinder reflector antennas, both located at a radio quiet site ($44^\circ 9'\text{N}$, $91^\circ 48'\text{E}$) in Hongliuxia, Balikun County, Xinjiang Autonomous Region, in northwest China. In order to avoid radio-frequency interference (RFI) generated by the correlator, the station house, which includes an analog electronics room, a digital correlator room (shielded from the analog room), and living quarters, is located 5.8 km (11.2 km by road) away from the telescope site. A power line and optical fiber cables about 8~km long connect the correlator building with the antenna array. 
Construction of the Pathfinder arrays was completed in 2016 and they are now taking data on a regular basis. This paper focuses on the dish array. Further details about the cylinder array appear in \cite{2016RAA....16..158Z,Cianciara2017,Tianlai,Zuo2019,Li2020a,Li2020b}.

\begin{figure*}
  \centering
  \includegraphics[width=0.57\textwidth,trim = 0 150 0 50, clip]{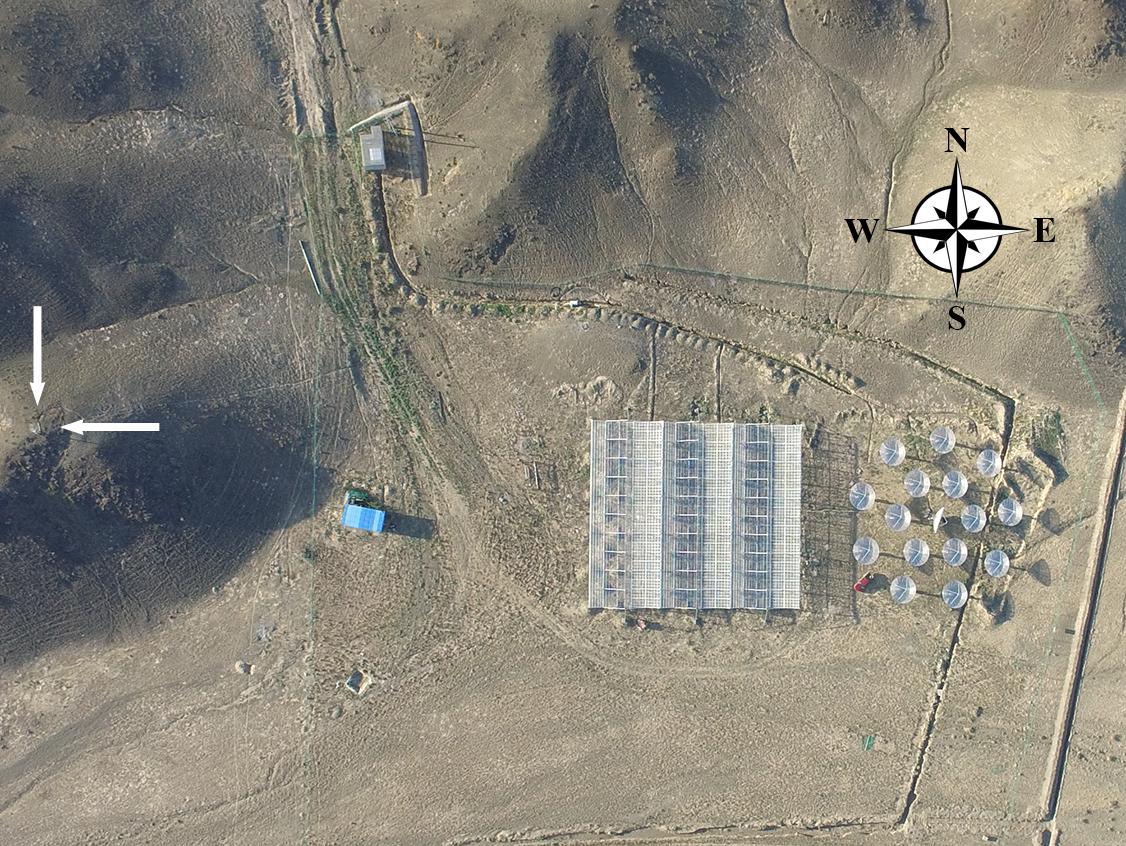}
\includegraphics[width=0.41\textwidth]{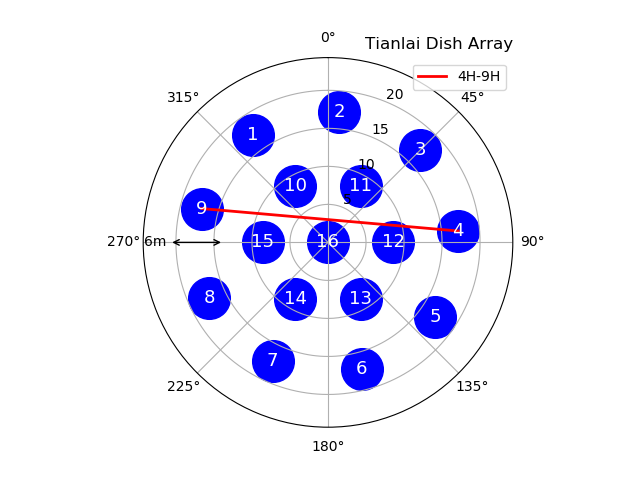}
\caption{ 
Left: Top view of the Tianlai Dish Array Pathfinder and Cylinder Array Pathfinder taken with a \texttt{DJI M600 Pro} drone at a height of 280~m above the ground.
The position of the calibration CNS is indicated by the white arrows on the left.  
The relative distance vector from the feed in dish 16 at the center of the array (when pointed toward the zenith) to the CNS is [-184.656,   13.915,  12.588] meters,  with x,y,z to the east, north, and zenith. The CNS is in the far-field of all dishes in the dish array.
Right: A schematic diagram of the Tianlai Dish Array; $0^\circ$ coincides with North.} The dishes are arranged in two concentric circles of radius $8.8\,$m and $17.6\,$m around a central dish. The dishes have dual-linear polarization feed antennas with one axis oriented parallel to the altitude axis (horizontal, H, parallel to the ground)) and the other orthogonal to that axis (vertical, V). 
For example, shown in red is one of the baselines that is studied later in this paper, the H polarization of dish 4 correlated with the H polarization of dish 9:  4H-9H. Other baselines used later in the paper use the same naming convention. 
\label{fig:arrayphoto_topview}
\end{figure*}


\begin{figure*}
  \centering
  \includegraphics[width=0.85\textwidth]{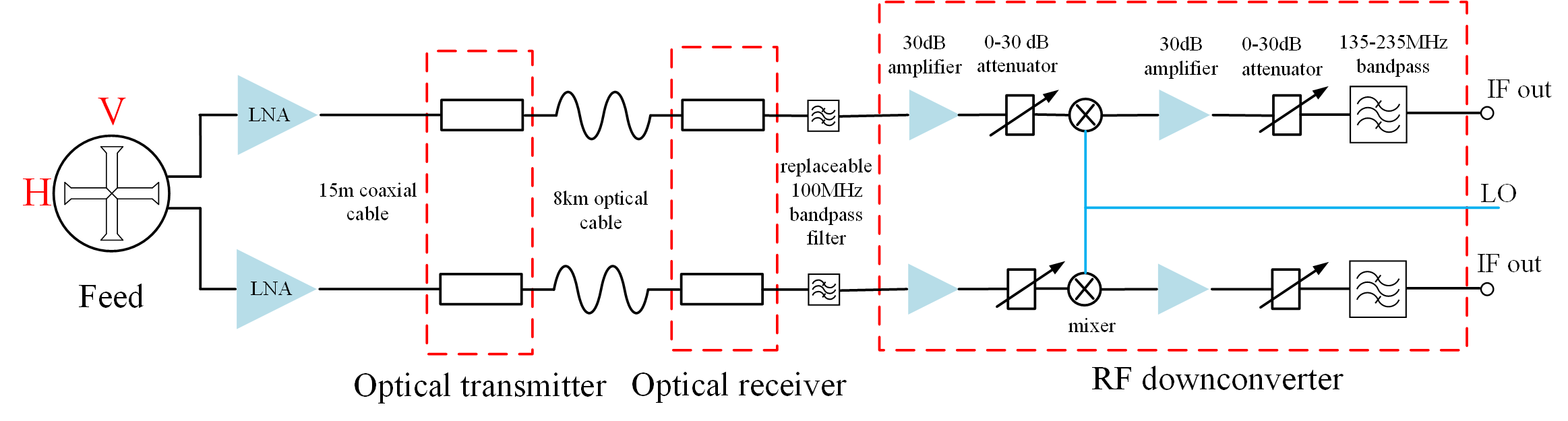}
\caption{Schematic of the RF analog system. 
}
\label{fig:schematic}
\end{figure*}

For each array, the feed antennas, amplifiers, and reflectors are designed to operate from 400~MHz to 1430~MHz, corresponding to $2.55 \geq z \geq -0.01$.  The instrument can be tuned to operate in an RF bandwidth of 100~MHz, centered at any frequency in this range by adjusting the local oscillator frequency in the receivers and replacing the band pass filters.  Currently, the Pathfinder operates at $700 - 800$~MHz, corresponding to HI at $1.03 \geq z \geq 0.78$. Future observations are planned in the $1330 - 1430$~MHz band ($0.07 \geq z \geq -0.01$) to facilitate cross correlation with low-z galaxy redshift surveys and other low-z HI surveys. 

The Tianlai Dish Array consists of $16$ on-axis dishes. Each has an aperture of $6\,\text{m}$. The design parameters of the dishes are shown in Table ~\ref{Tab:Tianlai_dish_feed}.  The dishes are equipped with dual, linear-polarization receivers, and are mounted on Alt-Azimuth mounts. One polarization axis is oriented parallel to the altitude axis (horizontal, H, parallel to the ground)) and the other is orthogonal to that axis (vertical, V) \cite{Zhang2020}.  Motors are used to control the dishes electronically. The motors can steer the dishes to any direction in the sky above the horizon. The drivers are not specially designed for tracking celestial targets with high precision.  Instead, in the normal observation mode, we point the dishes at a fixed direction and perform drift scan observations. The Alt-Azimuth drive provides flexibility during commissioning for testing and calibration. The dish array was fabricated by CASIC-23.\footnote{\url{http://www.casic23.com.cn}}

The dishes are currently arranged in a circular cluster (Figure \ref{fig:arrayphoto_topview}).  The array is roughly close-packed, with center-to-center spacings between neighboring dishes of about $8.8$~m. 
The spacing is chosen to allow the dishes to point down to elevation angles as low as $35^\circ$ 
without `shadowing' each other. One antenna is positioned at the center and the remaining $15$ antennas are arranged in two concentric circles around it. It is well known that the baselines of circular array configurations are quite independent and have wide coverage of the ($u$, $v$) plane. A comparison of the different configurations considered for the Tianlai Dish Array and the performance of the adopted configuration can be found in \cite{PAON4_Zhang_2016}.  The Tianlai dishes are lightweight and the mounts are detachable, so, in future, the dishes can be moved to new configurations if required. This paper describes observations with the current configuration. 

A schematic of the RF analog system can be found in Figure~\ref{fig:schematic}.  The whole system except filters has been designed to operate over a wide range of frequencies (400--1500~MHz).  The low noise amplifiers (LNAs) are designed to have low noise temperature (about 47~K at room temperature \citep{Li2020a}) and are mounted to the back of the feed antennas. The amplified RF signals pass through 15-meter long coaxial cables to optical transmitters mounted under the dish antennas. The RF signal amplitudes modulate optical transmitters so that the RF signals are converted to optical signals, which are then transmitted to the station house via 8~km optical fiber. At the RF analog system room in the station house, the optical signal is converted back to the RF electric signal. Replaceable bandpass filters with 100~MHz bandwidth 
are mounted between the optical receivers and analog downconverters.  An analog mixer then downconverts the RF signal to the 135--235 MHz intermediate frequency (IF) band. Finally, the IF signal is sent to the digital system 
through bulkhead connectors between the analog and digital rooms.  
The dishes currently observe in the frequency band 700--800 MHz
($1.03>z>0.78$) in 512 frequency channels ($\delta \nu=244.14\,\text{kHz}$,
$\delta z=0.0002$).

\begin{table}
\caption{Main design parameters of a Tianlai dish antenna. 
}
\label{Tab:Tianlai_dish_feed}
\centering
\begin{tabular}{l l}
\hline
Reflector diameter & 6~m \\
Antenna mount & Alt-Az pedestal \\
f/D & 0.37 \\
Feed illumination angle  &  $68^\circ$ \\
Surface roughness (design) & $\lambda/50$ at 21~cm\\
Altitude angle & $8^\circ$ to $88.5^\circ$\\
Azimuth angle &  $\pm 360^\circ$\\
Rotation speed of Az axis &  $0.002^{\circ}$\, $\sim$ $1^{\circ} /{\rm s}$\\
Rotation speed of Alt axis &  $0.002^{\circ} \sim 0.5^{\circ} /{\rm s}$\\
Acceleration &  $1^\circ/{\rm s}^2 $ \\
Gain(design) &  29.4+20log(f/700~MHz) dBi \\
Total mass & 800~kg \\
\hline	
\end{tabular}
\end{table}


The digital backend system 
of the dish array is a 32-input correlator that consists of three FPGA boards: two processing boards for signal sampling and processing, and one for control. The Analog to Digital Converters (ADC) in the processing boards convert the RF signal to time series data at a sampling rate of 250~MSPS and sampling length of 14 bits. Then the FPGA chips in the processing boards perform the FFT 
of the time series data. The two FPGA boards exchange half of the signal channels with each other through rapidIO cables, so all cross-correlations are computed in the FPGA boards while the computation loads on the  boards are balanced. Finally, the visibility from the dish array ($32$ auto-correlations and $496$ cross-correlations) are sent to a storage server by two ethernet cables and dumped to hard drives in \texttt{HDF5} format.


 


\begin{table*}
\centering

\caption{Observation log for the Tianlai Dish Array from 2016 to late 2019.}
    \begin{tabular}{r r r r r}
         \hline
         Data Set & Date & Calibration Sources & Targets & Length (hours)\\
         \hline
         Data 201605-06 & May 2016 & None & Cygnus A & 72\\
         CygnusANP 20170812 & Aug 2017 & Cygnus A & North Pole & 67 \\
         CasAs 20171017 & Oct 2017 & None & North Pole & 147\\
         CasAs 20171026 & Oct 2017 & None & Cassiopeia A & 290 \\
         3srcNP 20180101 & Jan 2018 & 3C48, Cassiopeia A, M1 & North Pole & 241 \\
         2srcNP 20180112 & Jan 2018 & 3C48, M1 & North Pole & 97\\
         IC443NP 20180323 & Mar 2018 & IC443 & North Pole & 181\\
         M87NP 20180407 & Apr 2018 & M87 & North Pole & 90\\
         2srcNP 20180416 & Apr 2018 & IC443, M87 & North Pole & 142\\
         3srcNP 20181212 & Dec 2018 & Cassiopeia A, 3C48, M1 & North Pole & 757\\
         1DaySun 20190113 & Jan 2019 & None & Sun & 48 \\
         3srcNP 20190128 & Jan 2019 & Cassiopeia A, 3C48, M1 & North Pole & 741\\
         3srcNP 20190228 & Feb 2019 & 3C123, M1, IC443 & North Pole & 764\\
         3srcNP 20190402 & Apr 2019 & M1, IC443, 3C273 & North Pole & 522 \\
         3srcNP 20190611 & Jun 2019 & M87, Hercules A, Cygnus A & North Pole & 737\\
         3srcNP 20190830 & Aug 2019 & 3C400, Cygnus A, Cassiopeia A & North Pole & 924\\
         3srcNP 20191022 & Oct 2019 & 3C400, Cygnus A, Cassiopeia A & North Pole & 302\\
         \hline
    \end{tabular}
    \label{tab:log}
\end{table*}

Calibration of the electronic gain of the receivers is crucial for any interferometer, and it is especially important for Tianlai to compensate for phase variation in the 8~km long optical cables. The absolute calibration of the system can be performed using bright astronomical standard calibration sources.  However, for small aperture arrays like the Tianlai Dish Array, there are not enough bright sources on the sky to meet the requirement of point source calibration, so we have designed a dedicated calibration noise source (CNS) to provide relative calibration. A broadband RF noise generator is placed in a thermostatically controlled environment and is supplied with regulated DC power to ensure the stability of the RF amplitude. The on-off timing of the CNS is controlled by a clock signal carried by optical fiber from the correlator 8~km away in the station house. 








\begin{figure}
  \centering
  \includegraphics[width=3.0in]{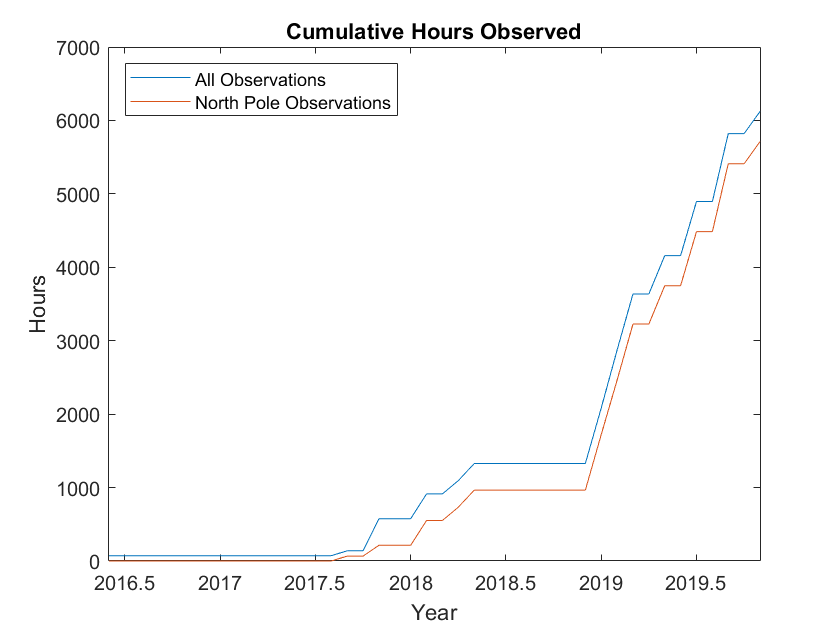}
\caption{ 
Accumulated observing time vs. date.}
\label{fig:ObservationDuration}
\end{figure}

\section{Observations} 
\label{sec:observations}

As of the end of 2019, we had collected about $6,200~{\rm hours}$ of observational data from the Tianlai Dish Array, including more than $5700~{\rm hours}$ of NCP data.
In Figure~\ref{fig:ObservationDuration} we show the accumulated observing time over the years. Details of individual runs are listed in Table~\ref{tab:log}.  Drift scans are performed at constant declination over several days at a time. These can be 
divide into two types: (1)24 hr observations at declinations away from the NCP, usually at the declination of bright sources: Cyg A ($+40^\circ 44'$), Cas~A ($+58^\circ 48.9'$), Tau A/M-1 ($+22^\circ 00'$) and also some high declination regions ($\sim 80^\circ$). (2) 24 hour observations at the NCP. Preceding each NCP observation, the antennas are pointed towards one or multiple strong radio sources for calibration.  The calibration sources for different NCP observations are listed in Table ~\ref{tab:log}. 

The visibilities from the dish array are averaged for a period of $1~{\rm s}$ (integration time). The data are stored for all $528$ correlation pairs (auto-correlation + cross-correlation) in $512$ different frequency channels. 
The data rate is about $\sim 175\,{\rm GB/day}$.
The weather data, which includes the temperature of the analog electronics room, site temperature, dew point, humidity, precipitation level, wind direction, wind speed, barometric pressure, etc., are stored separately during each run. These data can later be used for checking the correlation of different weather variables with the variation of electronic gain of the system. 


The CNS is turned on and off periodically. During 2017 the CNS switched on for 20~{\rm s} every 4~{\rm min}, so the fraction of noise-on time is $\sim 8.33\%$.
In 2018, the noise-on time was reduced to 4~{\rm s} per 4 min, which is 1.67\% of the observing time.

\begin{figure}
  \centering
  \includegraphics[width=0.99\columnwidth,trim = 10 50 30 50, clip]{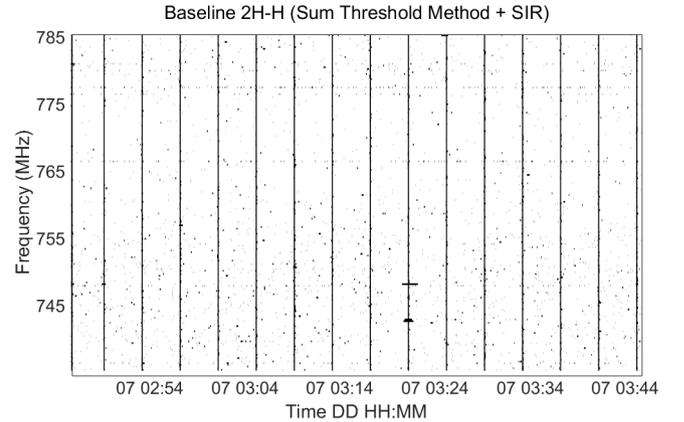}
\caption{ 
 Masking of the CNS and RFI after 
 applying both the sum threshold method and the SIR operator method. The vertical lines show times and frequencies masked when the signals from the CNS are used for calibration. Apart from the CNS we can see a very small amount of data at discrete frequencies and times masked as RFI. Because the amount of RFI is very small we have blown up a bandwidth of about 50~MHz where some RFI is visible. Some faint horizontal lines (at about 777~MHz and 767~MHz) are from intermittent RFI.}
\label{fig:mask_rf}
\end{figure}

In Figure~\ref{fig:mask_rf} we show the CNS and RFI mask derived from $1~{\rm hour}$ of nighttime data.  The periodic vertical stripes show the mask when the CNS is turned on, while the dots show the RFI. We use two different RFI cleaning methods (check Sec.~\ref{sec:analysis}).  Because the array is located in a radio quiet zone, we only lose about $0.6\%$ of data due to RFI contamination. 



\begin{figure*}
  \centering
  \includegraphics[width=.47\textwidth]{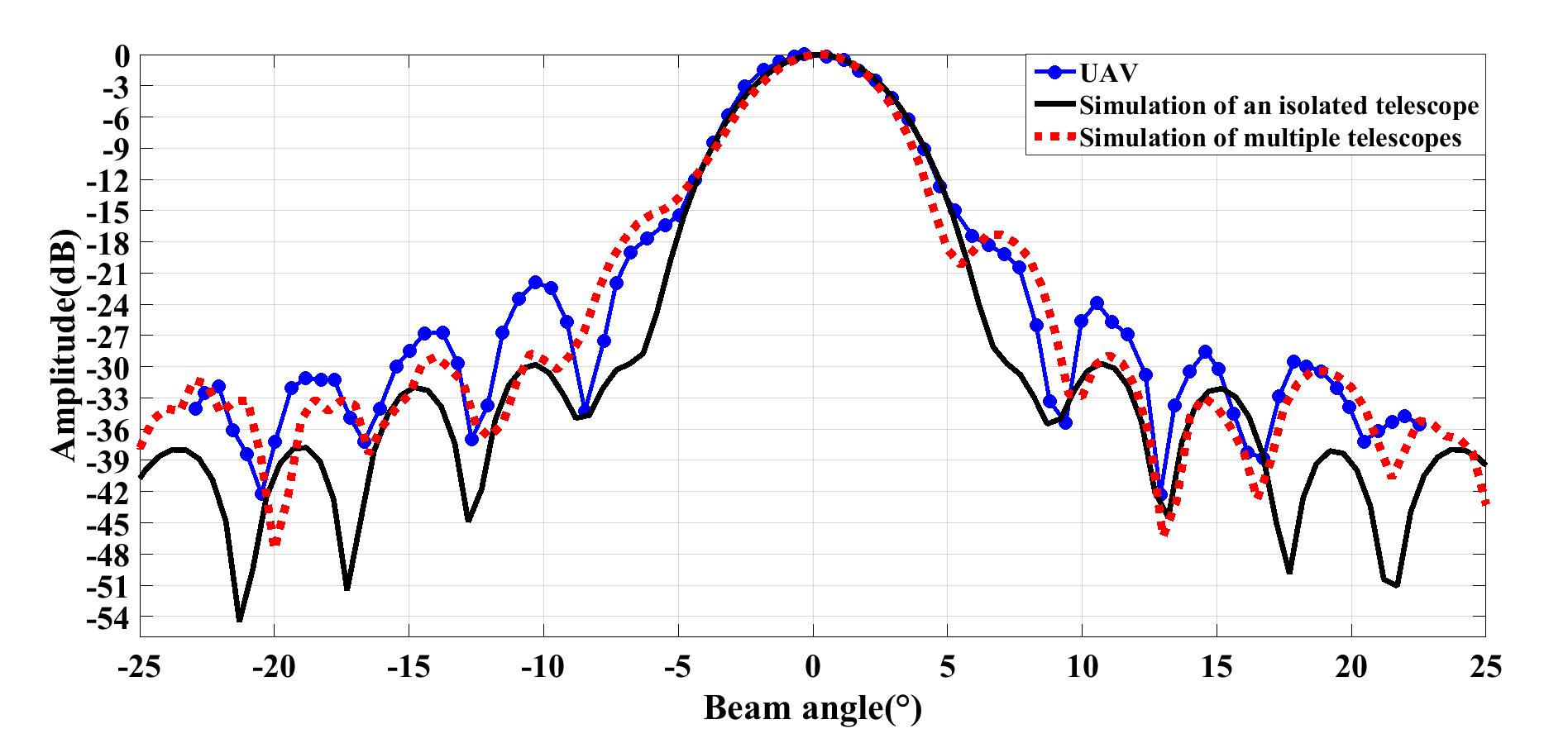}
  \includegraphics[width=.47\textwidth]{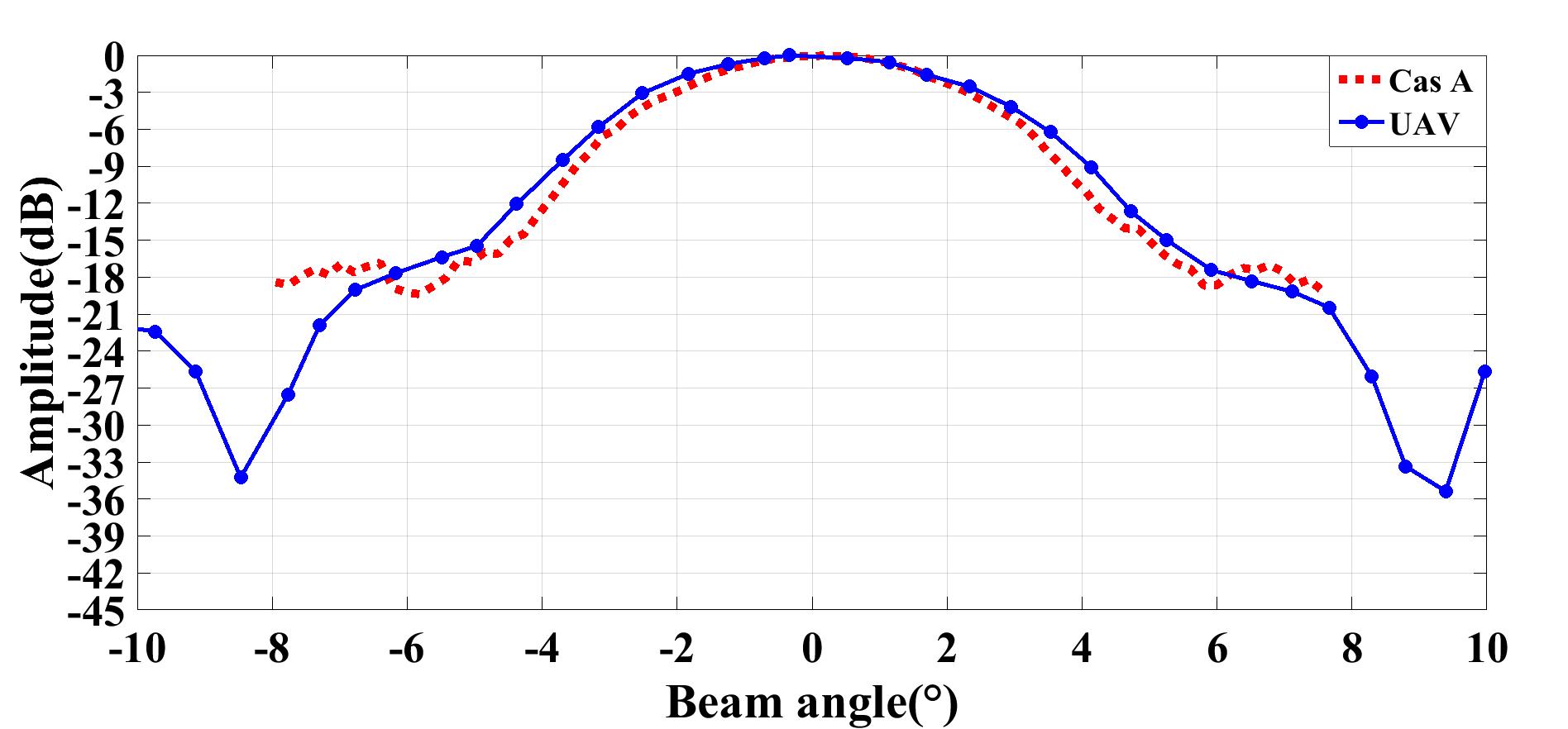}
  \caption{Measurements and simulations of the H-plane antenna pattern of the V-polarization of one dish. Left: Pattern measured with the UAV compared with two electromagnetic simulations. Right: Patterns measured both by an UAV  and by a transit of Cas~A. In both figures the UAV flies in the E-W direction and the measurements and simulations are performed at 730~MHz.}
  \label{fig:beam_vs_theta}
\end{figure*}

\section{Beam Patterns}
\label{sec:beams}

Separating the faint 21~cm signal from strong foregrounds requires exquisite knowledge of the frequency-dependent beam patterns of the antennas. Small imperfections in the antennas or changes in the environment (e.g., temperature, wind) can affect the beam patterns and introduce systematic errors in the measurement.  
In addition, future arrays with large numbers of antennas will likely require highly uniform antennas to exploit techniques such as redundant calibration and FFT correlation \citep{Tegmark2009,Tegmark2010,Sievers2017, Byrne2020}.  However, to our knowledge, detailed  requirements for the precision of knowledge of beam patterns and sidelobes and cross-polar response, their stability with time, their required degree of uniformity, and relative pointing accuracy have not yet been performed. Nevertheless, Shaw et al. \cite{Shaw2014} provide a relevant data point, showing that to control foreground contamination from mode-mixing, the dish beamwidths must be known and uniform to 0.1\%.
 
We have taken preliminary steps toward characterizing the beam patterns using transits of bright radio sources and scans with a radio source flown over the array on an unmanned aerial vehicle (UAV). We find reasonable agreement between the beam measurements and the electromagnetic models. The dirty maps of bright sources shown in Section \ref{sec:maps} do not require knowledge of the beams, however, the temperature calibration described in Section \ref{sec:calibrationabsolute}, which is used for the analysis of 10 nights of integration on the NCP in Section \ref{sec:NCP}, requires measuring the response of the instrument to astronomical calibration sources as well as knowledge of the directivity gain of the antennas. The latter requires knowing the beam pattern in all directions. Because we have not yet measured the beam patterns into the far sidelobes, for now we rely on the simulations for determining the directivity gain. Rough agreement between measurements and simulations in the main beam is encouraging, but extending measurements over the full beam remains an important research goal. Future deconvolved maps of the NCP will require knowledge of the main beam.
 
Furthermore, as is the case for other interferometers designed for HI intensity mapping, the primary beam patterns of the Tianlai dish antennas are affected by the presence of the other antennas in the array.  Both the UAV measurements and the electromagnetic simulations show an asymmetry in the beam patterns and an increase in the sidelobe levels compared to simulations of isolated antennas. This phenomenon has implications for the design of future HI arrays.

\begin{figure}
\includegraphics[width=0.49\textwidth]{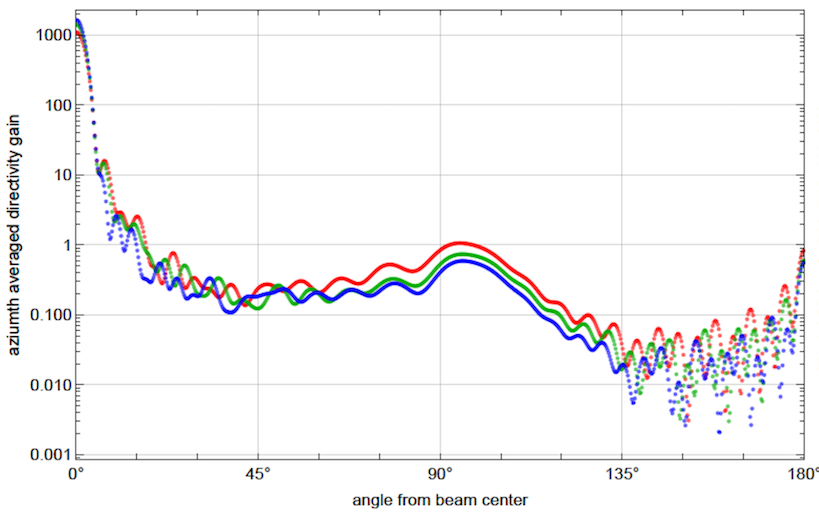}
\caption{Simulated beam patterns as a function of beam angle $\theta$ from the beam center of the antennas for 3 different frequencies, 700 (red), 750 (green) and 800~MHz (blue). Each plot shows the absolute co-polar directive gain in dBi, averaged over the azimuthal angle.  Angle $\theta$ is the polar angle, calculated from the center of the beam. The simulations show that the antenna sidelobes vary significantly as a function of frequency.}
\label{fig:simulated_beam}
\end{figure}

Figs. \ref{fig:beam_vs_theta} and \ref{fig:simulated_beam} summarize the antenna beam measurements and simulations. A detailed description of the UAV measurements and simulations appears in \cite{Zhang2020}. An UAV outfitted with a broadband noise source is flown in the far field along two paths over antenna \# 6, at the edge of the array:  a north-south path and an east-west path.  The E- and H-plane patterns are measured for each flight path over the 700-800~MHz band. As an example, Figure \ref{fig:beam_vs_theta} shows the H-plane measurement for the  east-west path at 730~MHz. Multiple flights are conducted at different RF power levels to map the beam into the far sidelobes.  Figure \ref{fig:beam_vs_theta} (Right) shows the beam pattern measured by the UAV and compares it with the profile of the main beam measured at the same frequency using the auto-correlation signal from one polarization during a transit of Cas~A. The Cas~A signal is not bright enough to measure the sidelobes of the antenna.
 
We also performed electromagnetic simulations of the feed antenna and the dish reflector using commercially-available software.\footnote{CST Studio Suite \url{https://www.3ds.com/products-services/simulia/products/cst-studio-suite/} and FEKO \url{https://www.altair.com/feko/}}  We overplot the simulated beam pattern with the UAV measurement in Figure~\ref{fig:beam_vs_theta} (Left). The patterns match well in the main lobe and the measured
position and width of each sidelobe shows good agreement with the simulation. However, the UAV measurements show a ``shoulder" at $\pm 6^\circ$, an asymmetry in the sidelobes, and a stronger signal in the sidelobes. 

We compare these measurements of the antenna beamwidths and sidelobes to electromagnetic simulations of the effects of errors in shapes of the dishes.  We have not yet performed measurements of the dish shapes (e.g. with photogrammetry), so rely on simulations of random errors in the reflector surface.  We consider random errors on both large scales and small scales. For large-scale errors (on the scale of 10's of cm), simulations were performed using two simulation packages, CST and FEKO.  Errors were simulated with amplitudes as large as $30$~mm, $7$ times larger than the surface roughness specification for the dish ($0.02 \lambda$ at $\lambda = 21$~cm), which was built with standard antenna construction techniques. These simulations widen the FWHM of the beam by less than $0.2^\circ$ and increase the level of sidelobes by less than $6$~dB.  We used similar simulations to investigate the effect of errors in the placement of the feed antennas. Simulated displacements of the feed antenna from the nominal focus by as much as +50~mm (away from the dish) and -10~mm (toward the dish)  can increase the FWHM of the dishes by about the amount we measure with the UAV, but this displacement is larger than the tolerance on the placement of the feed. These focus displacements have no significant effect on the sidelobes.   For small-scale errors, we estimate the effects of random surface deviations of $\lambda/50$ at $21$~cm, the design specification given for surface roughness of the dishes. Using antenna tolerance formulas\cite{Ruze66,Rahmat83}, we estimate the FWHM would increase less than $3\%$ and the sidelobe peak values would increase by less than 1~dB.  Hence, neither large-scale or small-scale errors in the dish surface or focus are consistent with the measurements of the main beam or sidelobes obtained with the UAV  (Figure \ref{fig:beam_vs_theta}).

A simulation that shows the same beam asymmetry and similar sidelobe levels as the UAV measurement includes dish \# 6 as well as 4 adjacent dishes in the array (dish \#5, \#7, \#13, and \#14). It is shown as the dashed red line in Figure \ref{fig:beam_vs_theta}. We conclude that these effects arise, at least partially, from scattering from other dishes in the array. Scattering from the ground and nearby hills is not included in the current simulations.  Simulating the full array and the ground requires more computing resources that we currently have available.  Ultimately, we plan to extend these simulations and measurements to all the dishes in the array, plus the ground, over the full range of frequencies. Simulations extended into the far sidelobes and backlobes of an isolated antenna  are shown in Figure \ref{fig:simulated_beam}.
 
\begin{figure}
  \centering
  \includegraphics[width=0.48\textwidth]{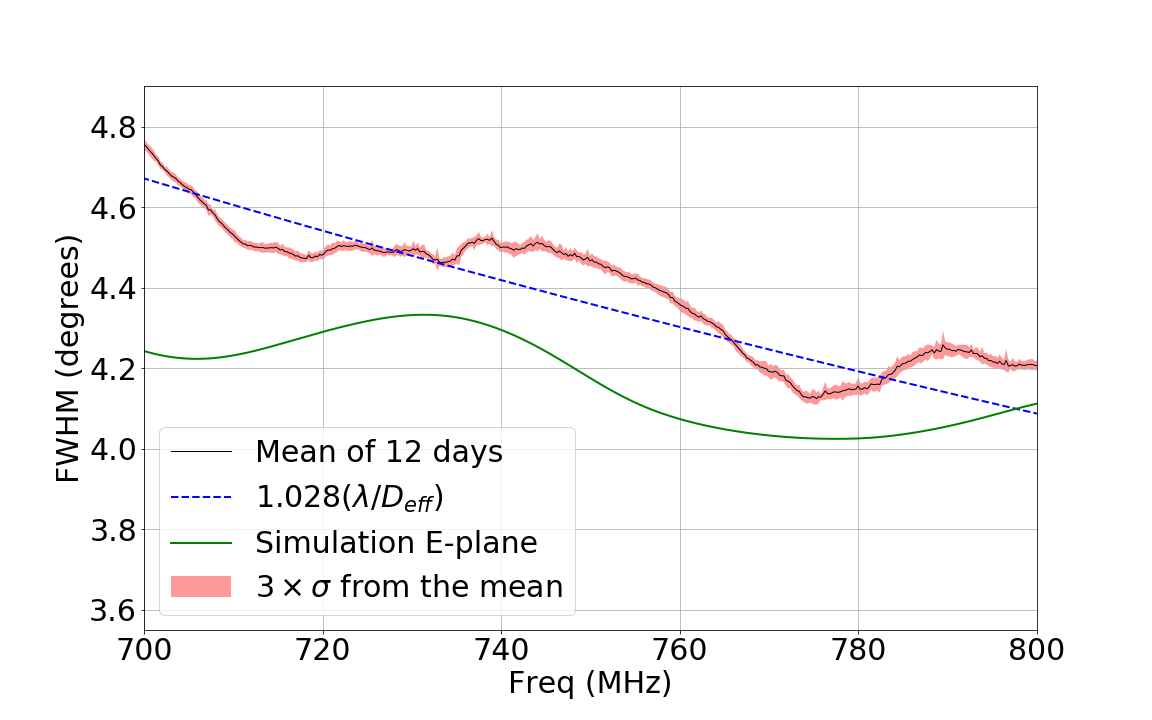}
  \includegraphics[width=0.48\textwidth]{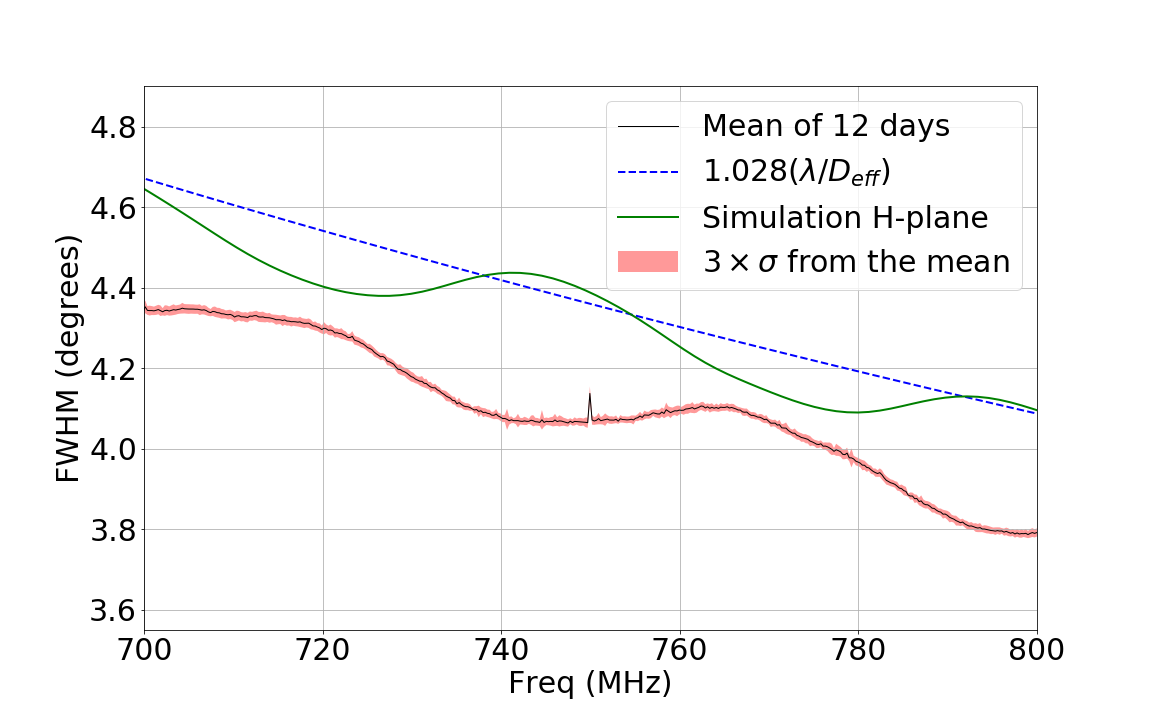}
  \caption{Plot of the mean FWHM of the main beam vs. frequency using daily transits of Cas~A over 12 days. The measurement is for a fairly typical baseline. The top figure represents the horizontal baseline (4H-9H), which primarily measures the E-plane of the antenna, while the lower figure represent the vertical baseline (4V-9V), primarily measuring the H-plane. The black line shows the mean in each frequency bin over this period and the red scatter plot shows the FWHM for each day.  The frequency binning is 244~kHz. The green line shows the expected FWHM vs. frequency based on an electromagnetic simulation.  For reference, the blue line shows the FWHM of a uniformly-illuminated Airy disk with an effective diameter $D_\mathrm{eff}$ that is 90\% of the actual 6 meter diameter.
}
  \label{fig:beamwidth_vs_f}
\end{figure}

As mentioned in Section \ref{sec:introduction}, knowledge of the beam patterns of each antenna as a function of frequency, and the stability of the pattern with time, are also important factors.  Figure~\ref{fig:beamwidth_vs_f} shows the FWHM of one cut through the beam pattern from a pair of antennas as a function of frequency. The pattern is measured repeatedly in the E-W direction by observations of the transit of Cas~A on 12 successive days starting on 2017/10/26. For each transit, the magnitude of one visibility vs. time is fit to a Gaussian.  The measured pattern is effectively the geometric mean of the patterns of two dishes, which are nominally coaligned. Day-to-day fluctuations of the FWHM are less than $1\%$.  The frequency dependence of the beam in the E- and H-planes is also calculated by the electromagnetic simulations; the simulated FWHM is co-plotted for comparison. We also plot the case of a diffraction - limited circular aperture ($1.028\lambda/D_{\rm eff}$ with $D_{\rm eff} = 0.9D$). (The $1.028$ prefactor comes from the FWHM of an Airy pattern from a uniformly illuminated disk.) The measured and simulated beamwidths differ by about $0.3 - 0.5$ degrees. This discrepancy originates in a corresponding difference between the measured and simulated beams of the feed antennas.  The broad ripples in the frequency-dependence of the beamwidth in Figure~\ref{fig:beamwidth_vs_f} are consistent with the appearance of standing waves between the feed antenna and the dish surface. The path length from the feed to the dish and back again is twice the $2.2$~m focal length, and should introduce ripples with a period of $68$~MHz, close to the observed period.  Ripples with similar amplitudes and periods appear in both the measurements and the simulations, but the phases of these do not line up in the $V$ baseline as well as they do in the $H$ baseline. We are trying to understand the reason for this discrepancy. Another known reflection in the system occurs at the ends of the $15$~m coaxial cable between the feeds and the optical transmitter (Figure 2). Ripples caused by this reflection correspond to a period of $7$~MHz, which is barely visible in the top plot of Figure \ref{fig:beamwidth_vs_f}, in addition to the dominant $68$~MHz ripple.

\begin{figure}
  \centering
  \includegraphics[width=0.47\textwidth]{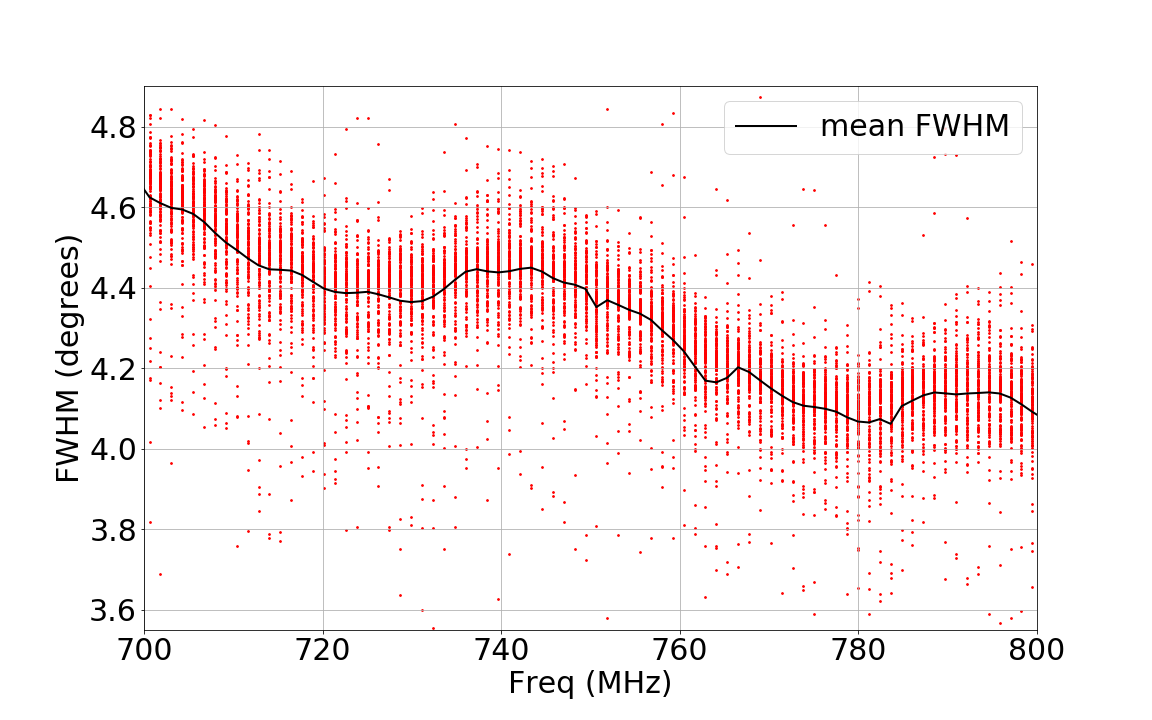}
  \caption{The black line shows the mean FWHM vs. frequency for 118 H-H baselines during a transit of M1 on 2018/01/02. The plot excludes auto-correlations, baselines with known faulty probes, and other outliers.}
  \label{fig:M1_beamwidth_vs_f}
\end{figure}

Another important characteristic of the antenna patterns is the uniformity of the different antennas. Here we estimate the uniformity of the beamwidths of the dishes in the Tianlai Dish Array by using transit observations of M1 to determine the effective beamwidth of all pairs of antennas (baselines). Figure \ref{fig:M1_beamwidth_vs_f} shows the mean value of the FWHM of 118 baselines in the array and the 1-sigma deviations from the mean as a function of frequency. The $1-\sigma$ deviations are $\sim 4\%$.

\begin{figure}
  \centering
  \includegraphics[width=0.47\textwidth]{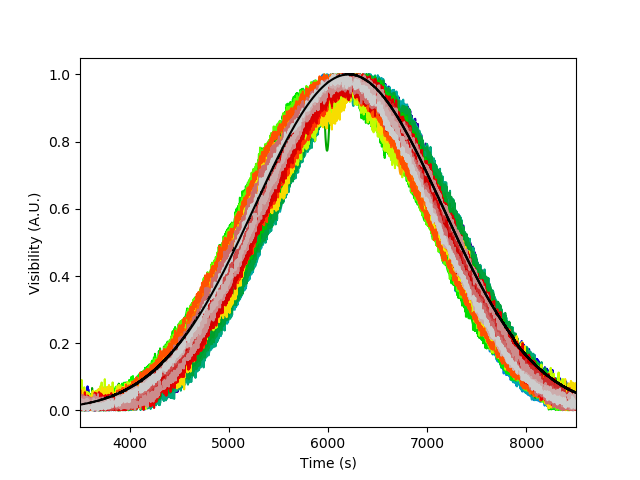}
  \caption{Variations of the modulus of the cross-correlation visibilities between the Tianlai dishes with time (in seconds) during a transit of Cas~A. Each of the 120 cross-correlations corresponds to one color. The black curve shows the expected response from a simplified simulation (from a 5.4~m diameter dish with a Gaussian beam). All (un-calibrated) cross-correlations have been renormalized to unity at their respective maxima. The observed time spread between the cross-correlations reflects the dishes' relative E-W pointing dispersion. The green dip at around 6000~s is an artifact from the CNS interpolation. Each curve is the average of 40 frequency channels (742.6172~MHz to 752.1387~MHz, with each frequency bin being 244 kHz), near the center of the RF band.}
  \label{fig:CasA_XCorr_mods}
\end{figure}

\begin{figure}
  \centering
  \includegraphics[width=0.47\textwidth]{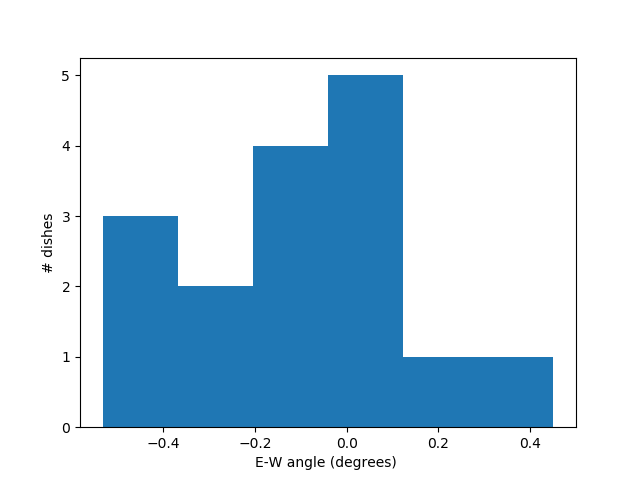}
  \caption{Extracted transit time differences w.r.t. to the expectation from a simplified simulation, for the 16 Tianlai dishes from the Cas~A transit of 2017/10/30. The $\sim 150$s FWHM of these time shifts from cross-correlation corresponds to an E-W pointing error of $\sim$ 0.47 degrees on the sky (not in RA). This indicates an E-W pointing error of $\sim$ 0.66 degrees for a single dish.  
  }
  \label{fig:CasA_DeltaTs}
\end{figure}

We have also studied the pointing accuracy of the dishes in the E-W direction. Using data from the Cas~A transit of 2017/10/30, we compare in Figure~\ref{fig:CasA_XCorr_mods} the variations of the modulus of the visibilities formed by the cross-correlation  between the dish signals. (In cross-correlation the Cas~A signal is almost undistorted by the diffuse Galactic signal; this is not the case for auto-correlations, which are affected by the diffuse background.)
To get a rough measurement of the E-W pointing accuracy, we have extracted the peak position from each of these cross-correlations by fitting a Gaussian curve. If the times of the peak response for two dishes are $t_1$ and $t_2$, the peak time of the cross-correlation will be their average peak time, $t_{12} = (t_1 + t_2) / 2$, so the variance $\sigma^2$ for the cross-correlation is only half of that for a single dish. The pointing of the antennas in the E-W direction can be regarded as a Gaussian distribution centered on the expected pointing. If the $\sigma^2$ is doubled, the FWHM will be $\sqrt{2}$ times larger, because $\mathrm{FWHM} \propto \sigma$. As illustrated in Figure~\ref{fig:CasA_DeltaTs}, the E-W pointing spread from cross-correlation indicates a 0.47 degree FWHM dispersion (on the sky), so the E-W pointing dispersion of single dish will be about 0.66 degree. 
This dispersion is roughly consistent with our current procedure for aligning the pointing of the dishes. We calibrate the absolute pointing of each dish by observing the shadow of the Sun projected onto the vertex of the parabolic reflector surface when the dishes are pointed at the Sun. We estimate this process can introduce pointing errors of about $0.2^\circ$.  In addition, backlash in the antenna gears introduces an additional error of about $0.05-0.2^\circ$, so loading by the wind on the reflectors will introduce an additional pointing uncertainty.

Pointing accuracy in the N-S direction is difficult to determine using transits of astronomical sources, as is dependence of the beam shape on elevation angle. (Elevation changes will introduce mechanical deflection of the dishes.) Currently, our absolute calibration procedure requires pointing the dishes toward radio sources at different elevation angles than our primary science target, the NCP.  In principle, full $2\pi$ beam patterns could be measured using the UAV \cite{Chang2015, Jacobs2017} to determine changes in the beam shapes and pointing accuracy as well as the far-sidelobe patterns. The UAV measurements described here concentrated on characterizing a single dish, but, given the relatively small size of the Tianlai arrays, we plan future measurements to map all the dishes in the array simultaneously when pointed at the NCP and at the elevation angles of calibration sources. 

Using the UAV, we have placed an upper limit on the cross-polar response of antenna \#6.  At the center of the beam, the cross-polar response is less than about $1 \%$ across the band. (See \cite{Zhang2020} for details of the measurement.) This is consistent with simulations, which predict about $0.4\%$ cross-polar response across the band.

\begin{figure*}
  \centering
  \includegraphics[width=0.80\textwidth]{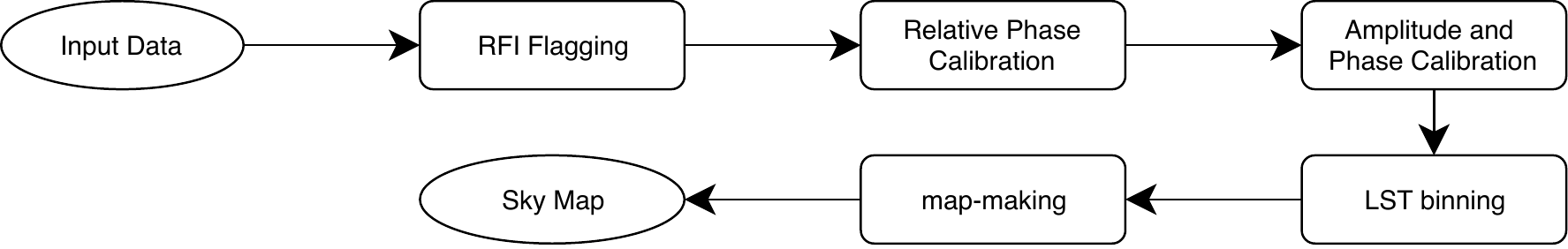}
  \caption{Schematic of the data processing pipeline implemented in the \texttt{tlpipe} package}
  \label{fig:tplipe_workflow}
\end{figure*}

\begin{figure} 
  \centering
  \includegraphics[width=3in]{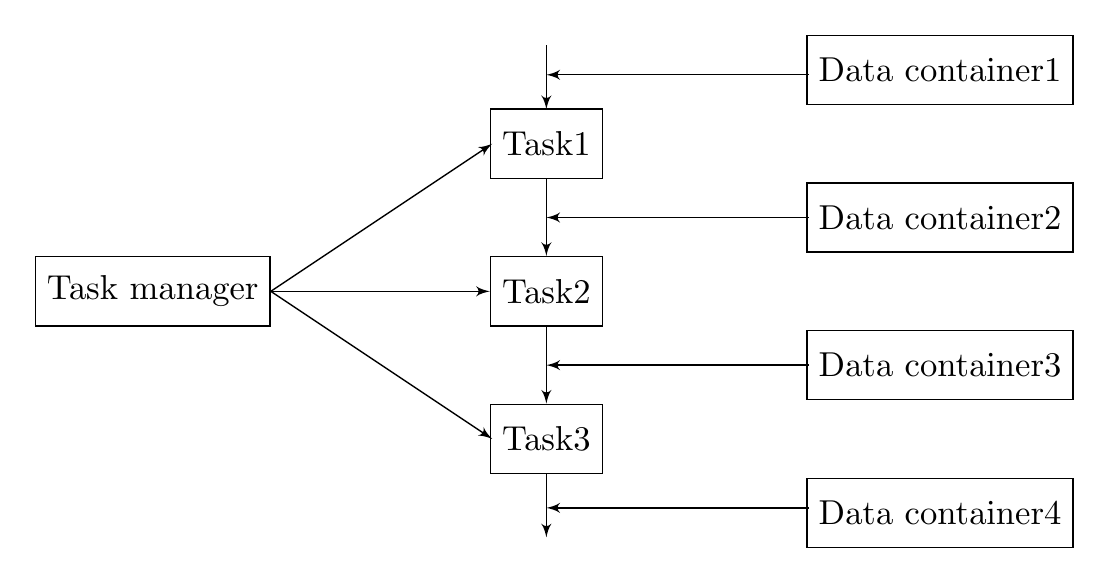}
  \caption{This schematic describes the relation between the three major components of the \texttt{tlpipe} data-processing pipeline.
 \texttt{tlpipe} implements two types of data container:  the RawTimestream and Timestream. The tasks take an instance of one of the two data container types as input, and produce an instance of one of the two data container types as output. The input and output can be the same instance (i.e. both RawTimestream or both Timestream), possibly with modified data and/or meta data, or different instances (i.e. input instance of RawTimestream, output instance of Timestream). Data are transferred from the input instance to the output instance, then the input instance is destroyed. The data container is the memory mapping of HDF5 files on disk. Tasks operate on the data contained in the data container in memory. The data volume is not multiplied on disk.}
  \label{fig:tplipe_dataprocessing}
\end{figure}

\section{Overview of (Offline) Analysis Process}
\label{sec:analysis}
For analyzing the Tianlai data we developed a Python pipeline named {\tt tlpipe} \footnote{\url{https://github.com/TianlaiProject/tlpipe}}. It is a collection of several stand-alone packages that can be used for reading the raw visibility data, masking, RFI flagging, calibration, data binning, map-making, etc. 
Multiple visualization and other utility packages have also been developed. The pipeline is written in a modular format and users can develop and add their own algorithms to the pipeline with little knowledge of how the 
rest of the pipeline works. 
Figure~\ref{fig:tplipe_workflow} shows the basic workflow of the \texttt{tlpipe} for making maps from the raw visibility data. A more detailed description appears in \cite{Zuo2020}.
Tianlai and \texttt{tlpipe} use the HDF5 formats consistently.  However, we have successfully imported data from other interferometers saved in formats different from the HDF5 format defined for Tianlai, and then processed them with \texttt{tlpipe}. 

\subsection{\texttt{tlpipe} data processing workflow}

We use Python-2.7 as the main programming language. This choice allows us to use its vast collection of scientific computing libraries. However, some of the performance-critical parts have been compiled in \texttt{C} by using Cython. Parallel processing has been implemented with the Message Passing Interface (MPI) framework. 

The basic workflow of \texttt{tlpipe} can be broken into 3 distinct components: task manager, tasks, and data container. 

\paragraph*{Task manager:} 
The task manager controls the overall flow of the pipeline and applies different Tasks to the data. 

\paragraph*{Tasks:}
\emph{Tasks} are independent codes that operate on the data in the data container. These tasks include CNS removal, RFI flagging, map making, and calibrations. We discuss some of the tasks in the next section. 

\paragraph*{Data container:}
The data container holds the data operated on by the pipeline. It includes an array of visibilities, a Boolean mask array corresponding to each visibility, and some supplementary data. 

The overall data processing workflow of \texttt{tlpipe} appears in Figure~\ref{fig:tplipe_dataprocessing}. 
The list of tasks is entered into a \texttt{*.pipe} file. The task manager takes the file as an input and applies the tasks to the data container in sequence. The tasks, in general, modify the visibility and the mask array in the data container.  

\subsection{Built-in data processing tasks}
\label{sec:tasks}

In the {\tt tlpipe} package, we have implemented more than 30 tasks. Users can write their own independent tasks to apply to the data container. Some of the built-in tasks in the \texttt{tlpipe} are as follows:

\paragraph*{Masking of the CNS}
The Tianlai data contain periodic calibration signals from the CNS which must me removed during analysis. A dedicated subroutine detects the presence of the CNS signal by measuring the difference between the amplitude of two consecutive data points in an auto-correlation channel and comparing it with the overall variance. The routine calculates the turn-on and turn-off times of the CNS and sets elements of a Boolean mask array to True when the CNS is on. In Figure~~\ref{fig:mask_rf} the masked CNS can be seen as periodic vertical straight lines. 

\paragraph*{RFI cleaning}
After masking the CNS we need to clean RFI from the data. Multiple RFI flagging algorithms are available in \texttt{tlpipe}, but a combination of the sum threshold method \citep{Offringa2010} and the scale-invariant rank (SIR) operator method \citep{Offringa2012} work best for the Tianlai data. Visibility data points flagged as RFI are recorded in the mask array.
The RFI masked from 1 hour of nighttime data using 
the sum threshold + SIR operator method is shown in Figure~\ref{fig:mask_rf}. 

\paragraph*{CNS calibration (relative calibration)}
Two methods for calibration are included in \texttt{tlpipe}. For absolute calibration of the amplitude and phase of the gain we use strong astronomical point sources (next section). However, as only a few bright sources are available and accessing them requires repointing the dishes, we also perform relative calibrations using a regularly broadcast CNS signal. The CNS calibration is primarily used to remove the phase variations over time but we are studying its use for amplitude calibration as well. 
 
We developed two different algorithms for the calibration using the CNS. The first task, \texttt{nscal}, uses the CNS to calibrate each visibility. For each baseline, it defines the visibility during the on and off cycles of the CNS to be $V^{\rm ON}_{a,b}=V_{a,b} + V^{\rm NS}_{a,b}$ and $V^{\rm OFF}_{a,b}=V_{a,b}$, where $V_{a,b}$ is the observed visibility from the sky and $V^{\rm NS}_{a,b}$ is the visibility from the CNS corresponding to the baseline ${a,b}$. The phase introduced by the CNS is then $\phi_{a,b} = {\rm Arg}(V^{\rm ON}_{a,b}-V^{\rm OFF}_{a,b})$.  Because the CNS phase is constant for a particular baseline, and the CNS amplitude is assumed to be constant,  the corrected visibility from the sky, after the CNS calibration, is $V^{\rm NS-Cal}_{a,b}=\exp(-i\phi_{a,b})V_{a,b}/|V^{\rm NS}_{a,b}|$, where the CNS signal is assumed to be much larger than the sky signal. In fact, because the CNS signal enters through the sidelobes of the beam, its amplitude is not very stable.  We use the CNS primarily for phase calibration.
Further details appear in \cite{Zuo2019}.

The second task, \texttt{nscalg}, uses the CNS to perform a global fit to the observed visibilites to determine an independent (complex) gain for each feed. Because only phase differences between feeds matter, the phase of feed 1 is fixed to be zero without any loss of generality. The gain amplitudes reported are relative to the (uncalibrated) CNS amplitude. This method is used in the CNS calibration results shown in Sec.~\ref{sec:calibration}. Because the gain changes with time, a spline is fit to the measured gains and used to interpolate between CNS calibration events. Because there are only $N$ gains but $N\times N$ visibilities, this process is not perfect. 

\paragraph*{Point source calibration (absolute calibration)}
\label{sec:PSC}
After the relative phase calibration using the CNS, transits of strong astronomical radio sources are used to make an absolute gain calibration that gives the actual amplitudes and phases of the complex gains for each feed. 
The solution is obtained by fitting the transit signal for each of the $N(N-1)/2$ visibilities (assuming known geometry) for each frequency, independently. 
The algorithm decomposes the $16\times 16$ visibility matrix for each polarization to yield the complex gains for the 16 feeds.  The H and V feed gains are determined independently. 
Calibration is provided in units of K or Jy.
 
Again, there are two calibration routines.  \texttt{PSCal} and \texttt{PScal2} both use the same robust principal value decomposition to determine the gain of each feed, but have some differences in the handling of outliers and diagnostic information.  See \cite{Zuo2019,Zuo2020} for details. The computed gain is applied to the entire data set until the next point source transits. 


\paragraph*{Map-making}
The built-in mapmaking code uses the $m-$mode analysis from \citep{Shaw2014,Shaw2013,PAON4_Zhang_2016}. 
There are also independently-written mapmaking codes which we use for data analysis. Details are discussed in Sec.~\ref{sec:maps}.

\paragraph*{Utility tasks and plotting tasks}
Apart from these standard tasks, \texttt{tlpipe} includes multiple utility packages and plotting tasks. The utility packages include codes such as those for removing contamination by the Sun from the daytime data. This technique uses an eigenvalue approach for removing the largest eigenvalue from the daytime data and can successfully remove $99\%$ of the solar contamination. It will be described in a future publication. Other utility packages include removing bad channels, daytime masking, etc. The plotting packages include codes for plotting waterfall plots, plotting time or frequency slices of the data, etc. 

\section{Calibration}
\label{sec:calibration}

As described in Section \ref{sec:analysis}, two complementary methods are used to calibrate the amplitude and phase of the electronic gains of the receivers.  Transits of point sources are used to obtain absolute gain and phase calibrations, every few days, up to a maximum of two or three times per 24 hours, while the CNS calibration procedure allows tracking every few minutes of the electronic phase calibration drift between the on-sky bright source calibration.  
(In fact, for observations of the NCP region, where there are no bright point sources, point source calibration requires repointing the dishes away from the pole. In the future, calibration from bright point sources that appear in the antenna sidelobes (see Figure \ref{fig:NCPDirtyMap}) may prove useful, but that topic is beyond the scope of this paper.)  In this section we apply these two types of calibration to the data and evaluate the stability of the array's gain (amplitude and phase) over time. 

\begin{figure*}
  \centering
  \includegraphics[width=3in]{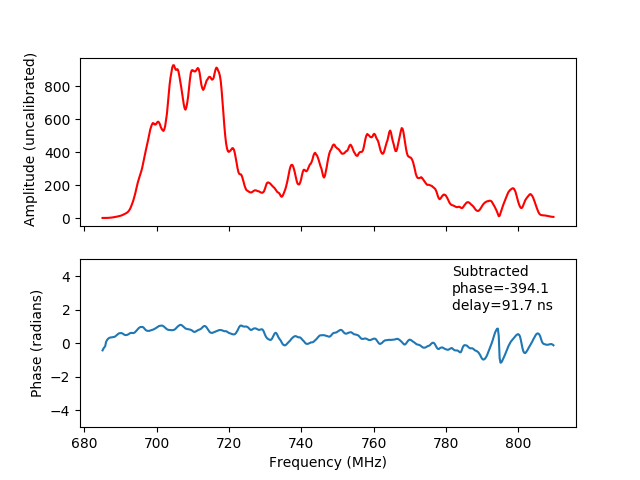}
  \includegraphics[width=3in]{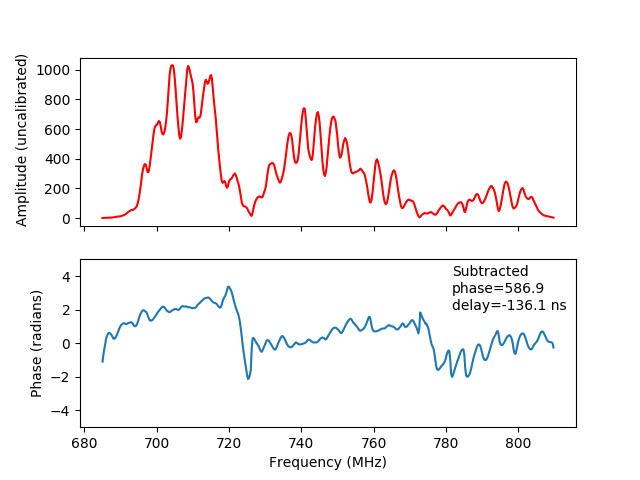}
  \caption{Left: Gain versus frequency for the cross-correlation of feeds 1H and 3H, measured using the CNS with \texttt{nscalg}. 
  The amplitude is in raw (uncalibrated) visibility units and the phase is in radians. Right: Gain versus frequency for the cross-correlation of feeds 3V and 16H.  The amplitude is in raw (uncalibrated) visibility units and the phase is in radians. The two baselines used in this plot are fairly typical. 
}
  \label{fig:gainvsfreq}
\end{figure*}

\subsection{Gain stability measured with the CNS}
\label{sec:calibration:stabilityNS}

The response of the dishes to the CNS is not a smooth function of frequency and the shapes of the response vary widely. Based on the comparison with astronomical point source calibration and normal observation data described in latter sections, we believe much of this frequency structure arises because, for most observing directions, the CNS is coupled to the antennas through their far sidelobes. 
Electromagnetic simulations of these far sidelobes demonstrate significant variation with frequency and position of each dish with respect to the CNS.  For these reasons the CNS is used only to calibrate the relative phases of the gains of the receivers, not their amplitudes.  We are investigating whether it could be used to calibrate the relative amplitudes as well. Figure \ref{fig:gainvsfreq} (Left) is a ``typical'' H-H cross-correlation and Figure \ref{fig:gainvsfreq} (Right) is a typical V-H cross-correlation. The plots show the cross-correlation of feeds in different dishes, but when the feeds are in the same dish the plots are similar. The phase plots have a phase and delay 
that is fit and subtracted from the visibility phase so that the residual phase is close to zero. 
Note that the amplitude of the response of the H-H and V-H polarizations are similar.  

\begin{figure}
    \centering
    \includegraphics[width=3in]{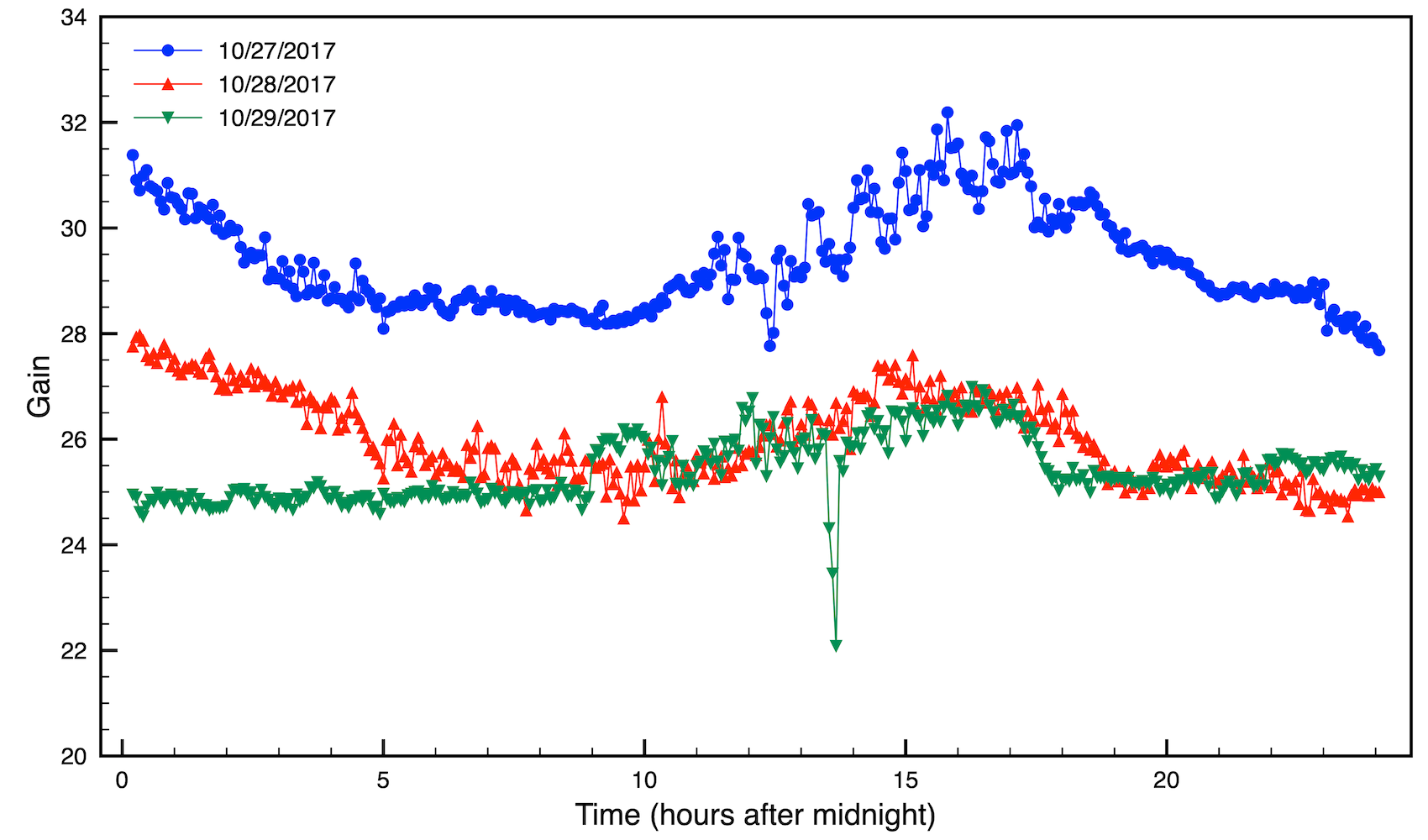}
    \includegraphics[width=3in]{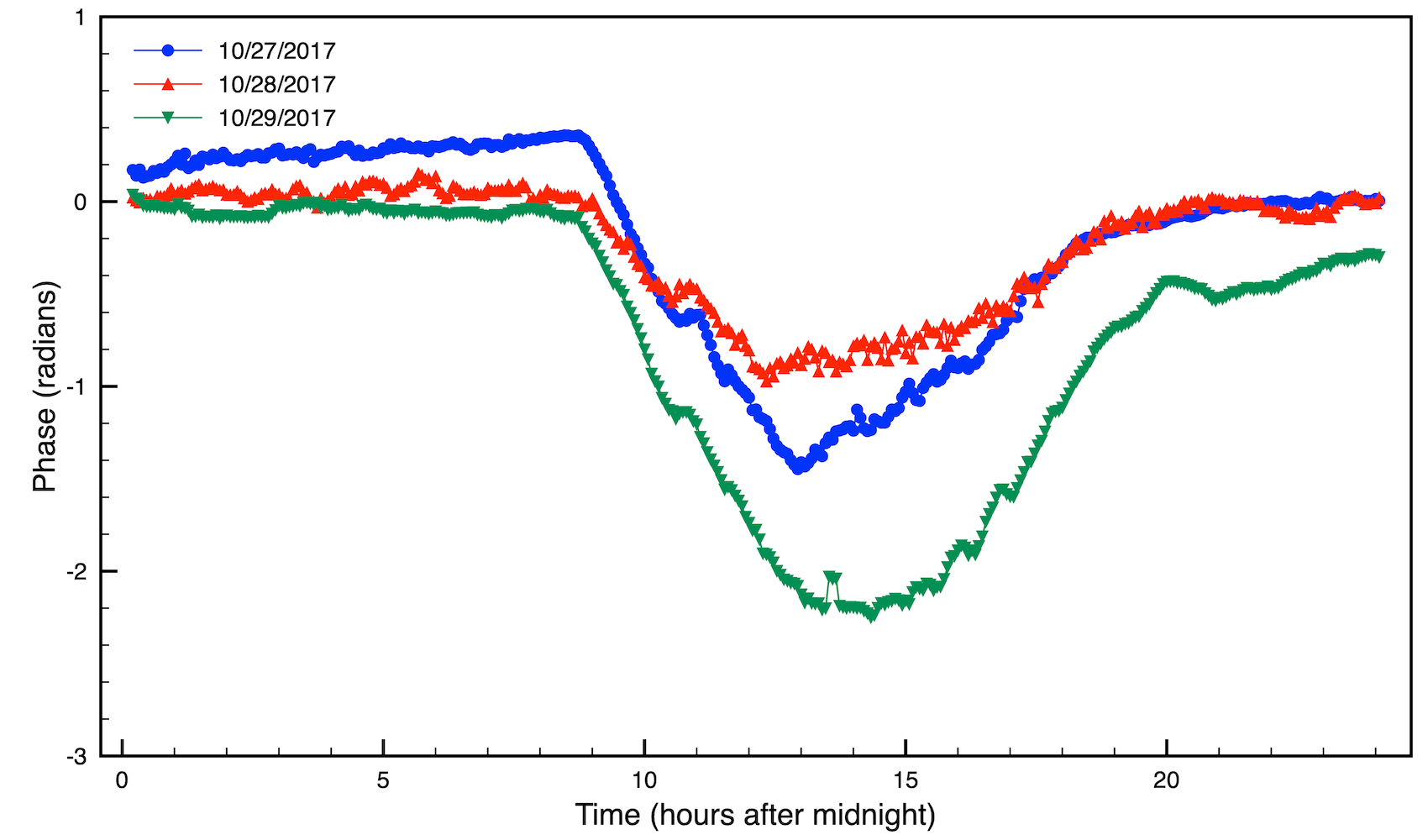}
    \includegraphics[width=3in]{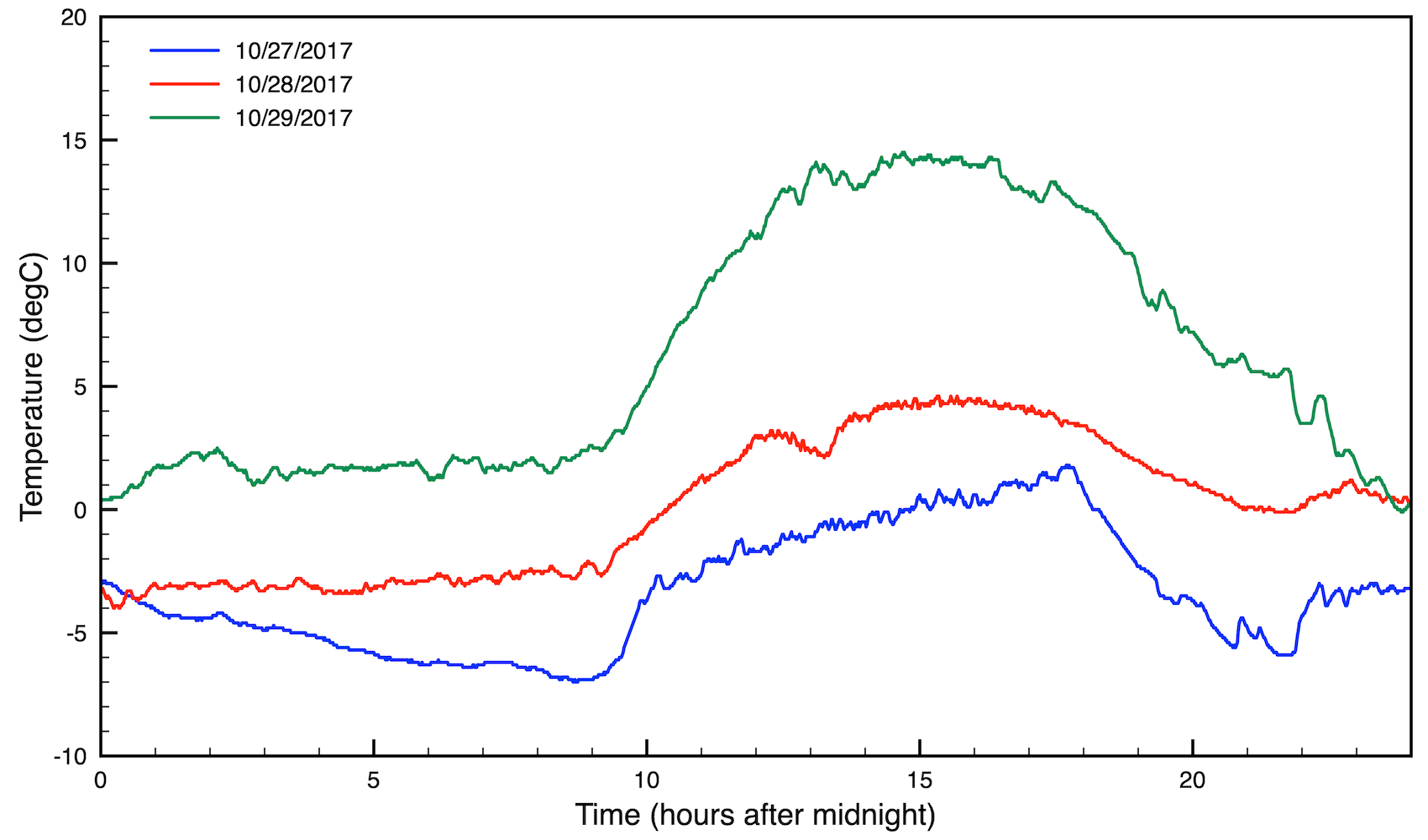}
    \caption{Gain versus time for feed 5V at 747.5~MHz for three days in October, 2017. Each color represents a different day. The gain amplitude (top) scale is uncalibrated and the gain phase (middle) scale is in radians. Site temperature for 3 days is shown in the bottom plot.}
    \label{fig:gainvstime10}
\end{figure}

Figure \ref{fig:gainvstime10} shows the gain as a function of time for feed 5V with respect to time for 3 consecutive days in October 2017. The gain is measured using the \texttt{nscalg} task described in the previous section.   The site temperature recorded for the same 3 days is shown in the bottom plot of Figure \ref{fig:gainvstime10}. 
The changes in gain amplitude and phase are correlated with each other and with temperature, particularly on short time scales, but the relationship is not 1-to-1. However, it is reasonable to expect that the temperature of the electrical components does not follow the site air temperature exactly. Indeed, there is a clear hysteresis behavior during the daylight hours. 
The gain amplitude and phase variations are too large to be caused by temperature changes in either the LNAs or the optical transmitter. Instead, they are likely caused by temperature changes in the fiber optic link between the receivers on the dishes and the correlator. The fibers are contained in cables suspended from telephone poles that traverse 8~km from the dish array to the correlator in the station house.  A phase shift of 1 radian could result from a temperature shift of $10^\circ$~C through a combination of effects: 1) if the fiber lengths were different by about 1\% and the expansion coefficient were $2\times 10^{-5}$ (typical for fiber optic material), or 2)
the temperature dependence of the index of refraction of the fibers varied by 4\% between fibers. Amplitude changes can also occur by changes in the bends in the fibers, which also could depend on temperature. Similar effects are seen in the Tianlai Cylinder Array, which uses an identical analog system; see \cite{Li2020a}.

\begin{figure}
  \centering
  \includegraphics[width=3in]{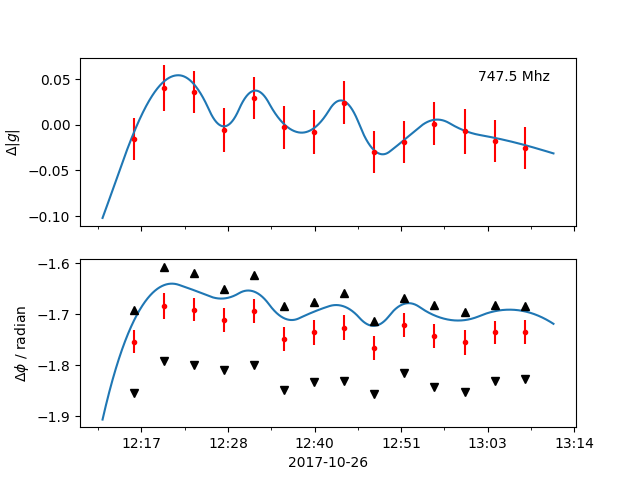}
  \caption{Test of the \texttt{nscalg} calibration task, in which a fit for the gain of each feed is determined using the CNS. These plots are of visibility versus time for the cross-correlation of feeds 1H and 5H.  The blue lines are the spline fit to the expected value of this visibility using the fitted feed gains. The actual response to the CNS pulses is shown as red points.  The amplitude scale is in arbitrary units and the phase change is in radians.   Also shown are the data for both the next higher and lower frequency bins. The fact that the adjacent frequencies show similar behavior suggests that the wiggles are ``real'', rather than noise, and probably due to a time delay.} 
  \label{fig:visvstime19}
\end{figure}

After applying \texttt{nscalg}, which determines the gains of the individual feeds, we can test the calibration process by computing the  visibilities from the fitted feed gains and comparing them to the observations.  Figure \ref{fig:visvstime19} shows such a test.  The observed 1H-5H visibility (red data points, measured when the CNS is on) is compared to the visibility that is expected using the fitted gains for feeds 1H and 5H.  The visibility from the fitted gains can, of course, only be determined when the CNS is active.  However, we can connect the points when the CNS is active with a spline curve to provide a calibration between the times when the CNS is active, as shown by the blue curve in Fig \ref{fig:visvstime19}. There is a small but significant offset of about 0.04 radians, or about $2\degr$, in the phase, but the amplitude is well described by the curve.  This plot is typical of a significant offset; other visibilities have similar offsets with the opposite sign and other visibilities show smaller or no offsets.  The errors are estimated from the variance of the noise signal over the 20 bins of 1~s that are measured while the noise signal is applied.  The amplitude data are much smoother than would be expected from the error estimate.  The apparent overestimate of amplitude error could be expected if most of the amplitude variation were due to fluctuations of the CNS amplitude.  However, the fitted gains are unaffected by any variation in the amplitude of the CNS because \texttt{nscalg} fits only the  relative gains of the feeds.  The phase errors are also measured from the variance of the noise signal but are not sensitive to CNS fluctuations since the visibility contains only the phase difference between feeds.
The fits for each frequency are independent of each other, so plotting adjacent frequencies is an indication that changes in time are not an artifact of the fitting process.

\subsection{Gain stability measured with point sources}
\label{sec:calibration:stabilityPS}
The stability of the instrument was studied by analyzing its response to Cassiopeia A (Cas~A) over 12 days. Cas~A dominates the radio sky in the northern hemisphere.  
The Tianlai array was pointed at a fixed declination of 58.8 degrees, the declination of Cas~A, and operated in driftscan mode. The data are listed as CasAs 20171026 in Table \ref{tab:log}.
We analyzed variations in the magnitude and phase of a typical visibility during repeated transits of Cas~A across the meridian.   The time-dependent response pattern follows the Gaussian profile of the main beam of the antennas shown in Figure \ref{fig:beam_vs_theta}.

\begin{figure}
  \centering
  \includegraphics[width=0.99\columnwidth,trim = 3 0 3 0, clip]{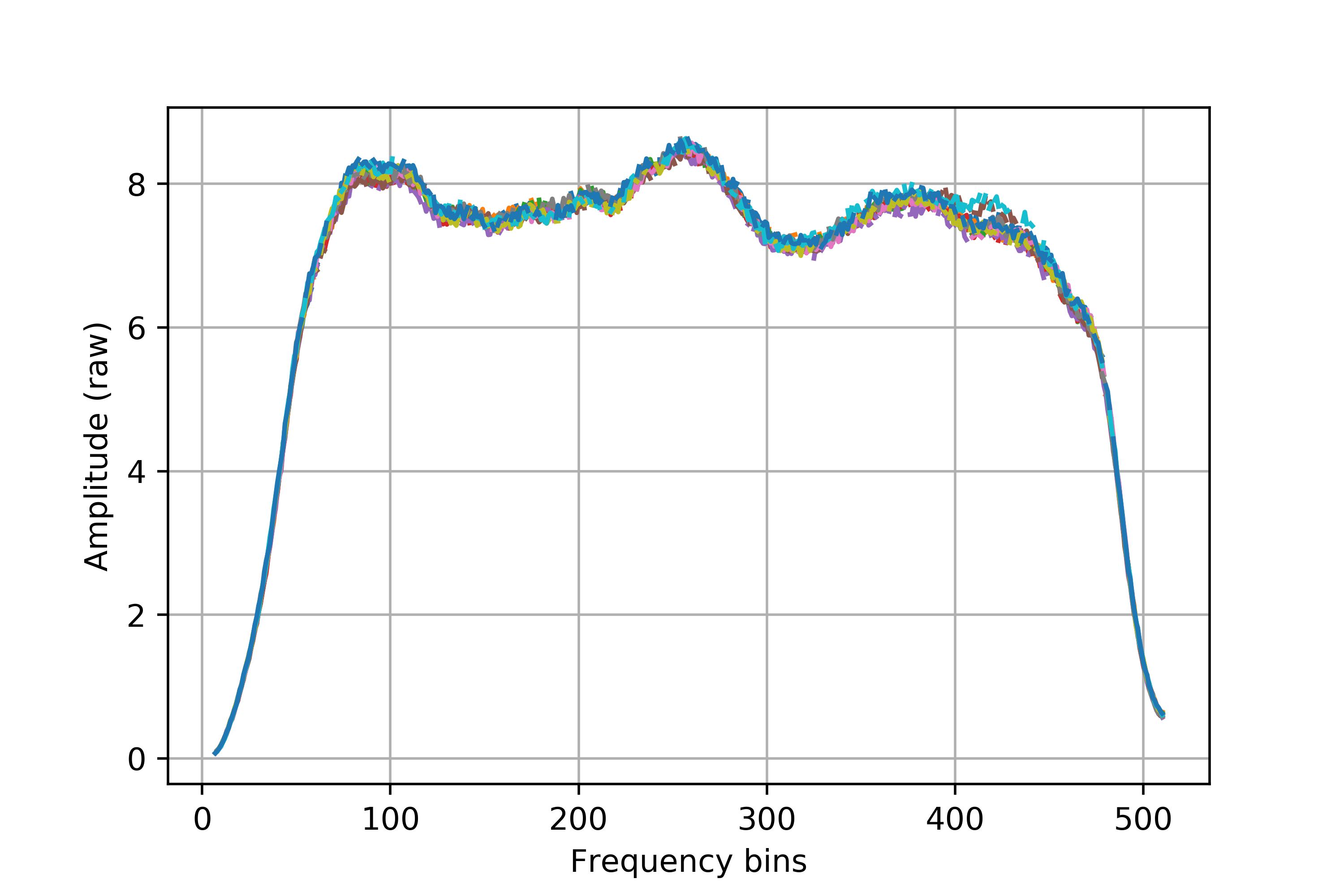}
  \includegraphics[width=0.99\columnwidth,trim = 3 0 3 0, clip]{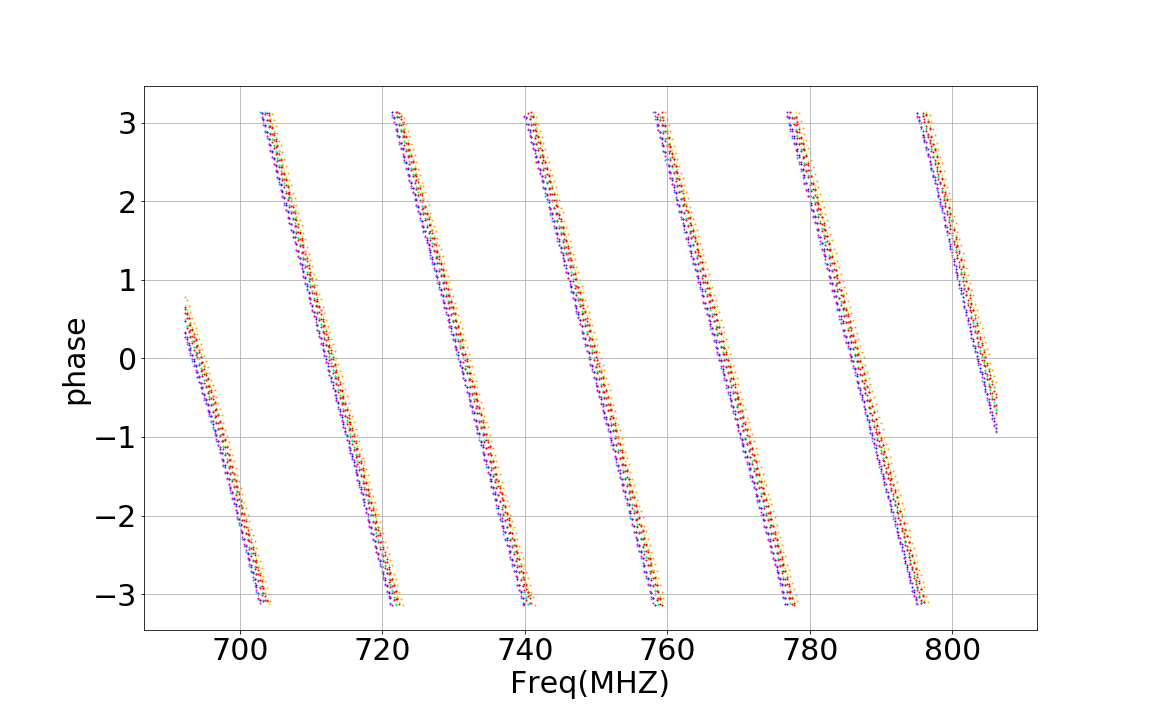}
\caption{ 
Uncalibrated gain amplitude (top) and phase (bottom) versus frequency during transits of Cas~A over 11 nights for baseline 4H-9H. Each colored curve represents the peak response during the transit for each night.}
\label{fig:CasA_gain_vs_freq}
\end{figure}

The amplitude and phase of the uncalibrated peak response for all frequency channels is shown in Figure~\ref{fig:CasA_gain_vs_freq}. 
The response for all 11 nights is plotted, showing that the gain amplitude and phase are quite stable over time. There is significantly less structure in the amplitude spectrum (Figure~\ref{fig:CasA_gain_vs_freq}) than in gain measurements made with the CNS (Figure \ref{fig:gainvsfreq}); as the signal from Cas~A enters through the main beam of the antennas, this suggests that the oscillating structure in the CNS case is probably a result of frequency dependence of the far sidelobes, through which the CNS signal enters.

\begin{figure}
  \centering
  \includegraphics[width=0.99\columnwidth,trim = 3 0 3 0, clip]{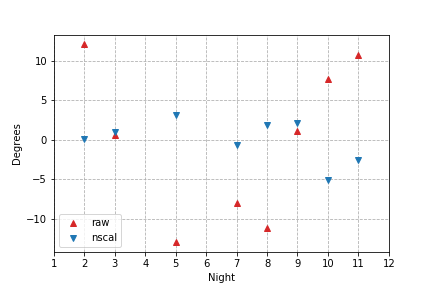}
  \includegraphics[width=0.99\columnwidth,trim = 3 0 3 0, clip]{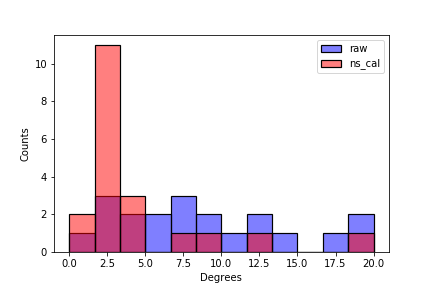}
\caption{ 
Top: Phase variation measured using \texttt{pscal} before and after calibration by the CNS using \texttt{nscal}. The phase variation is calculated as the deviation from the mean of 11 days at the peak of the Cas~A transit. Bottom: Histogram of the phase variation for 20 typical baselines, before and after running \texttt{nscal}. Application of \texttt{nscal} improves the phase deviation significantly. The magenta blocks show overlaps between the two histograms.}
\label{fig:CasA_gain_phs_vs_freq}
\end{figure}

We verify that the phase calibration performed by the CNS with the \texttt{nscal} task over 11 days is consistent with the absolute phase determined by repeated transits of Cas~A over the same period of time in Figure \ref{fig:CasA_gain_phs_vs_freq}. The error is at the level of a few degrees, limited by noise in the phase measurement. 

\begin{figure*}
  \centering
  \includegraphics[width=0.47\textwidth,trim = 0 0 50 0, clip]{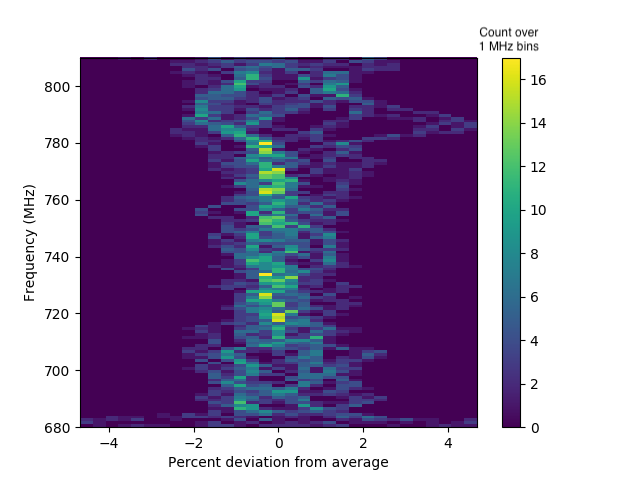}
  \includegraphics[width=0.47\textwidth,trim = 0 0 50 0, clip]{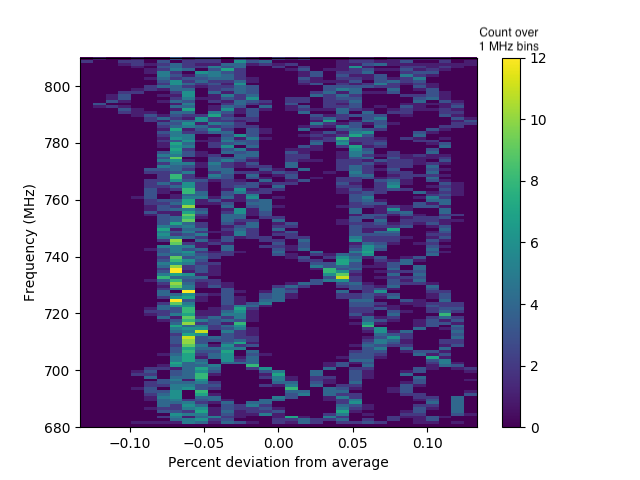}
  \includegraphics[width=0.47\textwidth]{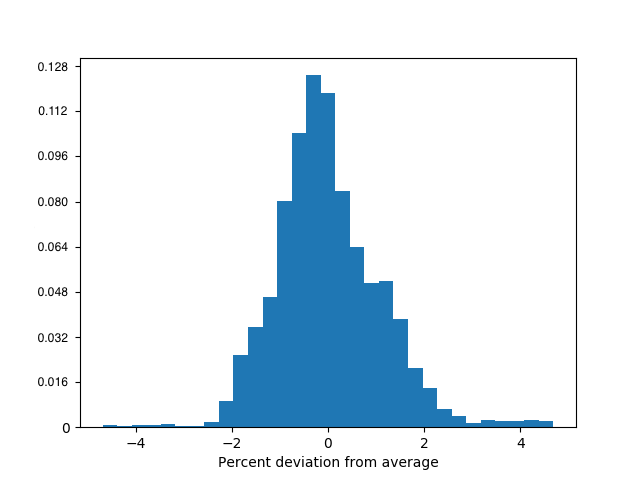}
  \includegraphics[width=0.47\textwidth]{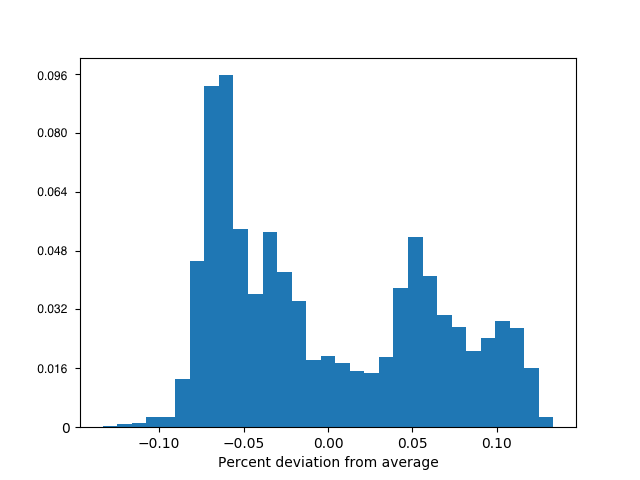}
\caption{ 
Histograms of the gain (left) and phase (right) of visibility 4H-9H measured over 12 days using Cas~A transits for calibration. No prior calibration procedures (e.g. \texttt{nscal}) have been applied. Left panels: Histograms of gain amplitudes.  Left Top: Histogram of gain amplitude variations vs frequency.  The color scale shows number of occurrences.  Frequency bins are 1~MHz wide.  Left Bottom: Same as top figure, but with all frequencies combined. Right panels: equivalent histograms of the variations in the slope of the phase {\it vs.} time.  The bimodal distribution in the phase slope is thought to come from temperature fluctuations, (Figure \ref{fig:gainvstime10}), which we observe to affect the slope of the phase. 
}
\label{fig:gain&phasehistograms}
\end{figure*}

Deviations of the uncalibrated gain from the mean values are shown in Figure \ref{fig:gain&phasehistograms}. The upper left plot shows fractional deviations in the amplitude of the gain compared to the mean for each frequency channel, with $1~{\rm MHz}$ resolution.  The lower left plot shows a 1-dimensional histogram made from the top plot, in which we combine all 512 frequency channels. These gain amplitude fluctuations (s.d. $\sim 1\%$) are about 1/3 of would be expected based on those seen in Figure~\ref{fig:gainvstime10}, where diurnal temperature swings of $10-15^\circ$~C appear to cause gain fluctuations of $\sim 10\%$. The ambient temperature during the 12 Cas~A transits 
varies from night to night with a standard deviation of $4.1^\circ$~C, so gain fluctuations with s.d.$\sim 3 \%$ would be expected.  Because the transits occur at nearly the same time, near midnight, other effects such as direct solar heating of the fibers, or wind, are less important.

For an East-West baseline, we expect the phase of the visibility to vary linearly with time during the times surrounding each Cas~A transit.  The linear coefficient is determined by the baseline geometry and the frequency. The right plots in Figure \ref{fig:gain&phasehistograms} show histograms of fractional deviations from the mean phase slope over 12 days. We believe these small changes in slope are from differential changes in the lengths of the long optical fibers with temperature. The phase of a given baseline has been observed to vary as the temperature changes, as shown in Figure \ref{fig:gainvstime10}. If the timing of the Cas~A crossing corresponds to a time of rapidly-varying temperature, the phase slope will deviate from the expected value by a small amount. The set of days observed in Figure \ref{fig:gain&phasehistograms} happens to include roughly an equal number of days with `large' vs. `small' temperature fluctuations (i.e. temperature changes throughout the day are greater or lower than $10^\circ$ C). This may explain the bimodal nature of the distribution of phase slopes. We know that the application of \texttt{nscal} can reduce the night-to-night phase variations (Figure \ref{fig:CasA_gain_phs_vs_freq});  \texttt{nscal} might reduce the scatter in the right-hand plots of \ref{fig:gain&phasehistograms} as well. 


\subsection{Absolute Calibration}
\label{sec:calibrationabsolute}

Absolute flux calibration is necessary to quantify the sensitivity and accuracy with which the dish array can measure sources dimmer than the calibrators, especially the very dim 21~cm emission which is the primary science target of the Tianlai project.  Initial calibration of dish observation runs (listed in Table~\ref{tab:log}) is obtained by point source calibration on the bright sources listed in that table. Absolute calibration is obtained by comparison to published (external) measurements of the flux of these calibrators.  Accuracy of this procedure is limited not only by the dish array but also by the comparison as discussed in Appendix~\ref{app::calibrators}, where the specific flux models used are also given.

While all of our runs span many days, we point toward calibrators only at the beginning of each run, when the Sun is down and we'd prefer that calibrators be near the zenith.  The brightest, Cyg~A and Cas~A, are always above the horizon but may be far from the zenith at the start of a run or even during the entire night.  This is one reason different calibrators are used for different runs.  Typically two or three calibrators are measured before a run and we may cross calibrate these. The brightest calibrators are easily detected by the interferometer even when they are far off axis, which will allow continuous ``real time'' calibration off of primary calibrators once we have an accurate off-axis beam model.  This will be an adjunct to sky calibration on sources nearer the center of the beam, for which we must develop an accurate sky model using our data.  Accurate absolute calibration of all of the observations will be a bootstrap process. 

One can calibrate visibilities in terms of temperature or flux density. For linear response 
\begin{equation}
V_{a,b,\alpha}^{\rm (raw)}=
g_{a,\alpha}^{(X)}\,g_{b,\alpha}^{(X)\,*}\,
V_{a,b,\alpha}^{(X)},
\label{eq:RawToCalibrated}
\end{equation}
where $a$ and $b$ are antenna indices, $\alpha$ is a frequency channel index (frequency $\nu_\alpha$), $g_{a,\alpha}^{(X)}$ is the complex gain, $V_{a,b}^{\rm(raw)}$ is the uncalibrated visibility value, and $X={\rm F}$ for flux density or ${\rm T}$ for temperature (which we will drop where we need not specify which). 

The temperature gain is
$g_{a,\alpha}^{\rm(T)} = \sqrt{8\pi\,k_{\rm B}/(\lambda^2 D(\nu_\alpha))}\, g_{a,\alpha}^{\rm(F)}$,
where $k_\text{B}$ is Boltzmann's constant and $D(\nu)$ is the beam directivity.  This ensures that $V_{a,a}^{\rm(T)}(\nu)=T_{\rm RJ}$ if the telescope is illuminated {\it only} by an isotropic blackbody with Raleigh-Jeans temperature $T_{\rm RJ}$.

Initial flux density calibration is straightforward using the point source calibration described above.  With accurate flux models of the primary calibrator one obtains an initial value of $g_{a,\alpha}^{\rm(F)}(\nu)$ for all antennas $a$ in all frequency channels. We will track subsequent gain variations using the CNS as well as the sky signal.  

Temperature calibration is derived from the flux calibration, $g_{a,\alpha}^{\rm(T)}$ from $g_{a,\alpha}^{\rm(F)}$.  This conversion requires knowledge of $D(\nu)$, which depends on the entire beam profile including the far sidelobes. Our current knowledge of these comes only from $4\pi$ electromagnetic simulations of the beam of an isolated dish which does not include interactions with neighboring dishes and other environmental factors which will be important far off axis.  The UAV data we have is not extensive enough to estimate a directivity and deviates significantly from the simulations (see Figure~\ref{fig:beam_vs_theta}).  Below we use   
$D(\nu)=1407+5.38\,((\nu/{\rm MHz})-750)$ which is a fit to the simulations in the band $\nu\in[700,800]\,$MHz.  The larger sidelobes found by UAV measurements suggest a smaller $D(\nu)$.  However $D(\nu)$ is bounded from below by the system temperature it implies (see section~\ref{sec:calibration:Tsys}) and it is not plausible that $D(\nu)$ is significantly smaller than 1000.  This might allow $V^{\rm(T)}$ values as much as $\sim40\%$ smaller than the ones quoted below. We quote uncertain temperature-calibrated rather than flux-calibrated visibilities because these quantities are easier to compare with expectations of diffuse or unresolved emission such as from 21~cm signal.

Uncertainties in the far off-axis beam and concomitant temperature calibration is not a major limitation in using the array to accurately map regions of the sky observed by the well understood central part of the beam when we use flux calibrated visibilities.  This is particularly true of observations of the NCP, where the regions of the sky which are far off axis remain far off axis at all times. A more problematic uncertainty in obtaining accurately calibrated visibilities comes from the residual drifts in gain identified in sections \ref{sec:calibration:stabilityNS} \& \ref{sec:calibration:stabilityPS}.  In future work we expect to be able to track these residuals using sky calibration.

\subsection{System temperature}
\label{sec:calibration:Tsys}


\begin{table}
\caption{We list here the average $T_{\rm sys}$ and the SEFD of the 27 fully functioning feed antennas during the 9 full nights of the 3srcNP 20180101 run.  The average is over the 700-800$\,$MHz band and excludes times when the CNS is transmitting. The entries in red are identified as ``hot antenna'' with $T_{\rm sys}$ more than 4 standard deviations above that of the 22 remaining feed antennas.}
\label{tab:SystemTemperature}
\begin{tabular}{|r|r|r|r|r|}
\hline
     & \multicolumn{2}{l|}{System temperature (K)} & \multicolumn{2}{l|}{SEFD (kJy)} \\ \hline
Dish & H-pol. & V-pol. & H-pol. & V-pol. \\ \hline
1    & -- & \red{107.5} & -- &  \red{16.6} \\ 
2    & 72.9 & 78.9 & 11.3 & 12.2 \\ 
3    & {\color{red} 87.0} &  \red{159.6} & {\color{red} 13.5} & {\color{red} 24.7} \\ 
4    & 78.3 & 76.0 & 12.1 & 11.8 \\ 
5    & 76.9 & 71.2 & 12.0 & 11.1 \\ 
6    & 77.6 & 72.2 & 12.0 & 11.2 \\ 
7    & --   & --   & --   & --   \\ 
8    & 76.1 & 81.5 & 11.8 & 12.6 \\ 
9    & 76.8 & {\color{red}209.4} & 11.9 & {\color{red} 32.8} \\ 
10   & 72.6 & 72.4 & 11.2 & 11.2 \\ 
11   & {\color{red}236.5} & 74.5 & {\color{red} 36.6} & 11.5 \\ 
12   & 79.5 & 77.3 & 12.3 & 12.0 \\ 
13   & 73.7 & 77.1 & 11.4 & 11.9 \\ 
14   & --   & --   & --   & --   \\ 
15   & 75.0 & 78.7 & 11.6 & 12.2 \\ 
16   & 75.9 & 70.5 & 11.8 & 10.9 \\ \hline
\end{tabular}
\end{table}

We define the system temperature by $T_{\rm sys}\equiv V_{a,a}^{\rm(T)}$ and corresponding system equivalent flux density by
${\rm SEFD}\equiv V_{a,a}^{\rm(F)}$; these are just the calibrated auto-correlations visibilities.  These quantities vary by only a few percent with frequency and time (except when the dishes are pointed toward a very bright source) since they are dominated by the receiver and ground temperature, whose fractional variation is small.  In Table~\ref{tab:SystemTemperature} we list the mean $T_{\rm sys}$ and SEFD for the nighttime data analyzed in section~\ref{sec:NCPASN}.  Out of the 32 feed antennas, 5 (1H, 7H, 7V, 14H, 14V)) are not functioning and 5 of the remaining 27 are identified as having abnormally large $T_{\rm sys}$, while the remaining have mean and standard deviation $(75.7\pm2.7)\,$K or for SEFD  $(11.7\pm0.4)\,$kJy.  The receiver noise temperature is dominated by the LNAs, which have a laboratory-measured noise temperature of $\sim35\,$K. The remaining $40\,$K should be mostly due to ground emission, $T_{\rm spill}$, which is approximately what is expected from the beam simulations.

If $D(\nu)$ were $40\%$ smaller than indicated by the simulations then $T_{\rm sys}\approx55\,$K requiring $T_{\rm spill}\lesssim20\,$K which isn't plausible given that smaller $D(\nu)$ is due to larger sidelobes, which would increase, not decrease, our expectations for $T_{\rm spill}$.

\subsection{Sensitivity}
\label{sec:sensitivity_originally_Tnoise}

System temperature is a useful quantity because it gives the minimum value of random noise (fluctuations) in the visibilities. If the illumination and receiver noise give   Gaussian random phase voltages then the ideal radiometer equation tells us the variance of the visibility is
\begin{equation}
\text{Var}(V_{a,b,\alpha})\equiv \langle V_{a,b,\alpha}\,V_{a,b,\alpha}^*\rangle
-|\langle V_{a,b,\alpha}\rangle|^2
=\frac{\langle V_{a,a,\alpha}\rangle\,
       \langle V_{b,b,\alpha}\rangle}
{\#_\text{sample}}
\label{eq:IdealRadiometerEquation}
\end{equation}
where $\langle\cdots\rangle$ is the expectation of visibilities averaged over realizations of voltage streams for fixed illumination pattern and $\#_\text{sample}$ ($=\delta\nu\,\delta t$ for continuous sampling) is the number of complex Fourier amplitudes averaged by the correlator to obtain a visibility in a pixel.  This variance provides a fundamental statistical limit on the accuracy of measurements of the illumination pattern. To the extent that the signal varies little between pixels one can average them into larger pixels, increasing $\delta\nu\,\delta t$ and decreasing the variance of the average.

\subsection{Sensitivity vs Integration Time}

\begin{figure}
 \centering
 \includegraphics[width=3.5in]{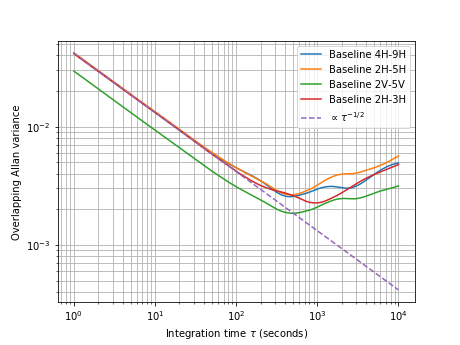}
 \caption{Overlapping Allan variance \protect\citep{riley2008handbook} vs. integration time, $\tau$, for four typical baselines centered at 747.5~MHz with bandwidth 0.244 MHz during nighttime only, for the real part of the visibility. The visibility is uncalibrated and the vertical axis is in arbitrary units. The intercept of the dashed line $\propto \tau^{-1/2}$ is adjusted so that it matches the variance trend. The plot shows that the noise integrates down as $1/\sqrt{\tau}$, as expected, for about 300 seconds. The imaginary part of the visibility shows similar behavior.}
\label{fig:rms_vs_inttime}
\end{figure}

The level of visibility fluctuations due to noise or system temperature (RMS) is shown in Figure~\ref{fig:rms_vs_inttime}.
The plot shows that the noise integrates down with integration time as expected for white noise, up to 300 seconds. Afterwards, it starts increasing due to rotation of the sky.

\section{Maps around sources}
\label{sec:maps}

The 16 dishes of the array  provide 16 auto-correlations and 120 cross correlation visibilities for each of the two linear polarisations (HH or VV), as well as 256 cross polarisation (HV) visibilities. To illustrate the array performance, we have reconstructed sky maps around a few bright point sources by combining single linear polarisation HH or VV signals. The sky maps shown here have been obtained through several algorithms which are briefly described in Appendix \ref{annex:maps}.

Figure~\ref{fig:CasArecmap_cor} (top part) shows the image of Cas~A, 
reconstructed using \texttt{QuickMap}, and data  from the 12-day October 2017 driftscan at the source declination. 
We have used a time interval of 4 hours of only one  of the 12 transits (2017-10-30) and a single frequency channel,
(244 kHz bandwidth)  at  747.5~MHz. The map shows a band of declination around Cas~A, from $53\degr < \delta < 63\degr$, covering 60 degrees in right ascension $ 300\degr < \alpha < 360$,  with $\sim 0.1\degr \sim 5$ arcmin resolution. 


\begin{figure}
  \centering
  \includegraphics[width=3.5in]{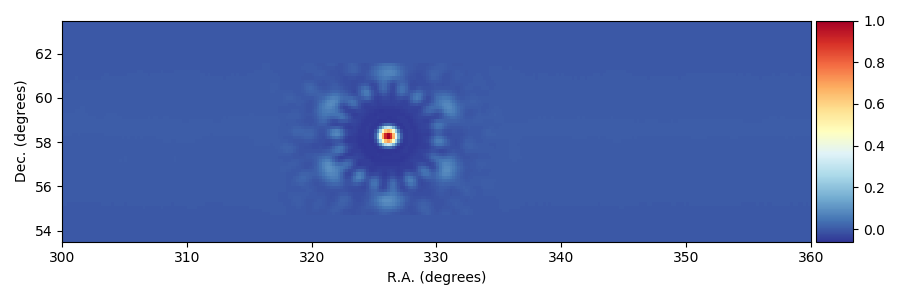}\\
  \includegraphics[width=3.5in]{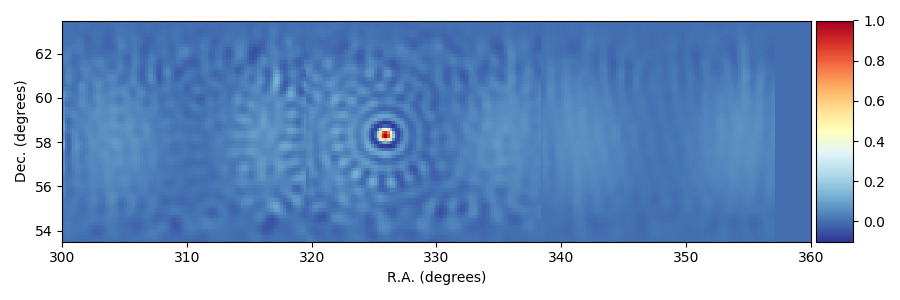}
\caption{ 
(top) Dirty image 
of Cas~A at 747.5 MHz, applying \texttt{QuickMap} to a drift-scan from the 2017/10/30 transit using 121 horizontal cross-correlation baselines and one auto-correlation. 
(bottom) Image of Cas~A reconstructed by pseudo-inverting the pointing matrix from the 2017/10/30 transit, in the frequency slice around 747.5 MHz. The method was applied to three successive rectangular areas before, during, and after the actual Cas~A transit to cover the same sky area as the dirty image shown on top. 
The maps include corrections for phase and relative gain.  The color scales are linear, in arbitrary units.}
\label{fig:CasArecmap_cor}
\end{figure}


The visibility data form a $(720 \times 121)$ complex array, where $720$ is the number of time samples from 4 hours of observation, each sample averaged over a 20 second time interval, and $121$ is the number of horizontal polarisation cross-correlations (H-H) plus one auto-correlation (16H).
A complex gain correction term has been computed, comparing the observed visibilities with the ones expected  for the Tianlai array geometry, and a simplified sky model with only one point source at the Cas~A position.

Using the same data set, we have used \texttt{BFMTV} to reconstruct a cleaner map. However, to limit the linear system size,  the four hours visibility data has been split into three parts, each covering about 80 minutes.  Three independent map tiles, each covering $\sim 20 \degr$ in RA, with some overlapping guard area, have been computed and assembled side by side to obtain the full map, covering 60 degrees in RA, as shown in figure \ref{fig:CasArecmap_cor} (bottom part). The main improvement in the map quality is the suppression of side lobes, which is clearly visible. On the other hand,  non-Gaussian beam features  induce low amplitude patterns in the \texttt{BFMTV} map, as a pure  Gaussian beam is assumed to build the pointing matrix. Contributions from sources outside the reconstructed map area are not handled in the version of \texttt{BFMTV} used here;  these show up as an additional fluctuation pattern, easily visible in the tile on the left side.  
\begin{figure*}
  \centering
  \begin{tabular}{cc}
  \includegraphics[width=2.5in]{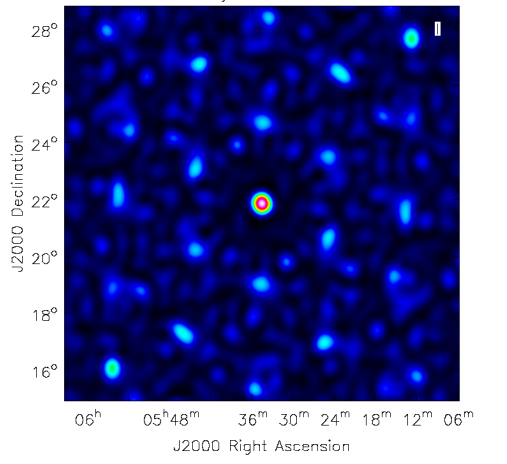}
  &
\includegraphics[width=2.5in]{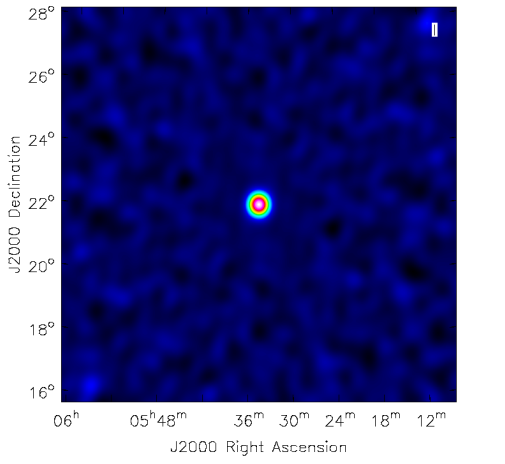}
  \end{tabular}
\caption{ 
(left) Dirty image of M1 using \texttt{CASA} and 1 hour of data, $(749.9\,\text{MHz} - 774.3\,\text{MHz})$. 
(right) 
Clean image of M1 using \texttt{CASA} and the same data as the left part.}
\label{fig:M1MapsCasA}
\end{figure*}

 We also made maps using the public \texttt{CASA}\footnote{Common Astronomy Software Applications package  : \url{https://casa.nrao.edu/}} software.  Figure \ref{fig:M1MapsCasA} show the dirty and clean images of M1 using 1 hour of data around M1's transit on 2018/1/1. 100 frequency channels $(749.9\,\text{MHz} - 774.3\,\text{MHz})$ and all baselines of HH polarization are used. We perform the phase, bandpass, and baseline amplitude calibration. Both images are made using \texttt{CASA}'s \texttt{tclean} task with the number of iterations of the clean algorithm set to 0 to obtain the dirty map and set to 100 to obtain the clean image, respectively.



\begin{figure}
  \centering
  \includegraphics[width=2.5in]{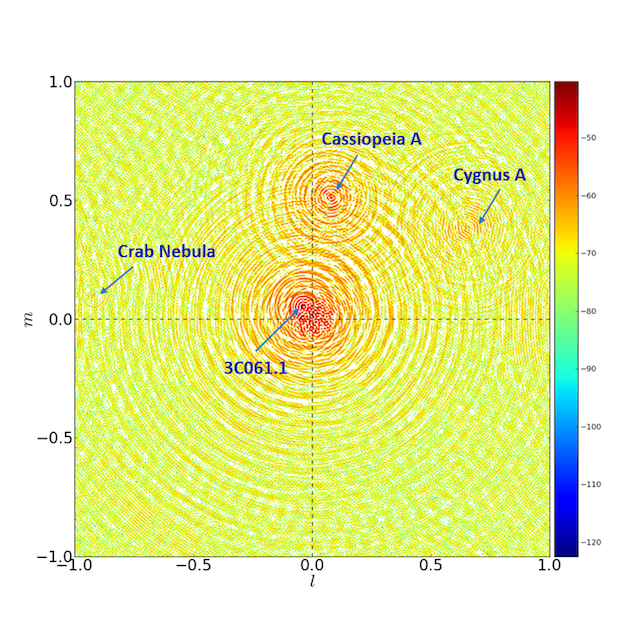}
\caption{ 
Dirty image of the NCP region from all HH and VV cross-correlation baselines and all frequencies $(700\,\text{MHz} - 800\,\text{MHz})$  using \texttt{TLdishpipe}. The color scale is in dB, with arbitrary normalization. The $l$ and $m$ coordinates are the Fourier conjugates to $u$ and $v$, respectively.}
\label{fig:NCPDirtyMap}
\end{figure}

Using {\texttt{TLdishpipe}}, a dirty images of 
the northern celestial hemisphere (NCP)  was constructed, shown in  
Figure~\ref{fig:NCPDirtyMap}. 
 146 hours of data over 10 nights starting on 2018-01-01  and all frequency channels (from 700 to 800~MHz  were included to construct this map. 

We can use these maps to check our understanding of the antenna patterns.  We compared the estimated signal magnitudes from Cas~A and Cyg~A in Figure \ref{fig:NCPDirtyMap} to what we would expect based on known source fluxes and the simulated beam pattern.  To enable the comparison, we averaged the simulated beam pattern in azimuth and in frequency, resulting in a single 
 pattern that depends only on the angle from the beam center of the antennas (Figure \ref{fig:simulated_beam}). Based on the simulated beam pattern, we would expect the flux from Cas~A to be $\sim$10 dB lower than the peak at the NCP, and that of Cyg~A to be $\sim$13 dB lower. This prediction for the flux of Cas~A is in agreement with results shown in Figure \ref{fig:NCPDirtyMap}, while the observed flux from Cyg~A is $\sim$7 dB below expectation. This suggests the simulated beam pattern in Figure  \ref{fig:simulated_beam} is accurate out to an azimuthal angle of 31.2$\degr$, while the average gain at 49.3$\degr$ is about 7 dB lower than the simulation suggests.
 

\section{NCP performance}
\label{sec:NCP}

The North Celestial Pole (NCP) region is selected as the first deep survey region of the dish array. For a Northern Hemisphere transit telescope with a limited field-of-view, the NCP is the only place on the sky that allows continuous observations.  The strategy of long time observation of the NCP gives the highest S/N visibilities within a given survey time.  
Surveying the smallest solid angle possible also yields the largest sample variance, which is a negative aspect of this strategy. At the time of this writing we have accumulated 3700 hours of integration with all dishes pointed directly at the NCP, although we only present a small fraction of it here.

Typical visibility amplitudes when pointed at the NCP are \(\sim500\,\text{mK}\) during daytime and \(\sim50\,\text{mK}\) during nighttime. With \(T_{\text{sys}}\sim 70\,\text{K}\) the noise temperature expected for cross-correlations with \(1\,\sec\) and \(244\,\text{kHz}\) sampling is \(T_{\text{sys}}/\sqrt{\delta\nu \text{$\delta $t}}\sim150\,\text{mK}\) (for the modulus), which would mean these {``}pixels{''} are receiver noise-dominated during nighttime. In sections 8.1-8.7 we present visibilities averaged into {``}1\,min-1\,MHz{''} pixels by taking the mean over \(\delta t=60 \sec\) and \(\delta \nu=4\times244\,\text{kHz}=0.977\,\text{MHz}\) pixels, reducing the noise  to $\sim 10\,\text{mK}$ level, so that nighttime pixels have S/N of a few. In sections 8.8-8.9 we revert to full $244\,$kHz frequency resolution because there we are studying the noise.  During \(1\,\min\) the Earth rotates by \(0.25^\circ\). The maximum angular resolution of the dish interferometer is \(\sim 0.35^\circ\), so sources near the Celestial Equator would not be greatly smeared during 1 minute and in-beam sources near the NCP have negligible smearing. For these rebinned pixels the S/N only vary rarely drops below unity at certain frequencies, sidereal times and baselines. 
 
\begin{figure}
  \centering
  \includegraphics[width=0.47\textwidth,trim = 10 0 0 0, clip]{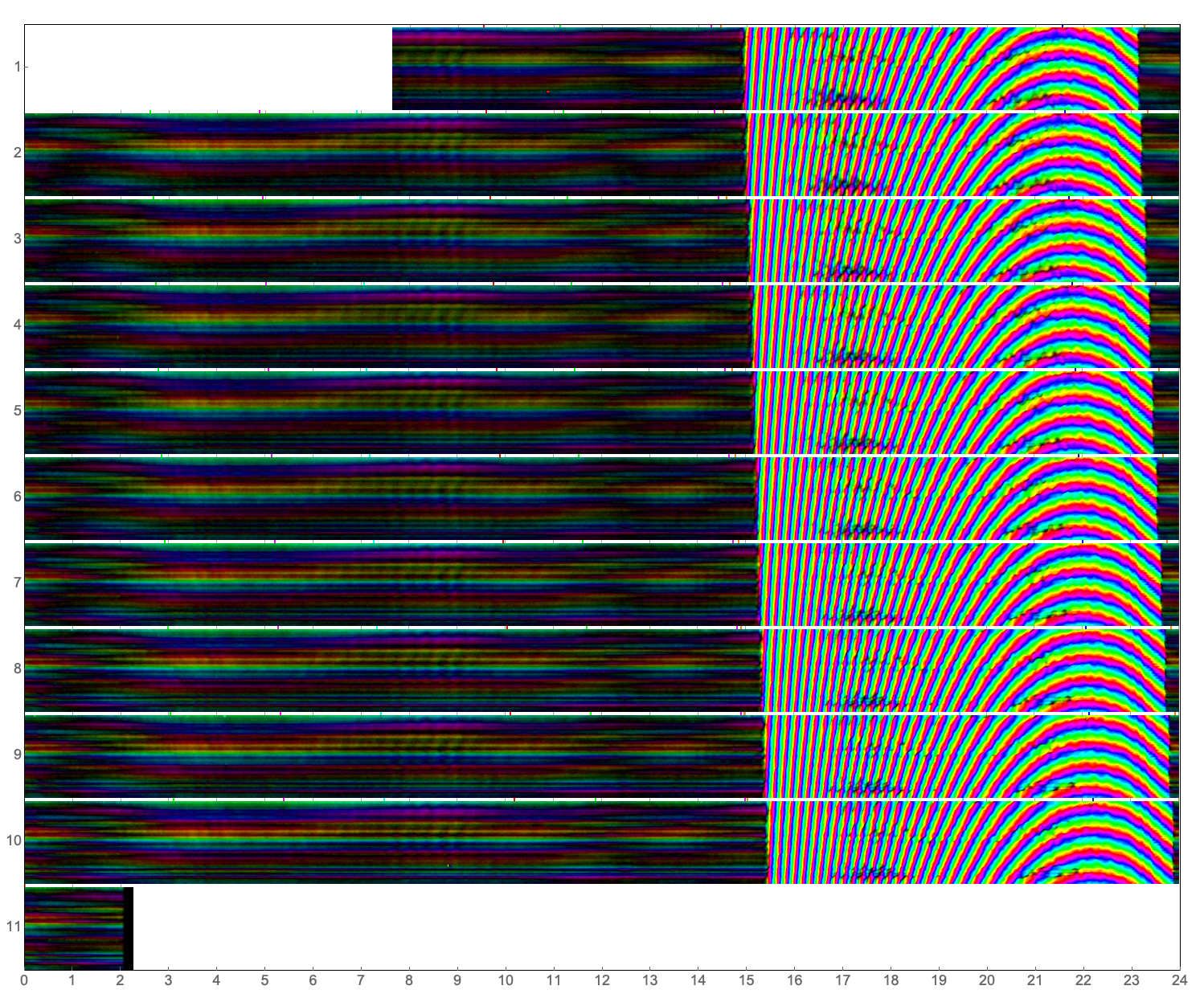}
\caption{A sample of visibilities (2V\(\times\)10V) from 234\,h of observations of the Tianlai dish array from 2018-01-02 to 2018-01-11.  The complex visibilities are represented by colors as described in Appendix~\ref{app::ColorRepresentation}. 
The visibilities are divided into 11 strips corresponding to sidereal days proceeding from top to bottom. The strips are aligned in local sidereal time (LST), in each strip time proceeds linearly from left to right (labeled on bottom) and frequency increases linearly from bottom (700~MHz) to top (800~MHz). What is shown is the mean of the visibility in 1\,min \(\times\) 1~MHz pixels after removing 7 second intervals around the CNS transmitting time. No RFI mitigation or corrections for gain drifts during the 11 days have been been made. 
}
\label{fig::FullDay02V10Vvisibilities}
\end{figure}

In this section we illustrate the NCP data with the visibility from a single baseline (2V $\times$ 10 V) from one run of 234 hours taken between 2018-01-02 and 2018-01-11. During this run only 13 of the dishes (78 of 120 baselines) were fully functional and fairly well behaved.  We focus on one particular visibility which we find to be illustrative of the typical behaviour of the interferometer. 
It has a baseline 5.9\,m East and 9.9\,m North or 11.5\,m total. Figure~\ref{fig::FullDay02V10Vvisibilities} graphically represents this visibility over the entire 234\,h run using the color representation specified in Appendix~\ref{app::ColorRepresentation}. For the purpose of this section precise absolute calibration is not important. Here we will use the initial point source calibration on Cas~A described previously and {\it not} track gain drifts after initial calibration.  Complex gains may drift over many days and nights of observation and in other contexts we plan to correct for this using the CNS and a sky model as described above. Here we wish to illustrate the smallness of the effects of these drifts on sky observations, so no further recalibration is applied. Also no specific RFI mitigation is applied, although at times we use median averaging over successive nights, which is a form of outlier rejection which would remove most RFI.

One clearly sees the much larger solar signal during daytime even though the Sun is \(107.5^\circ\) from the beam center. The day/night transition is fairly sharp, taking only a few minutes. One can see the daylight hours slowly drift to larger sidereal time over successive days as the Earth revolves around Sun. The {``}bow{''} pattern in the daytime phases is what one expects when a bright source passes directly above the direction of the baseline.  The irregularities in the Sun-dominated visibilities are due to interference with other bright off-axis sources, the complicated structure of the beam pattern far off-axis, and also the correlated noise described below.  Apart from the shift due to Earth's revolution, the Solar visibilities, including the irregularities, are highly repeatable.  During nighttime the visibility signal is smaller.

\begin{figure}
  \centering
  \includegraphics[width=0.48\textwidth,trim = 0 0 0 0, clip]{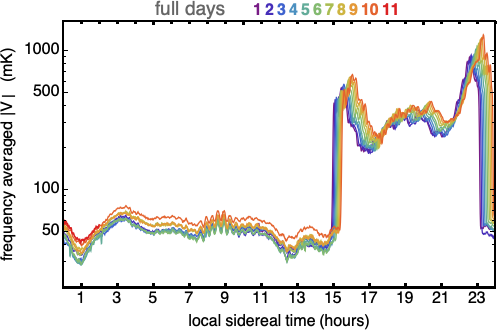}
\caption{The mean modulus of the visibilities of Figure~\ref{fig::FullDay02V10Vvisibilities} averaged in frequency for 1 minute pixels for the full sidereal day. The sequential sidereal days are represented in different colors from purple to red.} 
\label{fig::MeanVofLST}
\end{figure}

A more quantitative comparison of different sidereal days is given in Figure~\ref{fig::MeanVofLST}, which shows the frequency-averaged visibility modulus.  Daytime (roughly \(15^\text{h}\lesssim\text{LST}\lesssim 23^\text{h}\) at this time) is dominated by the Sun, which drifts to larger LST as expected. No LST drift is apparent in nighttime visibilities, i.e. the features do not move in LST.  The most obvious night-to-night variations are in the amplitude of the signal, which is a combination of gain drifts and varying contamination from correlated noise. We reiterate that no correction for time varying gain has been made here.

The Sun contributes from 300 to 1500\,mK and is peaked near sunset (\(15^\text{h}\)) and sunrise (\(23^\text{h}\)) as one would expect for vertical (V) feeds (for horizontal (H) feeds the solar signal peaks near midday). The Sun's motion on the the sky is clearly evident as the pattern shifts to the right on successive days.  Since the Sun's motion is mostly in the R.A. direction, and the beam is centered on the NCP, the Sun will \textit{approximately} trace the same path through the side lobe of the beam on sequential days but at different sidereal times.  Other day-to-day changes in the Sun signal are partly due to gain variation but also due to increasing declination of the Sun. The nighttime signal more accurately repeats every sidereal day although there are up to 20$\%$ on days 10 $\&$ 11. The nighttime signal is a combination of sky signal, which should depend only on sidereal time, and correlated noise, which is roughly constant. The night-to-night variation is a combination of gain drift and variations in the correlated noise.

\subsection{Nighttime Visibilities}
\label{sec:NCPnight}

\begin{figure}
  \centering
  \includegraphics[width=0.45\textwidth,trim = 0 0 0 0, clip]{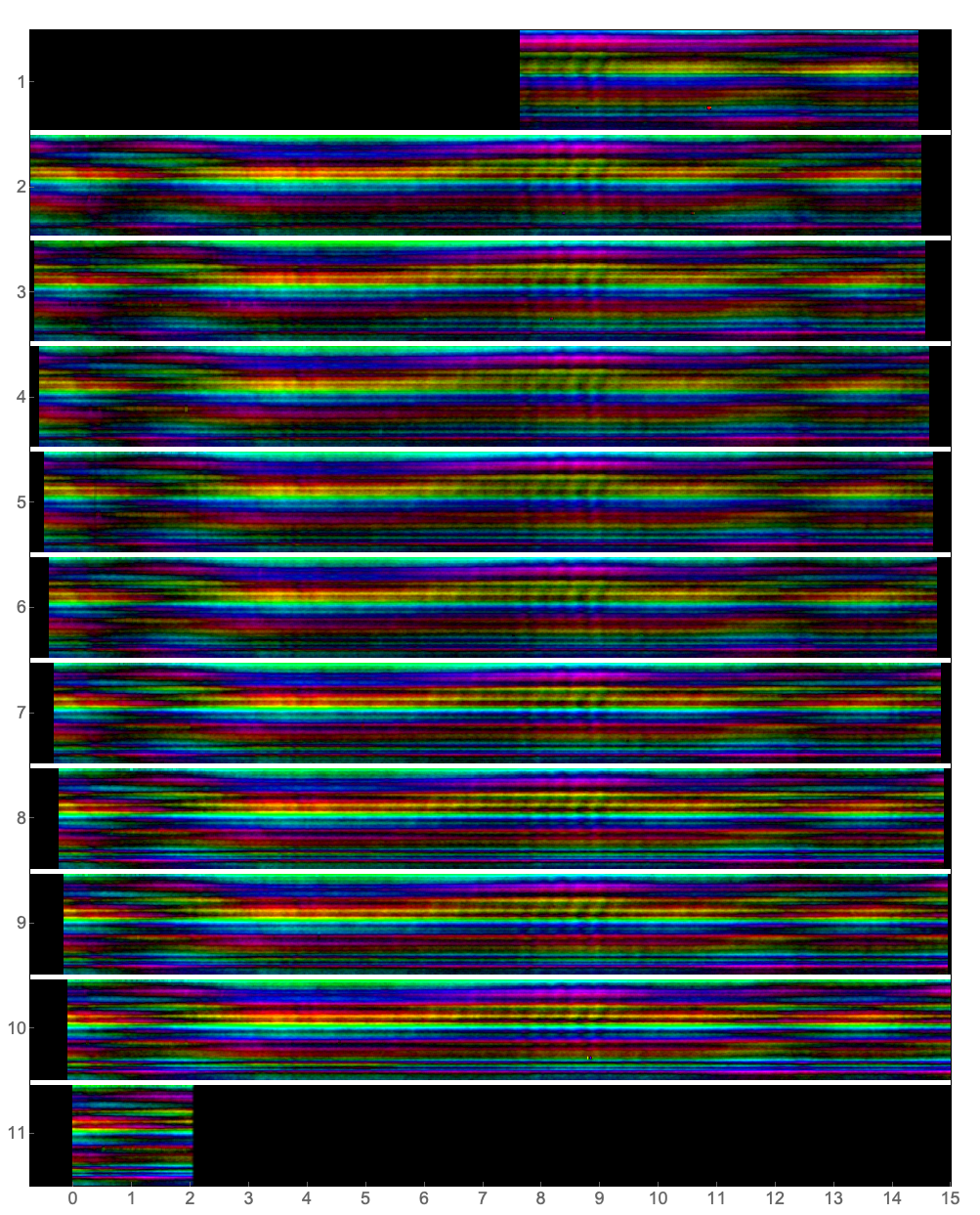}
\caption{Same asa in Figure~\ref{fig::FullDay02V10Vvisibilities} but only times when the Sun is below the horizon are shown, and the color brightness scale has been increased to better show the nighttime visibilities. The nighttime strips do not align in sidereal time since the revolution of the Earth around the Sun causes daylight and nighttime to drift to later sidereal times. The signal that is visible is a mixture of sky signal and correlated noise of similar magnitude.  This is typical of the dish array although some baselines have significantly less and some significantly more correlated noise.}
\label{fig::Nighttime02V10Vnosubtract}
\end{figure}

Figure~\ref{fig::Nighttime02V10Vnosubtract} shows the same visibilities as in Figure~\ref{fig::FullDay02V10Vvisibilities} except only the nighttime data are shown and the color saturation level has been adjusted to better represent the smaller nighttime signal. The dominant features are horizontal stripes with some temporal variation. This is what one would expect if, in addition to sky signal, the data contains a significant amount of constant noise which is correlated between the antennas, or ``correlated noise''. 


Figure~\ref{fig::Nighttime02V10Vnosubtract} also exhibits a few bright pixels such as near \(11^\text{h}\) on the 1st night and near \(9^\text{h}\) on the 10th night. 
These do not repeat with sidereal time, span a small frequency range, and may be external RFI.  RFI flagging will identify these and other less obvious RFI contamination. In this section we do not make use of RFI flagging, however, when we median average different nights, obvious outliers are suppressed.

\begin{figure}
  \centering
  \includegraphics[width=0.47\textwidth,trim = 0 0 0 0, clip]
  {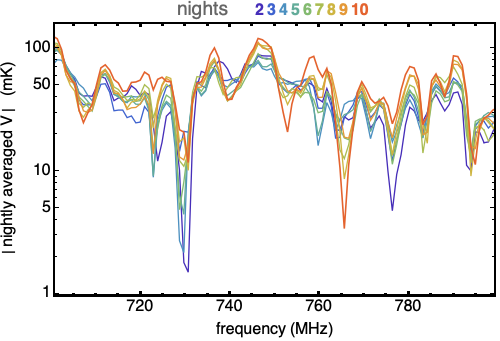}
\caption{The modulus of the nightly averaged (mean) visibilities for 9 nights in 977\,kHz pixels. The 1st and last night are not used since they cover much smaller intervals of LST. Between the 9 nights the nighttime LST interval varies due to Earth's revolution about the Sun.  Because this average is contributed to by both sky signal and correlated noise, to obtain the same contribution from the sky each night this average is computed over the common nighttime (\(23^\text{h}54^\text{m}<\text{LST}<14^\text{h}31^\text{m}\)), which is somewhat shorter than the full range of nighttimes (\(23^\text{h}15^\text{m}<\text{LST}<15^\text{h}01^\text{m}\)) over the 9 nights.}
\label{fig::MeanVofFrequency}
\end{figure}


\begin{figure}
\centering
\includegraphics[width=0.47\textwidth,trim = 0 0 0 0, clip]
{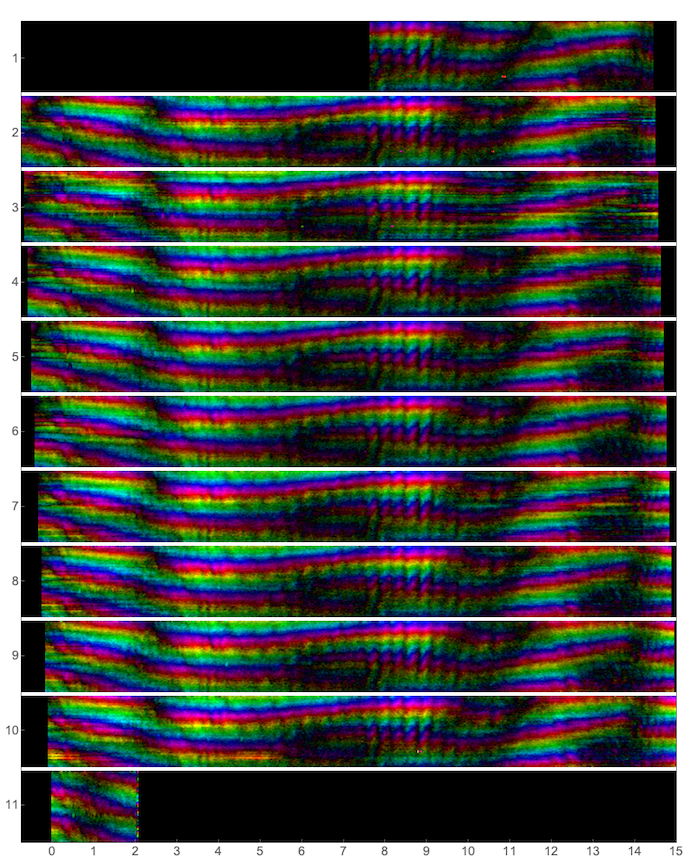}
\caption{Same as in Figure~\ref{fig::Nighttime02V10Vnosubtract} but with the nightly mean from Figure~\ref{fig::MeanVofFrequency} subtracted from each night. This signal accurately repeats every 24 sidereal hours. Also visible are a few bright pixels noted above and a few bright stripes of several hours duration which do not repeat every night (e.g. on day 10 near 715\,MHz appearing near \(0^\text{h}\) and between \(4^\text{h}\) and \(5^\text{h}\)). The bright stripes may be due to RFI but also may be an indication of variability of the correlated noise.}
\label{fig::Nighttime02V10Vsubtract}
\end{figure}

\begin{figure}
  \centering
  \includegraphics[width=0.47\textwidth,trim = 0 0 0 0, clip]
  {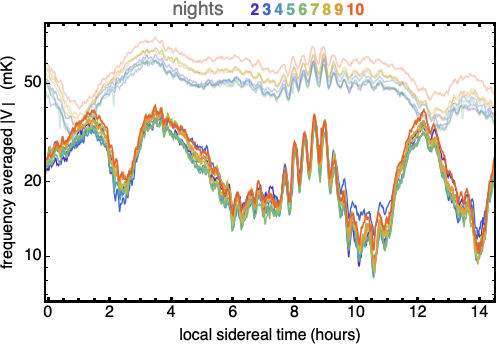}
\caption{The mean modulus of the visibilities of Figure~\ref{fig::FullDay02V10Vvisibilities} averaged in frequency for 1 minute pixels as in Figure~\ref{fig::MeanVofLST}. Here we restrict to common sidereal nighttime and nights 2-10. The top, lighter curves are the same as in Figure~\ref{fig::MeanVofLST}, while the bottom curves are after subtracting the nightly mean, whose modulus is shown in Figure~\ref{fig::MeanVofFrequency}.}
\label{fig::MeanAbsVofLSTnight}
\end{figure}

\begin{figure*}
  \centering
  \includegraphics[width=0.87\textwidth,trim = 0 0 0 0, clip]
  {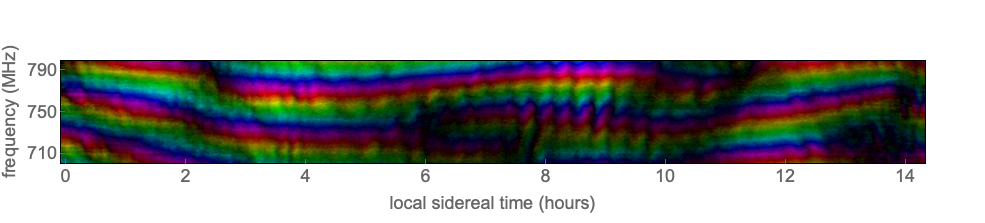}
\caption{The average sidereal night (ASN) visibilities for baseline 2V$\times$10V. This is the median average of 1\,min-1\,MHz visibilities at the same LST and frequency from the 11 nights shown in Figure~\ref{fig::Nighttime02V10Vsubtract}.}
\label{fig::ASNmedian}
\end{figure*}

\begin{figure*}
  \centering
  \includegraphics[width=0.87\textwidth,trim = 0 0 0 0, clip]
  {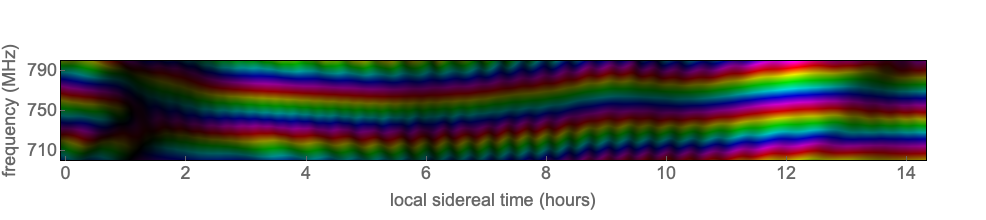}
\caption{Simulated visibility of Figure~\ref{fig::ASNmedian} based on the NVSS catalog using an Airy disk beam model (see Section~\ref{sec:ASNsimulated}).  Nightly mean subtraction has been applied.}
\label{fig::ASNsimulation}
\end{figure*}

\begin{figure*}
  \centering
  \includegraphics[width=0.87\textwidth,trim = 0 0 0 0, clip]
  {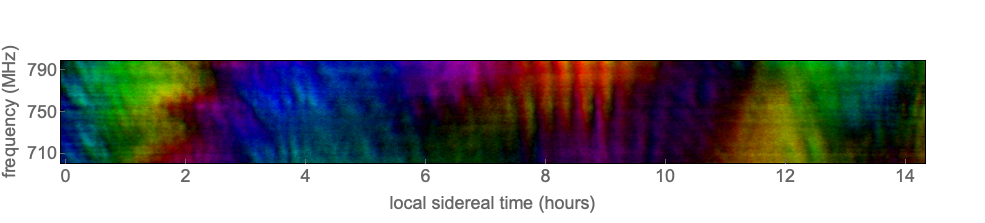}
\caption{The average sidereal night (ASN) for baseline 2V$\times$10V after polar dephasing, which removes the phase gradients due to the signal delay between the northern and southern feeds for sources near the NCP.}
\label{fig::ASNmedianDephased}
\end{figure*}

\subsubsection{Correlated Noise and Subtraction}
\label{sec:NCPcorrelatedNoise}

Correlated noise is identified as being nearly constant in time and not exhibiting the temporal fringe patterns one expects from Earth rotation. It is a type of RFI which may be natural, man-made or self-generated. The nighttime visibilities typically contain roughly equal contributions from sky signal and correlated noise with large variation (Section~\ref{sec:NCPotherBaselines}). Unfortunately, we have no other handle on the amount of correlated noise besides the visibilities themselves.  Self-generated correlated noise is likely dominated by ``cross-talk,'' which is radio emission from one feed being picked up by another.  Cross-talk has been seen by HERA \cite{Kern2019,Kern2020} and in Section~\ref{sec:NCPsimulated} we show the expected magnitude of cross-talk is comparable to the correlated noise we find.  While it is conceivable there is some leakage of signal between different channels during transmission to the correlator this is unlikely as the signal is transmitted via fiber optics. A potential environmental source of correlated noise is from ground emission: the array will image the $\sim300$\,K earth which is not uniform as hills define the Northern horizon.

Correlated noise that is constant in time can be removed by subtracting off the time-averaged mean visibility for each frequency. This would also completely remove any unpolarized sources which are exactly at the NCP, and partially remove sky signal from other directions. The controlled removal of a small fraction of the sky signal is easily modelled and accounted for when inferring the sky signal. Figure~\ref{fig::MeanVofFrequency} shows the modulus of the nighttime time-averaged (mean) visibility as a function of frequency for nights 2 through 10. This average is roughly similar between the different nights, but with significant (\(13\,\text{mK}\) rms) variation out of an RMS value of \(50\,\text{mK}\), or 25$\%$. The night-to-night variation is contributed to by both variation in the correlated noise and the gain, and is much larger than the system noise inferred by the system temperature.

In Figure~\ref{fig::Nighttime02V10Vsubtract} these nightly mean visibilities are subtracted, revealing a fringe pattern as expected for bright localized sources on the sky.  Detailed features are more apparent than in Figure~\ref{fig::Nighttime02V10Vnosubtract}, and these appear to repeat accurately each sidereal day. There are features which do not repeat: the few bright pixels noted above and a few bright stripes of several hours duration which do not repeat every night (e.g. on day 10 near 715\,MHz appearing near \(0^\text{h}\), and between \(4^\text{h}\) and \(5^\text{h}\)). These few non-repeating bright stripes may be due to external RFI, but also may be an indication of variability of the correlated noise. The median absolute visibility is 22\,mK, and the night-to-night median absolute deviation (a statistic which suppresses outliers) is 4\,mK, which is comparable to that expected from the system noise temperature. In this regard 2V$\times$10V is better than most baselines, where the night-to-night variation is significantly larger than the system noise. Our belief is that nearly all of the signal remaining in Figure~\ref{fig::Nighttime02V10Vsubtract} is sky signal.

\subsubsection{Night-to-Night Variation}
\label{sec:NCPvariation}

Figure~\ref{fig::MeanAbsVofLSTnight} gives a quantitative projection of Figure~\ref{fig::Nighttime02V10Vnosubtract} and  Figure~\ref{fig::Nighttime02V10Vsubtract}, before and after nightly mean subtraction.  This subtraction greatly reduces the night-to-night variation both in absolute terms and as a fraction of the remaining signal.  Subtraction of the nightly mean removes much of the correlated noise but also a significant fraction of the signal (gain times sky).  Since the sky signal should be the same at the same LST it does not contribute to night-to-night variation which can be due to variations in gain or in correlated noise. One would not expect that subtracting the nightly mean would decrease the fractional variation if the variation were only due to gain fluctuations, so we infer that much of what was subtracted is correlated noise. 

\subsubsection{Average Sidereal Night}
\label{sec:NCPASN}

One can average all the nights' visibilities into a single visibility which should have smaller noise than each of the individual nights. The averaging procedure used here is to take the median average of 1\,min-1\,MHz pixels at the same LST and frequency to create an {``}average sidereal night{''} or ASN.  This is what is shown in Figure~\ref{fig::ASNmedian} for the intersection of LSTs of the 9 complete nights. Median averaging suppresses the effect of outlying values, essentially removing non-repeating hot (or cold) pixels and stripes. There are no glaringly obvious {``}defects{''} in Figure~\ref{fig::ASNmedian}.

The visibility patterns of figures Figure~\ref{fig::Nighttime02V10Vsubtract} and Figure~\ref{fig::ASNmedian} give the visual impression of a wavy surface colored with nearly horizontal rainbow stripes, like ribbon candy or a flag fluttering in a breeze. The horizontal rainbow stripes indicates a vertical gradient in phase of the visibility or fringe pattern.  This is a consequence of the fact that most of the signal comes from near the NCP: the northern feed receives signals from the NCP before the southern feed, which leads to a phase delay which increases linearly with frequency. Note that if all the signal came from an unpolarized source precisely at the NCP then the visibility pattern \textit{before} nightly mean subtraction would be perfectly horizontal stripes, and the nightly mean subtraction would remove the entirety of the signal. Figure~\ref{fig::ASNmedian} only shows the remnant of this fringe pattern, which come from the sources located not precisely at the NCP. 
These waves are a superposition of slow waves with a timescale of hours and faster waves with a timescale of $\sim10\,$min.  The later we refer to as ``fast fringes''.  The variation in the time direction is due to rotation of the Earth. The slow waves are due to sources near the NCP which do not move rapidly on the sky, while the fast fringes are from bright far off-axis sources at low declination which move more rapidly as the Earth rotates (see section~\ref{sec:NCPoffaxis}).

\subsubsection{Fast Fringes and Bright Off Axis Sources}
\label{sec:NCPoffaxis}

To accurately identify all the sources contributing significantly to an ASN would require a more accurate beam model than we currently have. However, the beam pattern almost certainly does not vary as rapidly as the fast fringes evident in the ASN.  The fast fringes can only be from rapidly moving sources far from the NCP where the beam gain is low, ($\lesssim-30$\,dB smaller than at the beam center; which means they can only come from a few very bright far off-axis radio sources.  The lack of source confusion of very bright sources allows us to accurately identify the sources of these fast fringes even with only a single baseline using any of a variety of fitting or deconvolution techniques.

In Section~\ref{sec:NCPsimulated} we describe a simulation of the visibility from known radio sources which for 2V$\times$10V is shown in Figure~\ref{fig::ASNsimulation}. The measured and simulated visibilities have a very similar fast fringe pattern. In the simulation we can associate the dominant fast fringes with Cas~A and therefore infer that this is also the source of the fast fringes in the Tianlai observational data.

From Figures~\ref{fig::Nighttime02V10Vsubtract} \& \ref{fig::MeanAbsVofLSTnight} one sees that fast fringes are easily identified in only $\sim20\,$ min of data.  The ability to regularly isolate the contribution from individual well calibrated point sources such as Cas~A using only a single baseline provides us with a real-time calibration method for each baseline with which to supplement the CNS.

\subsubsection{Polar Dephasing}
\label{sec:NCPdephasing}

Since much of the sky signal should come from near the NCP one can adjust the phase center, as in a phased array, to point directly at the NCP, i.e. adjust the visibility phases by
$\Delta{\rm Arg}[V]=2\pi\,\frac{\nu}{c}\,\pmb{b\cdot \hat{n}}_{\text{NCP}}$ 
where $\pmb{\hat{\rm n}}_{\rm NCP}$ is the direction to the NCP and $\pmb{b}$ is the baseline.  In a horizontal (Earth) frame both $\pmb{b}$ and $\pmb{\hat{n}}_{\rm NCP}$ are constant in time and so is the correction to the phase gradient. For 2V$\times$10V the expected polar phase gradient corresponds to 2.4 stripes (full cycles through the phase spectrum) over 100 MHz bandwidth, which is just what one one sees in Figure \ref{fig::ASNmedian}.  Another visual comparison after the adjusting the phase center to the NCP is shown in Figure~\ref{fig::ASNmedianDephased}.  Nearly all the vertical phase gradients are removed, demonstrating that most of the signal does indeed come from sources near the NCP. What remains are slowly varying visibilities coming from sources near the NCP, which move slowly due to Earth rotation, plus more rapidly varying fringe patterns from bright sources far from the NCP.  

One usually phases an array to facilitate imaging of the region one is (electronically) pointing toward.  Our motives are somewhat different. Figure~\ref{fig::ASNmedianDephased} illustrates that the initial phase calibration is good enough to accurately point at the NCP.  This figure also illustrates the amount of mode mixing we have to contend with in Tianlai dish data. 

\subsection{Spectral Smoothness of Visibilities}
\label{sec:hi-k}

The 21~cm signal is much smaller than that of the {``}foreground{''} sources we have examined so far. One feature that differentiates the foregrounds from 21~cm is the foregrounds have a smooth broadband spectrum while 21~cm emission is not smooth and is in the form of Doppler shifted and broadened spectral lines from individual galaxies at different redshifts. This differentiating feature is confused by the phenomenon of {``}mode-mixing{''}, the fact that fine angular structures in the foreground emission will be aliased into relatively non-smooth spectral dependence of the visibilities due to the frequency-dependent angular response of the array. For example, while most of the frequency dependence (horizontal fringes) of Figure~\ref{fig::ASNmedian} has been removed by polar dephasing in Figure~\ref{fig::ASNmedianDephased}, there still remain horizontal components of the fast fringes from bright off-axis sources. While one can possibly subtract the fringe patterns of a few known bright sources, this would become intractable for the multitudes of sources which contribute to mode mixing at the level we are interested in. A variety of techniques have been proposed to project out mode-mixed foregrounds from the 21~cm signal, and we will use them in Tianlai in the future, but here we examine a more conservative approach: limiting analysis to frequency modes which are not significantly mixed with foregrounds at the level of the system noise temperature.  Here we quantify which modes these are. Foreground-contaminated frequency modes are sometimes said to be {``}in the wedge{''} and those not {``}outside the wedge{''}. Forecasts of the performance of intensity mapping often assume only modes outside the wedge are usable, so it is important to quantify where the wedge is!

There are various ways to quantify spectral smoothness of the visibilities. One is to decompose the visibility into frequency modes
\begin{equation}
V_{a,b,\alpha}=\sqrt{n_{\rm ch}}\sum _{n=0}^{n_{\text{ch}}-1} a_{a,b,n} U_{n,\alpha}
\end{equation}
where $\alpha$ indexes the equally spaced frequency channels, $n_{\text{ch}}$ in number.  $U_{n,\alpha}$ for fixed $n$ gives the spectrum of the mode which should be increasingly non-smooth in frequency
as $n$ increases.  It is convenient to take these modes to be orthonormal so that $U_{n,\alpha}$ is a unitary matrix and
\begin{equation}
\frac{1}{n_{\text{ch}}}\sum _\alpha |V_{a,b,\alpha}|^2
=\sum_{n=0}^{n_{\text{ch}}-1}|a_{a,b,n}|^2 .
\end{equation}
Thus $|a_{a,b,n}|^2$ gives the contribution of mode $n$ to the mean square $V_{a,b,n}$. A discrete Fourier transform is of this form but instead, we choose a polynomial-based decomposition where the frequency dependence of the modes is approximately described by Legendre polynomials
$U_{n,\alpha}{\propto\atop\sim} 
P_n\left[2\frac{\nu_\alpha   -\nu_{\rm min}}
               {\nu_{\rm max}-\nu_{\rm min}}-1
               \right]$.
Specifically we start with Legendre polynomials on a grid and Gram-Schmidt orthonormalize them.  For this ``Legendre decomposition'' $U_{n,\alpha}$ is real and orthogonal. Just as with Fourier transforms in the case of white noise each mode amplitude, $a_{a,b,n}$, is statistically independent with zero mean and identical variance $\langle|a_{a,b,n}|^2\rangle$.  These discrete Legendre polynomials are increasingly oscillatory (non-smooth) with increasing $n$ just as for a Fourier decomposition.

\begin{figure}
  \centering
  \includegraphics[width=0.47\textwidth,trim = 0 0 0 0, clip]{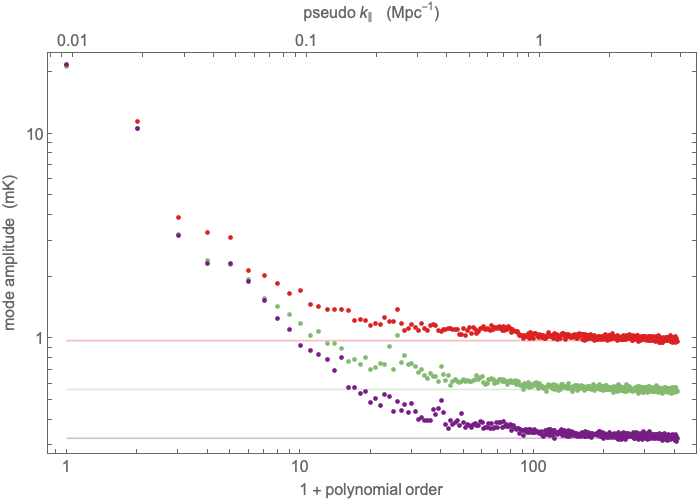}
  \caption{Plotted as a function of $n$ is the rms value of \(\left| a_n[\text{LST}]\right|\) averaged over all LSTs and 1, 3 and 9 successive nights of the 2V$\times$10V visibilities shown in Figure~\ref{fig::Nighttime02V10Vsubtract} after polar dephasing. Red points are for a single night, green three nights and purple nine nights. Polynomial order refers to \(n\). The top wavenumber scale gives the {\it approximate} $k_\parallel$ corresponding to a given $n$ for a cosmological 21cm signal. The color coded horizontal lines give the white noise spectrum for uncorrelated noise given by the system temperature measured during the same time interval. For small $n$ this signal converges after a few nights while for large $n$ it integrates down as $N_{\rm nights}^{-1/2}$ with amplitude very close to the system noise prediction.}
  \label{fig:LegendreSpectra}
\end{figure}

Applying this {``}Legendre{''} decomposition to the night-averaged and polar-dephased visibilities we compute the {``}$n$-spectrum{''}, which is the mean value of $|a_{a,b,n}|^2$ averaged over all time samples of an ASN.  Polar dephasing moves much of the $n\in[5,10]$ power into $n=0,1,2$ but does not change the total power.  In Figure~\ref{fig:LegendreSpectra} we plot these spectra for the visibility {\it mean} averaged over successive full nights. Here we have use full 244~kHz spectral resolution allowing us to measure the $n$-spectra up to $n=409$ rather than only $n=102$ for $977\,$kHz averaging.  We see that for $n=0,1,2$ the $n$-spectra seem to have {``}converged{''} in 1 night and for $n\lesssim8$ after $\sim3$ nights.  We would need more nights to see clear convergence at $n\gtrsim10$. For $n\gtrsim50$ the $n$-spectra are quite flat, close to white noise. The spectral flatness relies on limiting the band to $\nu\in[700,800]\,$MHz as the noisy band edges tilt the spectrum red-ward. For $n<30$ the spectra do fall off with increasing $n$ but not nearly as fast as one would predict for the very smooth spectra of optically thin synchrotron or free-fee emission. Much of this is due to mode mixing manifested by the fast fringes of bright off-axis sources which we could in principle subtract.  Incomplete removal of correlated noise is another possible cause for the slower than expected fall-off with $n$.  Applying the same procedure to the daytime data we find this white noise tail only extends to $n\gtrsim200$.

The $n>100$ white noise tail amplitude does {``}integrate down{''} $\propto N_{\text{nights}}{}^{-1/2}$ just as one would expect for (zero mean) noise which is uncorrelated between nights.  To illustrate this we also plot the $n$-spectra of the {``}system noise{''} predicted by the radiometer equation given the system temperature 
($\sqrt{T^{ 2{\rm V}}_{\text{sys}}\, T^{10{\rm V}}_{\text{sys}}}=75.5{\rm K}$)
measured from the auto-correlations.  For $n>100$ the system noise accounts for almost all of this power spectrum, leaving little room for contamination by sky signal or other sources of correlated noise.  Thus at the level of sensitivity obtained with 9 nights of data from a single baseline the radio emission from the sky does indeed have a smooth spectrum in that it does not contaminate the high-$n$ part of the spectra, leaving $\sim75\%$ of the $n$-modes apparently free from foregrounds. In section~\ref{sec:sensitivity0} we quantify just how free of foregrounds this region is.

Figure~\ref{fig:LegendreSpectra} uses mean averaging over different nights in contrast with Figure~\ref{fig::Nighttime02V10Vsubtract}, which uses median averaging.  While median averaging suppresses outlying visibility values as produced by RFI it has poorer noise performance.  Using median nightly averaging we find the $n$ spectra integrates down more slowly, as \(\sim N_{\text{nights}}{}^{-0.37}\).  The cause of this different scaling can be understood if one allows for night-to-night gain variations.  A median will choose for each pixel the central visibility value which can be from different nights for different pixels.  If the gain varies from night-to-night then neighboring pixels in the median average can take visibility values from different nights with different gains. Night-to-night variation in the gain will result in sharp features in the frequency spectra as well as the ASN visibility time series.  Such discontinuities would be introduced by any method which removes or suppresses RFI-flagged visibilities such that neighboring pixels sample different sidereal days with different weights.  Mean averaging gives equal weight to each night and therefore depends on the mean gain averaged over all nights, which is not expected to have sharp features in frequency or in time.  If RFI is rare then RFI flagging has an advantage over median averaging since RFI flagging affects only a small fraction of the data whereas median averaging will introduce discontinuities everywhere.  One can use the difference between mean and median averaging to quantify the level of night-to-night gain variation.

\subsection{Correlated Signal after Foreground Subtraction}
\label{sec:sensitivity0}


Foreground removal in hydrogen intensity mapping relies on the ability to separate rough spectrum 21~cm line emission from smooth spectrum foregrounds. In the previous section it was shown that an ASN mostly contains relatively smooth spectrum signal or, more specifically, that the non-smooth spectrum ($n>100$) part of the signal is very close to the system noise and contains very little sky signal or correlated noise.  In this section we constrain how much $n>100$ signal might remain in the ASN cross-correlations. One powerful and precise tool to measure this is the radiometer equation which gives the excess variance in the cross-correlations over that predicted from the auto-correlations.  

For a given antenna pair, $a$ and $b$ ($a\ne b$), we measure three visibilities: complex $V_{a,b}$ and real positive $V_{a,a}$ and $V_{b,b}$ at each time sample and frequency channel. Denote by $\langle\cdots\rangle$ the expectation of visibilities averaged over realizations of voltage streams for fixed illumination pattern.  Quantities without a $\langle\cdots\rangle$ are the measured values. Define {\it correlated signal} as {\it any} non-zero contribution to $\langle V_{a,b}\rangle$.  This may include not only sky signal, but also RFI, ground spill and correlated noise.  {\it Uncorrelated noise} is then any signal uncorrelated between antennas as is expected for receiver noise.  Since $\langle V_{a,b}\rangle$ is oscillatory and varies in an unknown way with time and frequency it is not amenable to precise estimation simply by averaging over long intervals of time ($t$) and/or frequency ($\nu$); in contrast to real positive definite quantities. One can however use the radiometer equation to estimate the rms of the oscillatory $\langle V_{a,b}\rangle$ through averaging positive definite quantities.  This method for determining $|\langle V_{a,b}\rangle|$ is useful when the signal is too small to determine by averaging $V_{a,b}$ over small intervals of $t$ and/or $\nu$. 

The ideal radiometer equation, valid when the voltages are given by a Gaussian random process stationary over the sample time, may be written
\begin{equation}
|\langle V_{a,b}\rangle|^2=
\langle|V_{a,b}|^2\rangle
-\frac{
\langle V_{a,a}\rangle\,
\langle\,V_{b,b}\rangle}
{N_\text{sample}} \ .
\label{eq:radiometerAlt}
\end{equation}
Here $N_\text{sample}$ is the number of complex Fourier amplitudes averaged by the correlator to obtain the visibilities.  The right hand side is the excess power above that expected for uncorrelated noise and contains only positive definite quantities allowing estimates of the left hand side by averaging. For each pixel define the excess power
\begin{equation}
W_{a,b}\equiv|V_{a,b}|^2
-\frac{V_{a,a}\,V_{b,b}}
{N_\text{sample}} \ ,
\label{eq:signalEstimator}
\end{equation}
For Gaussian illumination and $N_\text{sample}\gg1$
\begin{eqnarray}
\langle W_{a,b}\rangle&=&
|\langle V_{a,b}\rangle|^2 \nonumber \\
{\rm Var}(W_{a,b})&=&
3\,|\langle V_{a,b}\rangle|^4 \\
&&\hskip-60pt+4\,\langle V_{a,a}\rangle\,
                 \langle V_{b,b}\rangle\,
\frac{\langle V_{a,a}\rangle\,
      \langle V_{b,b}\rangle
 -2\,|\langle V_{a,b}\rangle|^2}
  {N_\text{sample}}\ . \nonumber
\label{eq:signalEstimatorStatistics}
\end{eqnarray}
Denote unweighted averaging over pixels (frequency and/or time) by $\overline{\cdots}$. Thus $\overline{W_{a,b}}$ is an unbiased estimator of 
$\overline{|\langle V_{a,b}\rangle|^2}$ with variance 
$\overline{{\rm Var}(W_{a,b})}/N_\text{pixel}$
where $N_\text{pixel}$ is the number of pixels averaged.  This assumes fluctuations are uncorrelated between pixels which is expected: in the time direction because the pixel duration is large compared to the light travel time across the telescope, and in the frequency direction because of the random phase approximation.

The sensitivity to small correlated signal that can be obtained from $\overline{W_{a,b}}$ is limited by the contribution of the uncorrelated noise to it's variance:
\begin{equation}
\delta V_{a,b}\equiv
\sqrt{
\frac{2\,\overline{\langle V_{a,a}\rangle\,
                   \langle V_{b,b}\rangle}}
     {N_\text{pixel}\,N_\text{sample}}
      } \,
\label{eq:signalEstimatorSSS}
\end{equation}
With continuous sampling
$N_\text{sample}=\delta \nu\,\delta t$ and
$N_\text{sample}\,N_\text{pixel}=
\Delta \nu\,\Delta t$
where $\Delta \nu$ and $\Delta t$ are the total bandwidth and integration time averaged over. For the ASN of section~\ref{sec:NCPASN} 
$\Delta \nu\simeq 100\,{\rm MHz}$,
$\Delta t\simeq 9\times14\,{\rm hr}$,
$\overline{V_{a,a}\,V_{b,b}}\sim (75\,{\rm K})^2$
so $\delta V_{a,b}\approx16\,\mu$K.

The correlated signal in time/frequency pixels is well above this sensitivity and large enough to estimate directly by averaging $V_{a,b}$ over a $\nu$-$t$ interval small enough that $\langle V_{a,b}\rangle$ would not vary significantly.  However this is not uniformly true of a Legendre decomposition of the visibility: for small $n$ the S/N is large while for large $n$ it is small and the excess power is a small fraction of the uncorrelated noise.

The radiometer equation transformed into $n$-modes is
\begin{equation}
|\langle a_{a,b,n}\rangle|^2
=\langle|a_{a,b,n}|^2\rangle
-\frac{(\langle V_{a,a,\alpha}\rangle\,
        \langle V_{b,b,\alpha}\rangle)_n}
      {n_{\rm ch}\,N_\text{sample}} \ .
\end{equation}
assuming random phases so that fluctuations in different frequency channels are uncorrelated. Here $\alpha$ is the frequency channel index and
$(f_\alpha)_n\equiv
\sum_\alpha\,{U_{n,\alpha}}^2\,f_\alpha$ is a frequency average since $\sum_\alpha\,{U_{n,\alpha}}^2=1$. This average is required to account for the frequency dependence of $T_{\rm sys}$.  One can measure the excess power
\begin{equation}
w_{a,b,n}\equiv|a_{a,b,n}|^2
-\frac{(V_{a,a,\alpha}\,V_{b,b,\alpha})_n}
      {n_{\rm ch}\,N_\text{sample}}\ .
\label{eq:signalEstimatorMode}
\end{equation}
which for $N_\text{sample}\gg1$ has expectation value
$\langle w_{a,b,n}\rangle=
|\langle a_{a,b,n}\rangle|^2$.
By measuring the average excess power one can estimate the average amount of correlated signal, 
$\overline{|\langle a_{a,b,n} \rangle|^2}$ for any $n$ since
$  \langle\overline{w_{a,b,n}}\rangle
=\overline{|\langle a_{a,b,n} \rangle|^2}$.
Here $\overline{\cdots}$ denotes an average over $t$ and/or $n$.  

We are interested to see if there is evidence for any correlated signal in the $n\ge100$ modes so we calculate
\begin{equation}
\sum_{n=100}^{n_{\rm ch}-1} \overline{w_{a,b,n}} = 1.1\,{\rm mK}^2
\end{equation}
for 1~min$\times$244~kHz pixels ($n_{\rm ch}=405$) of baseline 2V$\times$10V. This gives the contribution of $n\ge100$ modes to $\overline{|\langle V_{a,b,\alpha} \rangle|^2}$. While this excess power is well above the $\sim(20\,\mu{\rm K})^2$ sensitivity for this large a region of the $\nu$-$t$ plane it is a very small fraction, $<0.1\%$, of the $\sim(50\,{\rm mK})^2$ correlated signal that comes out the correlator. Thus the $n\ge100$ hi-pass filter has ``leakage'' of no more than -30dB of the total power.

The ($n>100$) $1.1\,\mathrm{mK}^2$ correlated signal is $4\%$ of the ($n>100$) $35\,\mathrm{mK}^2$ uncorrelated noise determined from the measured auto-correlations. If this excess power is uncorrelated between nights it will integrate down, only increasing the amount of observation time required to reach a given noise level by $\sim4\%$ above the ideal. Alternately, if this excess power comes from a sky signal fixed in sidereal time then it will not integrate down but will become more apparent with further observation.


Excess power at hi-$n$ may be from any combination of sky signal, ground spill, correlated noise or RFI; or it may be instrumental.  Since each of these these different contributions is expected to be uncorrelated with each other this puts an upper limit of $\sim1\,\mathrm{mK}^2$ on the $n\ge100$ power of each one of these components separately.

\subsection{Simulations of 2V$\times$10V}
\label{sec:NCPsimulated}

\subsubsection{Average Sidereal Night}
\label{sec:ASNsimulated}

\begin{figure}
  \centering
  \includegraphics[width=0.45\textwidth]{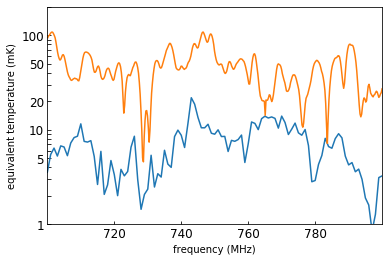}
\caption{ Comparison of the nightly mean spectrum with a model for correlated noise based on cross-talk. The orange curve is the average of the nightly mean visibilities shown in Figure \ref{fig::MeanVofFrequency} for baseline 2V$\times$10V.  The blue curve is an estimate of the correlated noise for this baseline assuming thermal noise from each antenna couples to the other with a coupling coefficient determined with an electromagnetic simulation.  The difference between these two curves is roughly consistent with estimates of the (constant) visibility from astronomical radio sources at the center of the beam.}
\label{fig::MeanVofFreqVSSimXTalk}
\end{figure}

One can use known radio sources to construct a rough model of the contribution of astronomical emission to the visibilities in the dish array.  The accuracy with which one can simulate these visibilities is limited by uncertainties in the radio sky, which is not well measured in our band, and by uncertainties in the beam model.  We have used the radio point source catalog of the NVSS (\cite{Condon1998}) to construct such a model which requires extrapolation of these 1.4\,GHz measurement to our 700-800\,MHz band.  The electromagnetic simulations of our beam do not currently have sufficient frequency resolution to be used for this purpose and instead we have used a simple Airy disk model with the main lobe matched to the dish FWHM discussed in Section~\ref{sec:beams}.  This is more appropriate than a Gaussian beam as it has comparable sidelobes to the electromagnetic simulations shown in Figure~\ref{fig:simulated_beam}.  For our fiducial baseline, 2V$\times$10V, the simulated nighttime visibility after mean nightly subtraction is shown in Figure~\ref{fig::ASNsimulation}.  This should be compared to our measurement shown in Figure~\ref{fig::ASNmedian}. The two exhibit a general qualitative similarity: the slow waves and fast fringes are of similar amplitude and are often in phase.  Some quantitative features are significantly different, which we attribute to inaccuracies of the Airy disk model and of the sky model.


\subsubsection{Cross-Talk}
\label{sec:NCPsimulatedXtalk}

Line-of-sight transmission between feeds will occur as this is not blocked by the dishes.  The beam gain in these directions is very small but the array is fairly compact, with dish centers placed as close as $\sim 1.5$ diameters from each other. We have made quantitative estimates of the transmission between terminals (ports) of the feeds on different dishes using electromagnetic simulations. By using the integral equation solver in CST Microwave Studio to model the entire array (dishes and ports), but allowing only one active port we compute the cross coupling between the active port to all the other passive ports as a function of $\nu$, while conserving computing resources.  We check that the results are robust to changes in parameters used by the simulation software (e.g. accuracy settings and mesh size).  Contrary to expectation, the coupling does not decrease markedly with baseline length. We believe this is related to the large number of scattering paths between antennas in the full array, in contrast to a single line-of-sight path for two isolated antennas, where simulations do show a decreased transmission with distance.  These results will be discussed further in a later publication and here we only present the results for our fiducial baseline 2V$\times$10V.

Making assumptions about how transmitted power is related to the observed system temperature, we estimate how much the cross-coupling contributes to the correlated noise in the corresponding visibilities.   Figure~\ref{fig::MeanVofFreqVSSimXTalk} shows the average of the modulus of the measured 2V$\times$10V visibility  over 9 nights (shown in Figure~\ref{fig::MeanVofFrequency}) alongside the corresponding quantity for the simulated cross-talk. The frequency structures of the two curves are qualitatively similar, but the simulated cross-talk is smaller.  Across the band, the average and standard deviation of the simulated cross-talk is $\sim15\,\pm8.3 $\,mK, which is $\sim35\,$mK below the observed nightly mean. It is expected that the nightly mean will differ from and be larger than the cross-talk since the former also includes other incident radiation such as the sky signal.

\subsection{Other Baselines}
\label{sec:NCPotherBaselines}

\subsubsection{H-H and V-V Visibilities}

\begin{figure*}
  \centering
  \includegraphics[width=0.87\textwidth,trim = 0 0 0 0, clip]
  {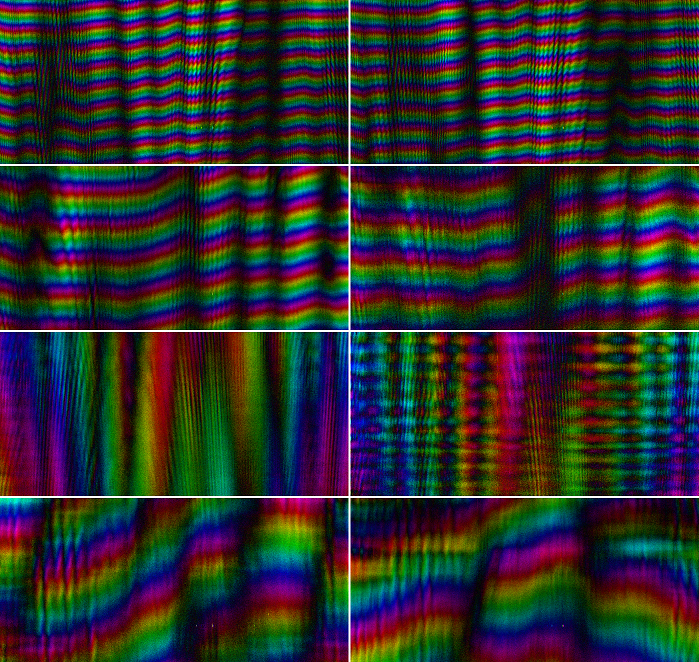}
\caption{Shown are the ASNs of two antenna pairs from each of four dish pairs from the same run as Figure~\ref{fig::ASNmedian}. The ASNs are 9 night mean averages with -1\,min-244\,kHz pixels and nightly mean subtracted.  From top to bottom the dish pairs, baseline lengths and angle from due North are: 2$\times$6, $34.6\,$m, $5^\circ$; 3$\times$13, $21.6\,$m, $22^\circ$; 4$\times$15, $26.3\,$m, $87^\circ$; 10$\times$16, $8.8\,$m, $30^\circ$. The left column of visibilities correlate horizontal H$\times$H antennas while the right column correlate V$\times$V antennas.}
\label{fig::SampleVisibilities}
\end{figure*}

\begin{figure}
  \centering
  \includegraphics[width=0.47\textwidth,trim = 0 0 0 0, clip]{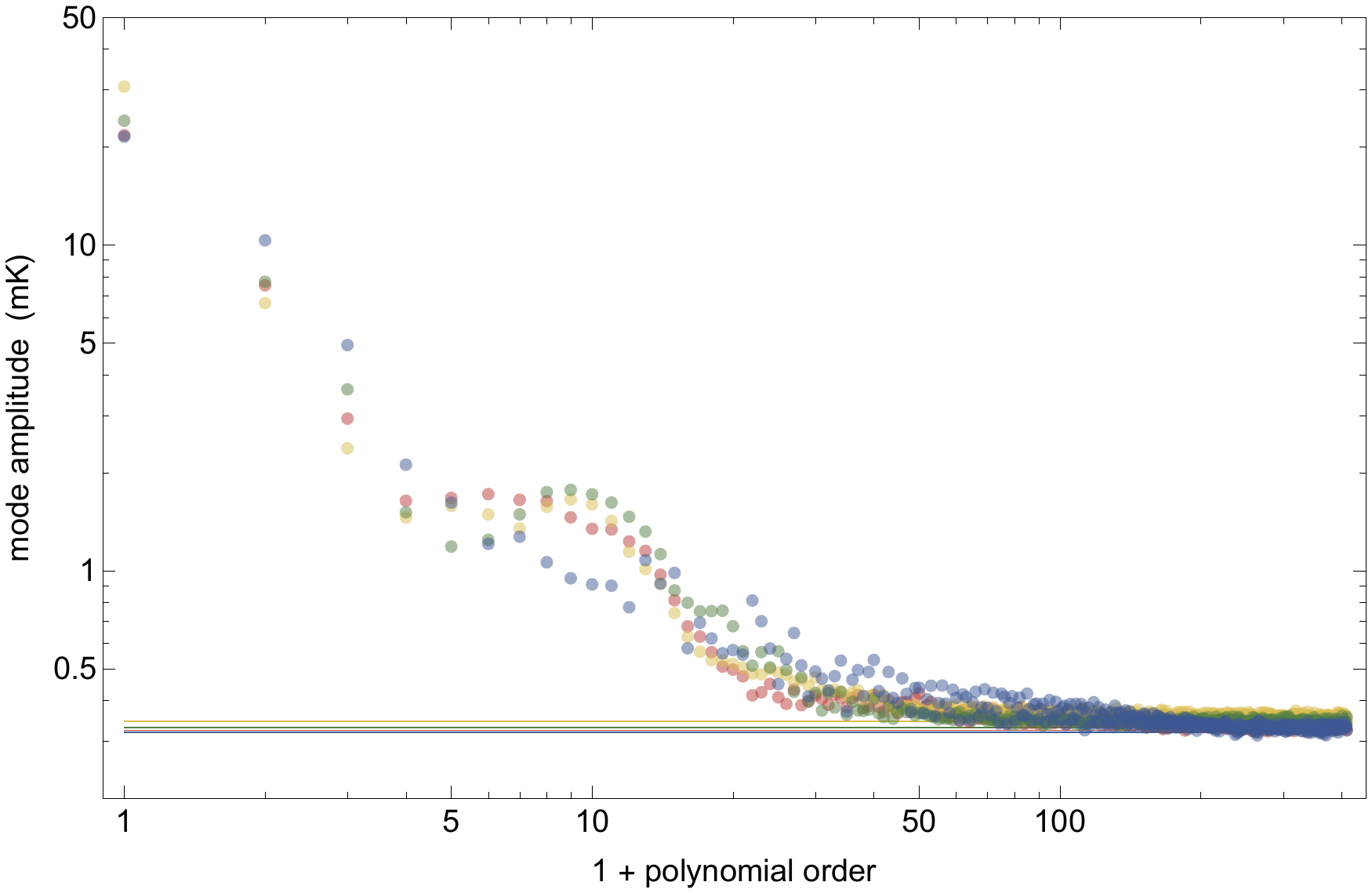}
\caption{Shown are the $n$-spectra of the ASN of the four H-H visibilities of Figure~\ref{fig::SampleVisibilities} after polar dephasing,  These are color coded as 2H$\times$6H (red), 3H$\times$13H (yellow), 4H$\times$15H (green) and 10H$\times$16H (blue).}
\label{fig::SampleLegendreSpectraHH}
\end{figure}

\begin{figure}
  \centering
  \includegraphics[width=0.47\textwidth,trim = 0 0 0 0, clip]{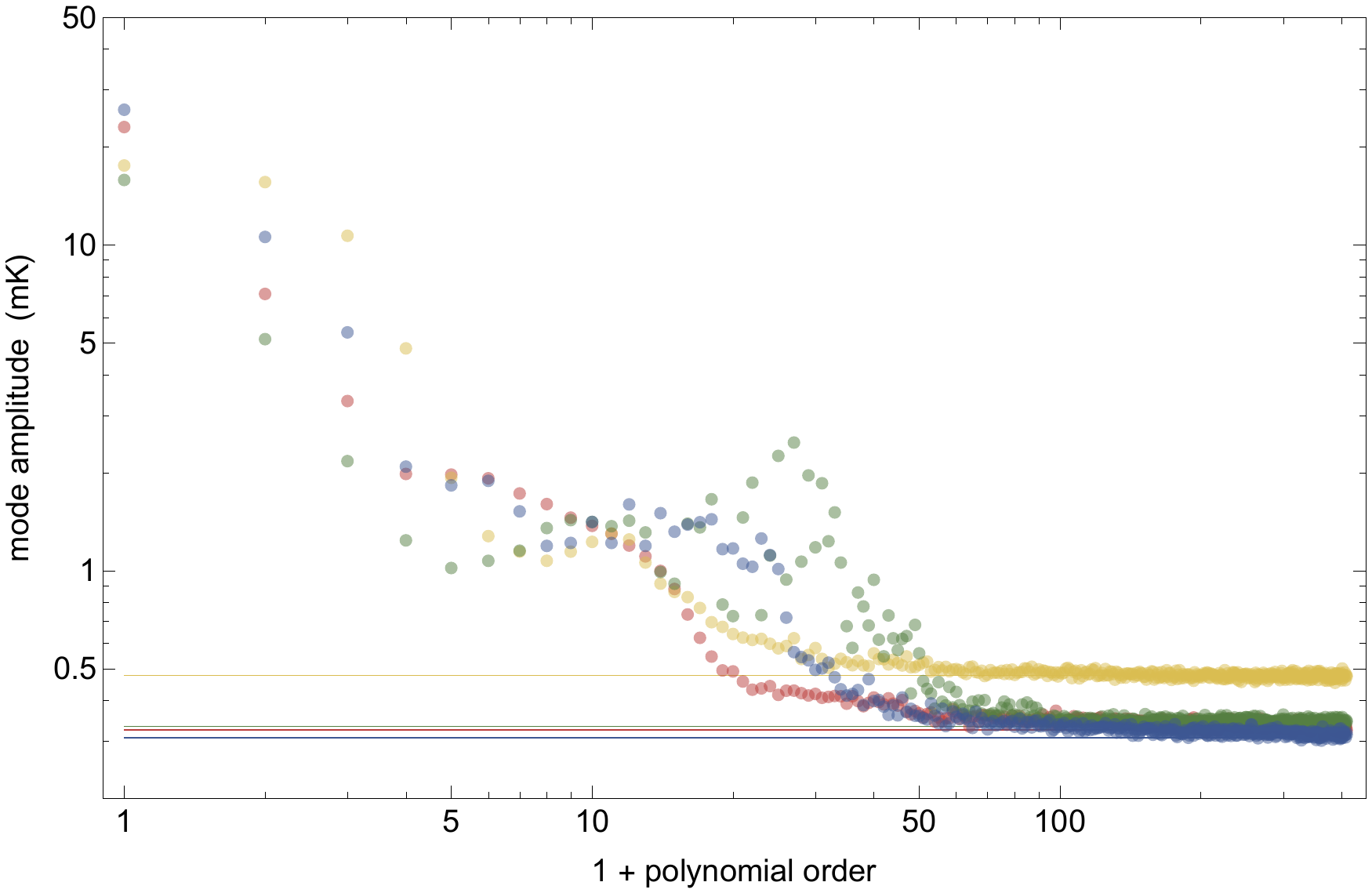}
\caption{Shown are the $n$-spectra of the ASN of the four V-V visibilities of Figure~\ref{fig::SampleVisibilities} after polar dephasing, color coded as  2V$\times$6V (red), 3V$\times$13V (yellow), 4V$\times$15V (green) and 10V$\times$16V (blue). 13V is a ``hot" antenna with abnormally large $T_\mathrm{sys}$ leading to larger noise for the yellow 3V$\times$13V visibility.}
\label{fig::SampleLegendreSpectraVV}
\end{figure}

For NCP observations we have so far concentrated on one baseline, 2V$\times$10V.  Here we briefly illustrate corresponding results for other baselines. The visibilities should differ according to their baseline $\bf{b}$ both in magnitude and direction.  There are 100's of baselines and here we only present a selection. Figure~\ref{fig::SampleVisibilities} shows the ASNs of H-H and V-V antenna pairs from four dish pairs. These are analogous to Figure~\ref{fig::ASNmedian} but differ not only in $\bf{b}$ but also are mean (not median) averages of the 9 nights and show smaller -1\,min-244\,kHz pixels. Mean averaging does not remove some of the ``defects'' (presumably RFI) in the visibility pattern while the $4\times$ smaller pixels means the noise is $2\times$ larger at the pixel scale. As expected, North-South baselines show horizontal striping while East-West baselines do not because there is little phase delay between the NCP signal arriving at the two dishes.  Another qualitative difference between these visibilities is that the fringe rates increase with $|\bf{b}|$ just as one would expect.  If all the incident radiation were unpolarized {\it and} the beams were unpolarized in all directions then one would expect the H-H and V-V to be identical. Our antennas beams are nearly unpolarized near the center but off axis this is far from the case so one does not expect identical H-H and V-V visibilities.  Offsets in the colored stripe pattern between H-H and V-V visibilities may be indicative of drifts in the complex gains, which were only normalized at the beginning of the run before pointing toward the NCP. The horizontal pattern in 4V$\times$15V that is not present in 4H$\times$15H appears to be a failure of mean nightly subtraction in removing correlated noise which apparently varied significantly over the course of the nighttimes.  Such a blatant failure is uncommon but less obvious artifacts such as some of the horizontal features in 10V$\times$16V are likely caused by the same failure.  We are exploring alternative methods to better deal with correlated noise.

The $n$-spectra of these eight ASNs are shown in Figures~\ref{fig::SampleLegendreSpectraHH}\&\ref{fig::SampleLegendreSpectraVV} similarly to Figure~\ref{fig:LegendreSpectra} except here we show only the spectra for the cumulative 9-nights.  These spectra differ markedly between the H-H and V-V feeds for the same dish pair.  This is not surprising since except at small $n$ these spectra contain a large contribution from fast fringes of bright sources which enter the beam far off axis where the beam is very polarized.  One of these sample visibilities includes the ``hot" antenna 13V which has more than double the typical $T_{\rm sys}$ of other channels.  Aside from increasing the noise a large $T_{\rm sys}$ does not seem to adversely affect the 3V$\times$13V visibility.  A larger problem manifests in the $20\lesssim n\lesssim80$ power in the 4V$\times$15V which has significant contamination from unsubtracted correlated noise. All of the $n$>100 spectra are very near to the white noise contribution from the system temperature, suggesting that none of these problems contaminate the hi-$n$ spectra.

We have also computed the excess correlated signal power defined in Section~\ref{sec:sensitivity0} averaged over the 9 full nights of nearly all the baselines of the 3srcNP 20180101 run but excluding auto-correlation baselines and baselines which contain any of the hot antennas identified in section \ref{sec:calibration:Tsys}. We find
\begin{equation}
\sum_{n=100}^{n_{\rm ch}-1} \overline{w_{a,b,n}} = (1.51\pm0.62)\,{\rm mK}^2,
\end{equation}
where the right-hand-side gives the mean and standard deviation of these excess powers for the baselines analyzed.  Whatever the source of this excess power it is relatively uniform across the array.



\subsubsection{Redundant Baselines}

\begin{figure}
  \centering
  \includegraphics[width=0.47\textwidth,trim = 0 0 0 0, clip]{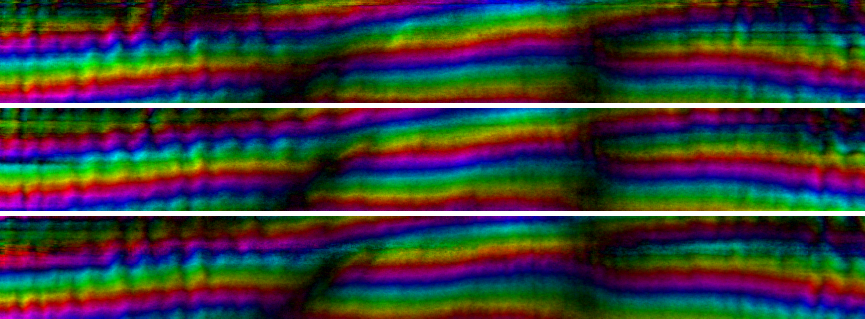}
\caption{Shown are the ASN of three nearly redundant baselines as in Figure~\ref{fig::ASNmedian}. The baselines and nominal separation of feeds in mm are, from top to bottom, \(12\text{V}\times 11\text{V}\): \(4391\,\text{E}-7596\,\text{N}-6\,\text{Z}\), \(13\text{V}\times16\text{V}\): \(4387\,\text{E}-7601\,\text{N}-5\,\text{Z}\) and \(16\text{V}\times 10\text{V}\): \(4424\,\text{E}-7604\,\text{N}+4\,\text{Z}\) (E=East, N=North, Z=vertical). The nominal separations are as surveyed after installation; because the feed positions are not continuously monitored there is some uncertainty in these numbers.  For example, if the dishes are not all precisely pointed in the same direction the separations will differ from the nominal values.}
\label{fig::RedundantVisibilities}
\end{figure}

The dish array has a few nearly redundant baselines. Ideally, redundant baselines, those having the same ${\bf b}$, should exhibit identical visibilities from the sky when correlating antennas with the same orientation, e.g. H-H or V-V.  This should be true if the field patterns of different antennas are identical.  Visibility differences can be a sign of differences between antennas, either intrinsically or due to interactions with neighboring dishes which are each arranged differently. Also contributing differences are relative gain drifts between different antennas as well as differences in residual correlated noise.  In Figure~\ref{fig::RedundantVisibilities} we display the ASNs of three such baselines after mean nightly subtraction.  Visually, they appear nearly the same; the major difference coming from a multiplicative offset: $V_{\rm13V,16V}\approx1.02\,e^{i\,0.63}\,V_{\rm12V,11V}$ and
$V_{\rm16V,10V}\approx1.00\,e^{i\,0.12}\,V_{\rm12V,11V}$.  The $\sim0.6\,$radian shift of the fringes is easily noticeable in the central panel relative to the other two.  This is evidence for a drift in the gain over the 10 days after the initial calibration.  It appears that the gain amplitude drifted by only a few percent while the gain phase changed by close to a radian.

For all three baselines the nighttime average mean square visibility is $\overline{|V_{a,b}|^2}\approx(30\,\mathrm{mK})^2$, which we can compare to the power in the difference in visibilities after applying a gain drift correction.  These residuals have mean square $\sim(10\,\mathrm{mK})^2$ in 1\,min-1\,MHz pixels.  This 9-fold decrease in the power is quite significant but is still much larger than the noise level of $\sim(5\,\mathrm{mK})^2$ for a 9-night average (the noise power in a difference is double that for individual baselines since the noise adds in quadrature).  This significant excess indicates there are $\sim10\%$ differences between antennas even after correction for gain drifts.

\subsection{NCP Summary}
\label{sec:NCPsummary}


We have examined in detail 9 full nights of data with the dish array pointed at the NCP. Typical auto-correlations are $T_{\rm sys}\approx75\,$K while cross-correlation visibilities have amplitude $\sim50\,$mK. The cross-correlations have a nearly constant frequency-dependent component (mostly ``correlated noise''), which we subtract, leaving a $\sim25\,$mK signal, consistent with known radio sources and which repeats accurately every night.  The system temperature predicts noise at the level of $\sim10\,$mK in 1\,min-1\,MHz pixels for 1 night or $\sim3\,$mK, averaging over 9 nights.  The night-to-night variation is larger than this by a factor of a few and this excess is likely dominated by night-to-night variation in the gain or imperfect correlated noise subtraction. In this analysis no time-dependent gain corrections have been applied and we expect to improve upon this using the CNS and sky calibration. 

In section \ref{sec:hi-k} we have decomposed the spectrum according to spectral smoothness.  The signal is contained mostly in the smoothest components (lo-$n$) and falls off with $n$ such that for $n>100$ the cross-correlations are precisely modeled by the random noise predicted from the auto-correlations by the radiometer equation.  Since $n$ extends to 409 this means that $\sim75\%$ of the data has been cleaned from nearly all foregrounds by this hi-pass filter.  Note that for cosmological 21cm emission this $75\%$ corresponds to cosmological comoving wavenumbers $k\gtrsim1\,h/$Mpc, which is a scale much smaller than where most of the interesting cosmological information resides. 

In section \ref{sec:sensitivity0} we explore in detail just how consistent the $n>100$ rms cross-correlations visibilities are with the noise predicted by the radiometer equation from the auto-correlations. We find only a 4\% excess in cross-correlation power. This places a limit on leakage of correlated noise or foreground signal into the white noise tail. This excess white noise is no more than $0.1\%$ (-30dB) of the total $(50\,\mathrm{mK})^2$ signal power.  Thus the $n>100$ hi-pass filtering is very effective in cleaning most of the smooth spectrum foreground signal from our data.  Such hi-pass filtering will only remove a small fraction of any 21-cm signal while significantly suppressing contamination by foregrounds.  This -30\,dB suppression is found for the 9 nights of integration time we have analyzed here and may be much greater in the full 200 nights of accumulated integration time.

\section{Conclusion}
\label{sec:conclusion}

Neutral hydrogen, ubiquitous throughout cosmic history since recombination, is a potentially powerful tool for cosmological observations. To achieve high precision measurements of the redshifted 21~cm emission, a number of dedicated 21~cm array concepts have been built or are under construction.  Even more ambitious arrays, involving tens of thousands of antennas \cite{Ansari2018,Slosar2019}, have been proposed. Dedicated 21~cm arrays have some common design features, determined largely by the nature of the observation and current technology. They all use large numbers of relatively small, inexpensive antennas. The antennas are either fixed, or can only move in elevation, and the observations use either a drift scan mode, or track the target with electronically steered beams 
With computing power following Moore's law scaling and price per computing operation and of data storage continuously dropping, this approach enables arrays of very large scales to be built with a very moderate cost. Based on simple forecasts, the projected capabilities of these arrays are very impressive (see, e.g. \cite{Seo2010,Ansari2012,Xu2015,Xu2016}). However, extracting scientific results from them poses a number of challenges. For example, these arrays produce a huge amount of data.  In addition, with small antennas and uncooled receivers the SNR of the raw data is relatively low; only a few bright sources are available for simple point source calibration, and as the antennas are not movable, calibration with strong point sources can only be performed occasionally. Furthermore, while for forecasting it is customary to assume that the antenna responses are identical, in reality each unit is somewhat different. It is crucial to develop technologies to handle the data from such arrays, and to test the key technologies in order to gain some concrete experience and expose possible problems. The Tianlai pathfinder arrays were built expressly for this purpose.

In this paper, we first described the overall architecture and design parameters of the Tianlai Dish Pathfinder Array (Sec. \ref{sec:instrument}). The hardware consists of 16 on-axis parabolic reflectors and feed antennas, the optical analog signal transmission system,  the down-converters which convert the RF signal to IF, and the digital FX correlator, which  produces the visibilities.  We also briefly introduced the observational data sets (Sec. \ref{sec:observations}), totalling 6,200 hours, a large fraction of which are deep integrations on the NCP.  RFI affects a very small fraction of the data and in our analysis so far we have not attempted to remove it.

In Sec. \ref{sec:beams} we studied the beam patterns of the dish antennas. We compared electromagnetic simulations in the 700--800~MHz frequency range to measured beam profiles in the E-W direction by analyzing auto-correlations and cross-correlations using the strong source Cas~A during transits of the meridian. The general shape of the beam is consistent with the simulation, but a few feeds are slightly misaligned, with an error of $0.66^{\circ}$ or 14\% of the FWHM. 
This is an example of the small non-uniformity which occurs naturally in the construction of large radio arrays.  Not taking into account such differences can induce errors in the final data.  Also, while we can use the motion of the sources introduced by Earth rotation to map the E-W profile easily, it is much harder to map the beam in the N-S direction. 

The analysis pipeline and tools we have developed were described in (Sec. \ref{sec:analysis}).

Calibration is a crucial step for the processing of interferometer array data, and we presented our approach in Sec. \ref{sec:calibration}. First, we performed bandpass calibrations for individual visibilities, and found relatively stable results.  More important is the calibration of complex gains. For an array with a large field of view and low sensitivity, such as the dishes, a special challenge is that on the one hand, there are few sources which are bright enough to be ``seen'' directly in individual visibilities, and on the other hand, it is very difficult to construct a sky model that has enough precision over the large field of view to include all sources that contribute to the visibility.  In our experiment we tried two methods of calibration.  We use a calibration signal broadcast from an artificial source (CNS) to perform what we call relative calibration. And, in what we termed absolute calibration, we use the transit of strong astronomical sources such as Cas~A to solve for the complex gain of each input channel, using the eigen-decomposition method we developed earlier \cite{Zuo2019}.  The results of the two methods are compared for the period when they overlap, and distributions of the complex gains are given. The distribution and variations of the gain are plotted, and the phase of the gain is found to be strongly correlated with the environmental temperature.  
In Sec. \ref{sec:calibration:Tsys}, we estimated the system temperature of the array feeds by calibrating on transits of Cas~A. These results were general agreement with the system temperature obtained for the Tianlai Cylinder Array, which has similar electronics. 
We also examined how the variance of the visibility scales with time. Higher sensitivity can be achieved with longer integration time within the range we tested.

Section \ref{sec:maps} presented maps of bright calibration sources using a variety of algorithms, including the standard radio interferometer analysis package, CASA, as well as several of our own design.  These maps confirm that the array hardware and calibration are functioning as expected. 

Section \ref{sec:NCP} described the analysis of long integrations on the North Celestial Pole (NCP), a region we continue to observe in order to integrate to low noise levels.  We find the data to be of very high quality, with the visibilities repeating day after day. There is clear contamination by the Sun during daytime, but in a very repeatable pattern.  Focusing on the nighttime data, we find and subtract a nightly mean value from each visibility;  we believe this quantity arises from ``correlated noise" that we suppose may be cross-talk between the antennas. The combination of visibilities from 9 nights of data gives us an average sidereal night that matches well with simulations that include a model for the antenna pattern and known radio point sources.  We clearly see the presence of Cas~A, which is well outside our main beam, and hope to use it as a continuous calibrator for future analyses. As a means to separate (smooth spectrum) foregrounds from potential HI signal, we perform a decomposition of the visibilities into spectral smoothness dependent modes.  As expected, the amplitudes of these modes are dominated by the smoothest modes and reach a floor consistent with uncorrelated (white) noise whose amplitude is described by the radiometer equation.  This noise floor integrates down as $N_{\rm nights}^{-1/2}$, as expected, and should decrease as $N_{\rm nights}$ increases.  The non-smooth spectrum component of the signal which, although comprising $\sim75\%$ of the data, is shown to contain no more than -30~dB of the total signal power.  This illustrates the effectiveness with which hi-pass filtering the spectra removes smooth spectrum sources.  We hope to improve on this with larger data sets.

In this work,  we have only studied a small fraction of the data set collected so far, and presented only some basic performance characteristics of the Tianlai Dish Array.
There remain many challenges, such as understanding the cross-coupling between feeds and removing foregrounds.  There are also many remaining tasks, such as the determination of the beam profile in the N-S direction, more precise calibration, map-making, etc.
Furthermore, we plan to retune the dish array receivers to lower redshift to overlap with an existing galaxy survey of the NCP (the North Celestial Cap Survey - NCCS) and the ongoing North Celestial Cap Redshift Survey (NCCRS) for cross-correlation analysis.   We also plan to perform detailed comparisons between the Tianlai Dish Pathfinder Array and the Tianlai Cylinder Pathfinder Array.  More thorough analyses of these issues will be investigated in a series of future works.

\section*{Acknowledgements}
We are thankful to Adam Beardsley, Richard Shaw, and Francisco Villaescusa-Navaro for their constant support, constructive comments, and other contributions. John Podczerwinski provided advice on electromagnetic simulations. We thank an anonymous referee, whose detailed suggestions improved the manuscript considerably.  The Tianlai array is operated with the support of NAOC Astronomical Technology Center. Work at UW-Madison and Fermilab is partially supported by NSF Award AST-1616554.  Fermilab is operated by Fermi Research Alliance, LLC, under Contract No. DE-AC02-07CH11359 with the US Department of Energy.  Work at UW-Madison is further supported by an NSF REU award, the Graduate School, the Thomas G. Rosenmeyer Cosmology Fund, and by student awards from the Wisconsin Space Grant.  Work at NAOC is supported by MOST grants 2016YFE0100300 and 2012AA121701, 2018YFE0120800, the NSFC grant 11633004, 11473044, 11761141012, 11653003, 11773031, CAS grant QYZDJ-SSW-SLH017, XDA15020200.  Part of the computation is performed on the Tianhe-2 supercomputer with the support of NSFC grant U1501501. Beam measurements with the UAV are supported by NSFC grant U1631118. 
Authors affiliated with French institutions acknowledge partial support from CNRS (IN2P3 \& INSU), Observatoire de Paris and from Irfu/CEA. 
This document was prepared by the Tianlai Collaboration includes personnel and uses resources of the Fermi National Accelerator Laboratory (Fermilab), a U.S. Department of Energy, Office of Science, HEP User Facility. Fermilab is managed by Fermi Research Alliance, LLC (FRA), acting under Contract No. DE-AC02-07CH11359.

\section*{Data Availability}
The data underlying this article will be shared on reasonable request to the corresponding author.


\bibliographystyle{mnras}
\bibliography{refs} 



\appendix

\section{Graphical Representation of Visibilities}
\label{app::ColorRepresentation}
We represent the complex visibilities graphically in terms of colors as illustrated in Figure A1. A visibility is a complex number depending on two real parameters, the real and imaginary part.  A complex number may be represented graphically with a single color since the human visual perceptual representation of colors is roughly a 3 dimensional space a 2 dimensional subspace of colors may be used.  The complex phase (or argument) of the visibility is of particular interest and exists on the unit circle which {``}wraps around{''} (every \(2\pi\)). The hue of a color also wraps red/yellow/green/cyan/blue/magenta/red so it is therefore natural to associate the complex phase with the hue. In the hue/saturation/brightness (HSB or HSV) representation of colors it is most natural to associate the magnitude of the visibility with the brightness. Hue zero is a red and we will use red to represent numbers with complex phase zero, i.e. positive real numbers. Black has zero brightness irrespective of the hue and we will use it to represent 0 whose argument is undefined.  There is no need to make use of the saturation. We generally use a linear relation between brightness and the modulus of the visibility adjusting the proportionality factor such that the smaller of the maximum modulus or 3 times the median modulus across the image saturates the brightness. Complex number with modulus greater than this maximum value will have saturated brightness. Thus this color representation has only a finite dynamic range for modulus. We find that this choice of saturation value results in a visually appealing representation of complex visibilities.

\begin{figure}
  \centering
  \includegraphics[width=0.47\textwidth]{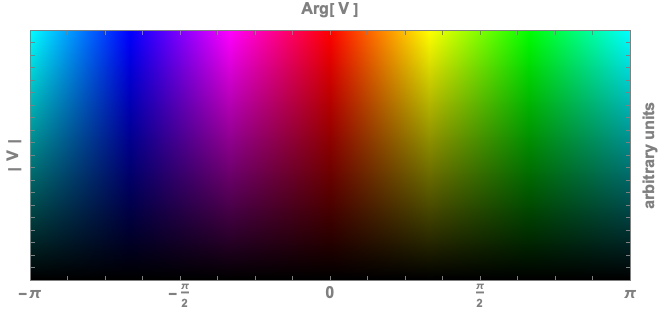}
\caption{The color palette used to represent complex visibilities in this paper.}
\label{fig:hsbrepresentation}
\end{figure}

\section{Absolute Calibration Considerations and Specifics}
\label{app::calibrators}

Absolute calibration is limited by:
\begin{itemize}
    \item {\bf statistical/systematic errors:} external measurements are not perfectly precise;
    \item {\bf spectral uncertainties:} external measurements occur in different frequency bands;
    \item {\bf variability:} external measurements were not taken at same time and fluxes may have varied;
    \item {\bf angular resolution:} external measurements are taken by interferometers with better angular resolution than Tianlai which cannot avoid blending nearby uncalibrated emission with a calibration source. 
\end{itemize}
If one is careful none of these is liable to be an important limitation, for the following reasons:
\begin{itemize}
    \item {\bf statistical/systematic errors:} these are small and have been quantified for the 4 brightest calibrators (Cyg~A, Cas~A, Tau~A = M1, Vir~A = M87) [\cite{Trotter2017}];
    \item {\bf spectral uncertainties:} we will use published interpolations between accurate well-sampled measurements [\cite{Perley_2017,Trotter2017}].  Given the smooth spectrum nature of most radio emission these interpolations are expected to be accurate.
    \item {\bf variability:} This is only an issue for variable sources, which should not include radio galaxies (Cyg~A, Vir~A = M87, Her A, 3C123). Varying flux can be stochastic (3C48) or fading (Cas~A, Tau~A [\cite{Baars1977,Reichart2000,Trotter2017}]). Temporal extrapolations of the measured fading flux of Cas~A and Tau~A [\cite{Trotter2017}] will be used and are expected to be accurate.  Our observations span years and we hope to directly measure the fading flux of Cas~A relative to that of Cyg~A.  Stochastic variability remains a worry especially since stochasticity is not necessarily known.
    \item {\bf angular resolution:} the low angular resolution of the Tianlai pathfinder means that most sources appear point-like so radio substructure is not an issue.  Contamination of the flux of neighboring sources is an issue for this low resolution telescope but less so for the brightest calibrators which are locally more dominant; Cyg~A and Cas~A in particular.  We plan to cross-calibrate the ``flux in beam'' of dimmer calibrators with the brightest ones.
\end{itemize}

Specifically we will use the spectral and temporal model fits to the flux densities of the four brightest calibrators from \cite{Trotter2017}: 
\begin{eqnarray}
f_\nu^\text{CygA}\hskip-10pt&=&\hskip-10pt
\frac{2896\,\text{Jy}}{f^{0.961}}
e^{ -0.0572\,(1+0.057\text{ln}f)\,(\text{ln}f)^2 } \nonumber \\
f_\nu^\text{CasA}\hskip-10pt&=&\hskip-10pt
\frac{2662\,\text{Jy}}{f^{0.7125}}
e^{ -0.00205\,(1+0.488\text{ln}f)\,(\text{ln}f)^2
-(1-0.142\,\text{ln}f)\frac{\Delta t_{2018}}{114\,\text{yr}} } 
\nonumber\\
f_\nu^\text{TauA}\hskip-10pt&=&\hskip-10pt
\frac{947\,\text{Jy}}{f^{0.227}}
e^{ -0.00658\,(1+0.789\text{ln}f)\,(\text{ln}f)^2 
-\frac{\Delta t_{2018}}{987\,\text{yr}} }\nonumber \\
f_\nu^\text{VirA}\hskip-10pt&=&\hskip-10pt
\frac{366\,\text{Jy}}{f^{0.867}}
e^{ 0.0077206\,(1-1.783\text{ln}f)\,(\text{ln}f)^2 }
\nonumber
\end{eqnarray}
These are expressed in terms of times and frequencies relevant to this paper: $f\equiv\frac{\nu}{750\,\text{MHz}}$ and $\Delta t_{2018}$ is the time since 2018-01-01. The Cas~A expression applies only after 1997-04-04.  This Cas~A flux model is used for the results in this paper.

\section{Map making tools used in this paper}
\label{annex:maps}

We give in this Appendix some details on the map making tools used in this paper. 
\subsection{\texttt{QuickMap}}
\texttt{QuickMap} \footnote{ \texttt{QuickMap}, \texttt{BFMTV} , m-mode map making and a set of simulation tools
  are grouped in the JSkyMap software package available from { \tt  https://gitlab.in2p3.fr/SCosmoTools/JSkyMap } }
is the simplest method used here and correspond to carry out beam forming using instantaneous visibility measurement. It is well suited to reconstructing sky maps using drift scan visibilities at mid latitudes. For each time step, a one dimensional map along
the declination direction, in the meridian plane is obtained by a linear combination of  all available baselines, including observations
at several adjacent declination. A map corresponding to a strip of sky along the right ascension direction is then obtained
by assembling these 1D maps which corresponds to beam forming. A brief description of this method can be found in 
\citep{PAON4_Zhang_2016}

\subsection{\texttt{BFMTV}}

\texttt{BFMTV} stands for {\bf B}rute {\bf F}orce {\bf M}ap making from {\bf T}ransit {\bf V}isibilities and is program able
to reconstruct an optimal map, from a set of time sampled visibilities obtained in drift scan mode. Indeed, assuming known beams associated to each feed, the observed set of time dependent visibilities are linearly related to the unknown sky pixels, and measurement
noise $n_{ij}^k$ for each frequency channel $\nu$. Arranging time dependent measured visibilities $\left[ {\cal V}_{ij}^k \right]$,
as well as sky pixels  $\left[ S^n \right] $as vectors, this linear relation can be written in matrix form:
\begin{eqnarray}
  \left[ {\cal V}_{ij}^k\right] & = & \left[ \left[ \mathbf{A} \right] \right] \times  \left[ S^n \right]  + \left[ n_{ij}^k\right]
\label{eq:mapmaking}                                      
\end{eqnarray}

The $ \left[ \left[ \mathbf{A} \right] \right] $ matrix, sometimes called the pointing matrix contains the array geometry,
individual feed angular response (beam) and the sky scanning strategy. It is possible to obtain an {\it optimal} (if all elements, e.g. pointing, beam, noise... are perfectly known) solution for the sky,
by solving the above equation, for example by computing the pseudo-inverse of the $  \mathbf{A} $ matrix, knowing the 
noise covariance matrix. However, the size of this matrix can easily reach $ 10^4 \times 10^4$, when processing a
small subset (~ 1 hour) of Tianlai dish array data, and $ 10^6 \times 10^5$ for the polar cap data, making this method
highly computing intensive for large data sets.

The m-mode map making addresses this problem by writing the large $ \left[ \left[ \mathbf{A} \right] \right] $ matrix
as a bloc diagonal matrix, taking advantage of azimuthal symmetry in spherical geometry and full 24 hours drift
scan observation in transit mode. However, due to the pollution of day time Tianlai data, we have not yet applied
the m-mode map making to the data set discussed here.

\subsection{\texttt{CASA}}

Considering planar sky geometry, valid for sky maps $S(\alpha, \delta)$ subtending a small solid angle, one can consider the
sky Fourier modes $ {\cal S}(u,v)$  where $(u,v)$ are the conjugate Fourier variables associated with the two angular
coordinates $(\alpha, \delta)$.  It is well known that a visibility from a baseline $(\Delta x, \Delta y)$
can be considered as the measurement of a specific Fourier sky mode
$(u_0,v_0) = \left( \frac{\Delta x}{\lambda} ,   \frac{\Delta y}{\lambda} \right) $.
A solution for the unknown sky  $S(\alpha, \delta)$, corresponding to the unknown vector $ \left[ S^n \right] $ in equation \ref{eq:mapmaking}
can be obtained by computing an inverse Fourier transform on the  Fourier modes corresponding to the observed visibilities.
This method, which can deal with large data sets thanks to the efficiency of FFT algorithms,
is implemented in many radio interferometer data processing packages, such as
\texttt{CASA}\footnote{Common Astronomy Software Applications package: \url{https://casa.nrao.edu/}}.
However, as many Fourier modes are often missing, the obtained sky maps, called dirty images present usually many artefacts.
A subsequent cleaning and deconvolution step is applied to obtain the clean image. A widely used \texttt{CLEAN}  method, 
is an iterative method, based on the identification of the brightest source at each step on the dirty image,
subtracting then its contribution (see for example \citep{CLEANCornwell} and \citep{CLEANHogbom}).

\subsection{\texttt{TLdishpipe}}

The \texttt{TLdishpipe} package is a Tianlai implementation of the above algorithm, although the CLEAN step has not yet been
developed. This is a fast interferometeric data imaging tool which performs  parallel processing of different baselines independently,
saving the intermediate data to disk. It reduces the requirements on the computing system, especially memory, and allows flexibility in combining different baselines. 
The calibrated visibility data are gridded in the $(u, v)$ plane by a two-dimensional histogram method and weighted by visibility.
Finally, the dirty image is obtained by the two-dimensional fast Fourier transform of the gridded visibility  data in the $(u,v)$ plane.


\bsp	
\label{lastpage}
\end{document}